\documentclass[useAMS,usenatbib]{mn2e}

\usepackage{epsfig}
\usepackage{amsmath}
\usepackage{amssymb}
\usepackage{ifthen}
\usepackage{txfonts}
\usepackage{rotating}
\usepackage{url}
\usepackage{varioref}
\usepackage{verbatim}
\usepackage{latexsym}
\usepackage{graphicx}

\DeclareSymbolFont{cmletters}{OML}{cmm}{m}{it}
\DeclareMathSymbol{v}{\mathalpha}{cmletters}{"76}

\usepackage{multirow}



\DeclareSymbolFont{cmletters}{OML}{cmm}{m}{it}

\DeclareMathSymbol{v}{\mathord}{cmletters}{"76}

\def\be{\begin{equation}}
\def\ee{\end{equation}}
\newcommand{\kb}{ k_{\rm B} }

\newcommand{\msun}{{\rm M_{\odot}}}

\newcommand{\bR}{{\bf{R}}}

\newcommand{\rholab}{{\rho}}

\newcommand{\uvec}{{{u}}}
\newcommand{\bvec}{{{b}}}
\newcommand{\Bvec}{{{B}}}

\newcommand{\xvec}{{\underline{x}}}

\newcommand{\alf}{Alfv\'en}

\newcommand{\erg}{{\rm\,erg}}

\newcommand{\gdet}{\sqrt{-g}}

\newcommand{\cut}[1]{\hbox{}}

\DeclareSymbolFont{cmletters}{OML}{cmm}{m}{it}
\DeclareMathSymbol{v}{\mathalpha}{cmletters}{"76}

\usepackage{subfigure}
\usepackage{ifthen}
\usepackage[usenames,dvipsnames]{color}
\usepackage{graphicx}

\newcommand\aj{\rmfamily{AJ}}%
\newcommand\araa{\rmfamily{ARA\&A}}%
\newcommand\apj{\rmfamily{ApJ}}%
\newcommand\apjl{\rmfamily{ApJ}}%
\newcommand\apjs{\rmfamily{ApJS}}%
\newcommand\apss{\rmfamily{Ap\&SS}}%
\newcommand\aap{\rmfamily{A\&A}}%
\newcommand\mnras{\rmfamily{MNRAS}}%
\newcommand\prd{\rmfamily{Phys.~Rev.~D}}%
\newcommand\pasj{\rmfamily{PASJ}}%
\newcommand\jqsrt{\rmfamily{J.~Quant.~Spec.~Radiat.~Transf.}}%
\defcitealias{bz77}{BZ77}

\newcommand{\emath}{{\rm e}}
\newcommand{\eff}{{\eta}}

\newcommand{\Qone}{Q_{\theta,\rm MRI}}
\newcommand{\Qx}{Q_{x,\rm MRI}}

\newcommand{\Qtwo}{{S_{d,\rm MRI}}}

\newcommand{\Qthree}{{Q_{\phi,\rm MRI}}}

\newcommand{\rhorest}{{\rho}} % rest-mass density
\newcommand{\ug}{{u_{\rm g}}} % gas internal energy density
\newcommand{\expf}{{\rm e}^{-\xi}}

\def\bE{\bar{E}}
\def\bR{\bar{R}}
\def\bu{\bar{u}}

\title[DC and Synchrotron in Super-Eddington Flows] % short
{Double Compton and Cyclo-Synchrotron in Super-Eddington Disks, Magnetized Coronae, and Jets} % long

\author[J.~C.~McKinney, J.~Chluba, M.~Wielgus, R.~Narayan, A.~Sadowski]
{Jonathan C. McKinney$^1$\thanks{\hbox{E-mail: jcm@umd.edu~(JCM)}},
Jens Chluba$^2$, Maciek Wielgus$^{3}$, Ramesh
Narayan$^4$,\newauthor Aleksander Sadowski$^5$,
\\
$^1$University of Maryland at College Park, Dept. of Physics, Joint Space-Science Institute, 3114 Physical Sciences Complex, College Park, MD 20742, USA\\
$^2$Jodrell Bank Centre for Astrophysics, University of Manchester, Oxford Road, Manchester M13 9PL, UK\\
$^3$Copernicus Astronomical Center, ul. Bartycka 18, PL 00-716 Warszawa, Poland\\
$^4$Harvard-Smithsonian Center for Astrophysics, 60 Garden St., Cambridge, MA 02134, USA\\
$^5$MIT Kavli Institute for Astrophysics and Space Research,77 Massachusetts Ave, Cambridge, MA 02139, USA
}

\begin{document}
\date{Accepted 2016.  Received 2016; in original form 2016.}
\pagerange{\pageref{firstpage}--\pageref{lastpage}} \pubyear{2016}
\maketitle

\newcommand{\bea}{\begin{eqnarray}}
\newcommand{\eea}{\end{eqnarray}}

\newcommand{\AS}[1]{\textbf{\color{Magenta}#1}}
\newcommand{\RN}[1]{\textbf{\color{Green}#1}}
\newcommand{\AT}[1]{\textbf{\color{Blue}#1}}
\newcommand{\YZ}[1]{\textbf{\color{Orange}#1}}
\newcommand{\koral}{\texttt{koral}}
\newcommand{\harmrad}{\texttt{harmrad}}
\newcommand{\harm}{\texttt{harm}}

\def\bE{\bar{E}}
\def\bR{\bar{R}}
\def\bu{\bar{u}}

\newcommand{\MBH}{{M}}
\newcommand{\MBHO}{{M_i}}
\newcommand{\Mdot}{{\dot{M}}}
\newcommand{\Mdotedd}{{\dot{M}_{\rm Edd}}}

\label{firstpage}

\begin{abstract}

We present an extension to the general relativistic radiation
magnetohydrodynamic code HARMRAD to account for emission and
absorption by thermal cyclo-synchrotron, double Compton,
bremsstrahlung, low-temperature OPAL opacities as well as Thomson and
Compton scattering. We approximate the radiation field as a
Bose-Einstein distribution and evolve it using the radiation
number-energy-momentum conservation equations in order to track photon
hardening. We perform various simulations to study how these
extensions affect the radiative properties of magnetically-arrested
disks accreting at Eddington to super-Eddington rates.  We find that
double Compton dominates bremsstrahlung in the disk within a radius of
$r\sim 15r_g$ (gravitational radii) at a hundred times the Eddington
accretion rate, and within smaller radii at lower accretion
rates. Double Compton and cyclo-synchrotron regulate radiation and gas
temperatures in the corona, while cyclo-synchrotron regulates
temperatures in the jet.  Interestingly, as the accretion rate drops
to Eddington, an optically thin corona develops whose gas temperature
of $T\sim 10^9$K is $\sim 100$ times higher than the disk's black body
temperature. Our results show the importance of double Compton and
synchrotron in super-Eddington disks, magnetized coronae, and jets.

\end{abstract}


\section{Introduction}

Black hole (BH) accretion flows become radiation-dominated and
geometrically thick once the luminosity $L\gtrsim 0.3 L_{\rm Edd}$,
where $L_{\rm Edd}\approx 1.3\times 10^{46} [M/(10^8\msun)]\,$erg/s is
the Eddington luminosity for solar mass $\msun$ and BH mass $M$.  Such
accretion flows tends to have temperatures of $10^9\gtrsim T\gtrsim
10^6$ Kelvin \citep{abr88}, where bremsstrahlung (free-free, $ep\to
e'p'\gamma$) is important.

Double (radiative) Compton (DC, $\gamma_1 e\to \gamma'_1\gamma_2 e'$)
\citep{1981ApJ...244..392L,1981MNRAS.194..439T,1983ASPRv...2..189P,1984MNRAS.209..175S}
can dominate free-free while regulating temperatures so that pairs
remain sub-dominant (see figure 1 in \citealt{1981MNRAS.194..439T}).
While double Compton has received attention in cosmology to determine
distortions to the cosmic microwave background
\citep{2012MNRAS.419.1294C,2014MNRAS.440.2544C} and been applied to
gamma-ray bursts \citep{2013ApJ...764..143V,2015ApJ...802..134B}, it
has received little attention in accretion theory except by
\citet{1996ApJ...470..249P}.

For photon number density $n_\gamma$, ion number density $n_i$,
electron mass $m_e$, speed of light $c$, Boltzmann's constant $k_b$,
and temperature $T$, the rates of each process show that DC dominates
free-free if
\begin{equation}\label{crit1}
\frac{n_\gamma}{n_i}\gtrsim 1.3\left(\frac{m_e c^2}{k_b T}\right)^{5/2} ,
\end{equation}
\citep{1981MNRAS.194..439T,1984MNRAS.209..175S}, which has the
correction by \citet{1984MNRAS.209..175S} of $1.3$ instead of the
value $0.1$ by \citet{1981MNRAS.194..439T}.  In this paper, we find DC
dominates free-free for the mean opacity when
\begin{equation}
T > 7\times 10^7{\rm K} \rho^{2/11}
\end{equation}
where $\rho$ is mass density in cgs units.  For BH X-ray binaries
accreting at ten times the Eddington rate, one has $\rho\approx
10^{-3}{\rm g}{\rm cm}^{-3}$ giving equality at $T\approx 2\times
10^7$K, while simulations show that $T\sim 5\times 10^7$--$10^8$K
\citep{2015arXiv150804980S} indicating double Compton dominates
free-free even if not yet included. For super-massive BHs relevant for
tidal disruption events (TDEs) accreting at the Eddington rate, one
has $\rho\sim 10^{-9}{\rm g}{\rm cm}^{-3}$ giving equality at
$T\approx 1.4\times 10^6$K, which is reached within tens of
gravitational radii.

Still, it remains uncertain which opacity dominates the mean flow
behavior.  The Novikov-Thorne solution \citep{1973blho.conf..343N} or
slim disk solution \citep{abr88} could be used as a guide to how
important DC is, but we find the temperature sensitivity makes the
conclusions not robust -- especially as the slim disk solution is
modified by winds, vertical structure, and magnetic fields.  Also,
unlike free-free, double Compton is entirely dependent upon the
radiation temperature for non-relativistic electrons, so non- local
thermodynamic equilibrium (LTE) effects are important to treat. This
motivates using simulations to consider both opacities with separate
gas and radiation temperatures.

In addition, cyclo-synchrotron (hereafter, synchrotron) is expected to
be an important source of opacity or emission for low-luminosity
accretion flows or in regions where non-thermal electrons are present.
In super-Eddington accretion flows, one expects synchrotron to be
unimportant in the disk.  However, the atmosphere above the disk
(i.e. corona) and jet can contain a strong magnetic field at low
densities where synchrotron can dominate
\citep{1997MNRAS.291..805D,2011PhPl...18d2105U}.  In addition,
magnetically-arrested accretion disks (MADs) are much more strongly
magnetized than weakly magnetized disks that are usually considered
\citep{2003ApJ...592.1042I,2003PASJ...55L..69N,2011MNRAS.418L..79T}.
Magnetically-supported atmospheres can harden the spectrum by
producing a more extended vertical disk \citep{2006ApJ...645.1402B},
and synchrotron can provide an abundant source of low-energy soft
photons that can undergo inverse Compton before being absorbed. So the
magnetic field strength can regulate photon hardening in disk
atmospheres and magnetized jets.

In this paper, we consider the competition between free-free,
bound-free, bound-bound opacities from OPAL
\citep{1996ApJ...464..943I}, double Compton, and synchrotron in MAD
type black hole accretion flows.  We also consider separate absorption
and emission mean opacities to focus on emission rates in the energy
and momentum equations, rather than the Rosseland mean that only
applies in the diffusion limit for the momentum equation.  The
absorption mean handles the effect of irradiation and how the
absorption of radiation is affected by the photon distribution being
different than that given by Planck at the local temperature
\citep{2016arXiv160503184S}. Such non-LTE effects lead to a
significant correction to the mean opacities and can qualitatively
change the outcome \citep{2003ApJ...594.1011H}. Additionally, we
evolve the photon number separately from the photon energy
\citep{2015arXiv150804980S}, with separate number and energy
opacities, in order to track photon hardening due to inverse
Comptonization. The photon distribution is assumed to be
Bose-Einstein, but with a low-energy transition to Planck when
absorption is faster than inverse Compton.

These physical effects are included within our general relativistic
(GR) radiative magnetohydrodynamics (MHD) (GRRMHD) code HARMRAD
\citep{2014MNRAS.441.3177M} that uses the M1 closure.  Similar GR
\citep{2012ApJS..201....9F,2013ApJ...764..122T,2014MNRAS.439..503S,2014MNRAS.441.3177M,2014ApJ...796...22F,2015ApJ...807...31R,2016ApJ...826...23T}
and non-GR \citep{2014ApJ...796..106J} schemes have been developed.

In \S\ref{sec:eom}, we outline our equations of motion, in
\S\ref{sec:sims} we present our GRRMHD simulations, and in
\S\ref{sec:summary} we summarize our results.  For the appendices, in
\S\ref{sec:be}, we discuss our use of the Bose-Einstein distribution,
in \S\ref{meanopacities} we discuss how we compute the mean opacities,
in \S\ref{sec:ff} we discuss free-free and related low-temperature
opacities, in \S\ref{sec:synch} we discuss synchrotron opacities, and
in \S\ref{sec:compton} we discuss Compton scattering and double
Compton opacities.

\newcommand{\expfun}[1]{{{\rm e}^{#1}}}

\newcommand{\Op}[1]{\hat{\bf #1}}

\newcommand{\nmi}{n_{-}}

\newcommand{\npl}{n_{+}}

\newcommand{\ngamma}{n_\gamma}

\newcommand{\vrel}{|v_{\rm rel}|}

\newcommand{\xt}{\tilde{x}}

\newcommand{\xe}{x_{\rm e}}

\newcommand{\xc}{x_{\rm c}}

\newcommand{\Absatz}{\vspace{2ex}}

\newcommand{\id}{{\,\rm d}}

\newcommand{\beq}{\begin{equation}}   %

\newcommand{\eeq}{\end{equation}}   %

\newcommand{\vc}[1]{\ensuremath{\mathbf{#1}}}

\newcommand{\Div}[1]{\vek{\nabla} \! \cdot \vek{#1}}   %

\newcommand{\Rot}[1]{\vek{\nabla} \times \vek{#1}}

\newcommand{\beqa}{\begin{eqnarray}}   %

\newcommand{\eeqa}{\end{eqnarray}}   %

\newcommand{\beal}{\begin{align}}

\newcommand{\enal}{\end{align}}

\newcommand{\bspl}{\begin{split}}

\newcommand{\espl}{\end{split}}

\newcommand{\bsub}{\begin{subequations}}

\newcommand{\esub}{\end{subequations}}

\newcommand{\bmulti}{\begin{multline}}   %

\newcommand{\beqm}{\begin{mathletters}}   %

\newcommand{\eeqm}{\end{mathletters}}   %

\newcommand{\Doll}[1]{\bf#1\rm}

\newcommand{\Glossar}[2]{\bf#1 \rm\quad #2}

\newcommand{\Abkuerz}[1]{#1}

\newcommand{\Abschnitt}[1]{\subsection*{#1}}

\newcommand{\Abst}[1]{\,#1}

\newcommand{\kB}{k_{\rm B}}

\newcommand{\me}{m_{\rm e}}

\newcommand{\Ne}{n_{\rm e}}

\newcommand{\Np}{n_{\rm p}}

\newcommand{\Nb}{n_{\rm b}}

\newcommand{\Te}{T_{\rm e}}

\newcommand{\Tg}{T_{\gamma}}

\newcommand{\The}{\theta_{\rm e}}

\newcommand{\Thg}{\theta_{\gamma}}

\newcommand{\mprot}{m_{\rm p}}

\newcommand{\sigT}{\sigma_{\rm T}}

\newcommand{\sigKN}{\sigma_{\rm K-N}}

\newcommand{\rC}{r_{\rm c}}

\newcommand{\nPl}{n_{\rm Pl}}

\newcommand{\nBE}{n_{\rm BE}}

\newcommand{\nW}{n_{\rm W}}

\newcommand{\rhoPl}{\rho_{\rm Pl}}

\newcommand{\rhoBE}{\rho_{\rm BE}}

\newcommand{\NPl}{N_{\rm Pl}}

\newcommand{\NBE}{N_{\rm BE}}

\newcommand{\vek} [1]{\mbox{\boldmath${#1}$\unboldmath}}

\newcommand{\matr} [1]{\mbox{$\hat{#1}$}}

\newcommand{\SP}[2]{\mbox{$\vek{#1}\cdot\vek{#2}$}}

\newcommand{\pd}{\partial}

\newcommand{\pAb}[2]{\frac{\displaystyle\pd #1}{\displaystyle\pd #2}}

\newcommand{\PAb}[3]{\frac{\displaystyle\pd^{#3} #1}{\displaystyle\pd {#2}^{#3}}}

\newcommand{\tAb}[2]{\frac{\displaystyle d #1}{\displaystyle d #2}}

\newcommand{\TAb}[3]{\frac{\displaystyle d^{#3} #1}{\displaystyle d {#2}^{#3}}}

\newcommand{\Abl}[2]{\frac{{\rm d} #1}{{\rm d} #2}}

\newcommand{\AKc}[1]{{\rm C}_{#1}}

\newcommand{\AKs}[1]{{\rm S}_{#1}}

\newcommand{\sgn}[1]{\,{\rm sgn}\,{#1}}

\newcommand{\Mitw}[1]{\left<#1\right>}

\newcommand{\Vv}[1]{\mbox{${#1}$}}

\newcommand{\Vvek}[2]{\mbox{${#1}^{#2}$}}

\newcommand{\VvN}[1]{\mbox{${#1}_{0}$}}

\newcommand{\VvNN}[1]{\mbox{${#1}_{0}^2$}}

\newcommand{\VSP}[2]{\mbox{$\Vv{#1}\Vv{#2}$}}

\newcommand{\STV}[1]{\mbox{$\tilde{#1}$}}

\newcommand{\figref}[1]{Abb.\ref{#1}}

\newcommand{\tabref}[1]{Tab.\ref{#1}}

\newcommand{\kapref}[1]{Kap.\ref{#1}}

\newcommand{\deltaD}[1]{\mbox{$\delta( {#1} )$}}

\newcommand{\Aut}[1]{\it{#1}\rm }

\newcommand{\AutP}[1]{\it{#1}\rm }

\newcommand{\AutPet}[1]{\it{#1 et al.}\rm }

\newcommand{\AutPno}[1]{{#1}\rm }

\newcommand{\AutPnoet}[1]{{#1 et al.}\rm }

\newcommand{\Vvekm}[1]{\mbox{$\Vvek{#1}{\mu}$}}

\newcommand{\Vvekcv}[2]{\mbox{${#1}_{#2}$}}

\newcommand{\Vvekcvm}[1]{\mbox{$\Vvekcv{#1}{\mu}$}}

\newcommand{\pot}[2]{#1 \times 10^{#2}}

\newcommand{\ngam}{n_{\gamma}}

\newcommand{\tC}{t_{\rm C}}

\newcommand{\tK}{t_{\rm K}}

\newcommand{\tge}{t_{\gamma \rm e}}

\newcommand{\teg}{t_{\rm e\gamma}}

\newcommand{\texp}{t_{\rm exp}}

\newcommand{\ThCMB}{\Theta_{2.7}}
\newcommand{\Thz}{\theta_{z}}
\newcommand{\Yp}{Y_{\rm p}}

\newcommand{\Obhh}{\Omega_{\rm b}h^2}

\newcommand{\Omhh}{\Omega_{\rm m}h^2}

\newcommand{\zeq}{z_{\rm eq}}

\section{Equations of Motion}\label{sec:eom}

The conservation laws are
\begin{eqnarray}
\label{eq.rhocons} \hspace{1in}(\rho
u^\mu)_{;\mu}&=&0,\\\label{eq.tmunucons}
\hspace{1in}(T^\mu_\nu)_{;\mu}&=&G_\nu,\\\label{eq.rmunucons}
\hspace{1in}(R^\mu_\nu)_{;\mu}&=&-G_\nu,
\end{eqnarray}
where $\rho$ is the gas density in the comoving fluid frame, $u^\mu$
is the gas four-velocity as measured in the ``lab frame'', and
$T^\mu_\nu$ is the MHD stress-energy tensor in this frame,
\begin{equation}
\label{eq.tmunu} T^\mu_\nu = (\rho+u_{\rm g}+p_{\rm g}+b^2)u^\mu
u_\nu + (p_{\rm g}+\frac12b^2)\delta^\mu_\nu-b^\mu b_\nu,
\end{equation}
$R^\mu_\nu$ is the stress-energy tensor of radiation, $G_\nu$ is the
radiative four-force describing the interaction between gas and
radiation, $u_{\rm g}$ and $p_{\rm g}$ are, respectively, the internal
energy and pressure of the gas in the comoving frame, and $b^\mu$ is
the magnetic field 4-vector \citep{2003ApJ...589..444G}. The magnetic
pressure is $p_b=b^2/2$ in our Heaviside-Lorentz units.  The ideal
induction equation and entropy equations are also used
\citep{2014MNRAS.441.3177M}. We assume collisions keep
gas and electron temperatures similar \citep{2016arXiv160503184S}.

\subsection{Radiative four-force}

We use a covariant formalism for computing the interaction due to absorption,
emission, and scattering \citep{2014MNRAS.439..503S,2014MNRAS.441.3177M}
via a 4-force between the gas and radiation of
\begin{equation} \label{eq.Gcon}
G^\mu = - (\kappa_{\rm a}+\kappa_{\rm s}) R^{\mu\nu} u_\nu -\left(\kappa_{\rm
  s} R^{\alpha\beta} u_\alpha u_\beta + \lambda\right) u^\mu,
\end{equation}
where $\kappa_{\rm a}$ is the energy absorption mean opacity in the
fluid frame (in which ions and electrons are isotropic), $\lambda$
is the fluid-frame total energy density loss rate before absorption
(which includes changes in energy while conserving photon number),
and $\kappa_{\rm s}$ is the fluid-frame energy scattering mean
opacity.

\subsection{Photon Number Evolution}

If the radiation is Planckian, then the chemical potential $\mu=0$ and
the radiation temperature is derived from only the fluid-frame
radiation energy density via $T_\gamma = (E/a)^{1/4}$ where
$E=R^{\mu\nu} u_\mu u_\nu$, which can differ from the gas temperature
$T_{\rm g}$ or electron temperature $T_{\rm e}$.

A more general photon distribution, like our choice of a Bose-Einstein
distribution (see \S\ref{sec:be}), can be considered by
simultaneously evolving the number density of photons in the radiation
frame ($n_r$, for radiation isotropic in the frame with 4-velocity
$u_r^\mu$) via
\begin{equation}
(n_r u_r^\mu)_{;\mu} = \dot{n} ,
\end{equation}
where the fluid-frame number density of photons is $n = n_r (-u_r^\mu
u_\mu)$ and $\dot{n}_r=\dot{n}$ by Lorentz invariance, with
\begin{equation}\label{dotn}
\dot{n} = -(\kappa_{na}c) n - \lambda_n ,
\end{equation}
for a number absorption mean opacity $\kappa_{na}$ and total number
emission rate $\lambda_n$ before absorption.  This photon number
conservation equation approximates the Kompaneets equation most
accurately when thermal Comptonization is included and when gas and
radiation temperatures are similar \citep{2015arXiv150804980S}.

\subsection{Opacities and Rates}\label{sec:overall}

We obtain fits to several different energy-weighted and
number-weighted opacities (computed as in \S\ref{meanopacities}),
including OPAL opacities for solar abundances (energy mean
$\kappa_{\rm a,eff}$ and number mean $\kappa_{\rm an,eff}$)
in \S\ref{sec:ff}, synchrotron (energy mean $\kappa_{\rm a,syn}$ and
number mean $\kappa_{\rm an,syn}$) in \S\ref{sec:synch}, and Compton
scattering (scattering energy mean $\kappa_{\rm s}$, thermal
Comptonization energy exchange rate $\lambda_c$, and double Compton
energy mean opacity $\kappa_{\rm a,dc}$ and number mean opacity
$\kappa_{\rm an,dc}$) in \S\ref{sec:compton}.

Then, the energy mean energy absorption opacity is
\begin{equation}
\kappa_a = \kappa_{\rm a,eff} + \kappa_{\rm a,syn} + \kappa_{\rm a,dc} ,
\end{equation}
and the number mean number absorption opacity is
\begin{equation}
\kappa_{an} = \kappa_{\rm an,eff} + \kappa_{\rm an, syn} + \kappa_{\rm an, dc} .
\end{equation}

From these opacities, the energy and number emission rates before any
absorption are obtained in \S\ref{emission}, giving energy emission
rate $\lambda_e$ and number emission rate $\lambda_n$ such that
\begin{equation}
\lambda = \lambda_e + \lambda_{\rm c} ,
\end{equation}
and $\lambda_n$ includes all number emission.

We do not include thermal pairs relevant when $k_b T_{\rm e}\gtrsim
m_e c^2$ as only the jet can be that hot and there one requires pair
production (see, e.g., appendix B in \citealt{2012MNRAS.419..573M}).

\section{Simulations}\label{sec:sims}

Our GRRMHD simulations evolve accretion flows around black holes with
black hole mass of $\MBH=10M_{\odot}$.  All models have identical
initial conditions, except some have different initial density values
due to an overall rescaling of the density in order to vary the final
mass accretion rate $\dot{M}$.  For different models, we vary the
choices of opacity and how the temperature of radiation is evolved,
and we investigate how the opacity and radiation temperature evolution
affect the accretion flow behavior and properties.

\subsection{Diagnostics}
\label{sec:diagnostics}

For various quantities $R$, we consider time-averages ($[R]_t$),
spatial averages, and their spatial distributions.  Diagnostics are
computed from snapshots produced every $\sim 4r_g/c$ for $r_g\equiv
GM/c^2$ with gravitational constant G.

\subsubsection{Fluxes}

The stress energy tensor $T^\mu_\nu$ includes both matter (MA)
and electromagnetic (EM) terms:
\begin{eqnarray}\label{MAEM}
{T^{\rm MA}}^\mu_\nu &=& (\rho + u_{\rm g} + p_{\rm g} ) \uvec^\mu \uvec_\nu + p_{\rm g} \delta^\mu_\nu \nonumber ,\\
{T^{\rm EM}}^\mu_\nu &=& b^2 \uvec^\mu \uvec_\nu + p_b\delta^\mu_\nu - \bvec^\mu \bvec_\nu ,
\end{eqnarray}
Here, $u_{\rm g}$ is the internal energy density and $p_{\rm g}=(\Gamma-1)u_{\rm g}$ is
the ideal gas pressure with adiabatic index $\Gamma$. The
contravariant fluid-frame magnetic 4-field is given by $\bvec^\mu$,
which is related to the lab-frame 3-field via $\bvec^\mu = \Bvec^\nu
h^\mu_\nu/\uvec^t$ where $h^\mu_\nu = \uvec^\mu \uvec_\nu +
\delta^\mu_\nu$ is a projection tensor, and $\delta^\mu_\nu$ is the
Kronecker delta function.  The magnetic energy density ($u_b$) and
pressure ($p_b$) are $u_b=p_b=\bvec^\mu \bvec_\mu/2 = b^2/2$.  The
total pressure is $p_{\rm tot} = p_{\rm g} + p_b$, and plasma $\beta\equiv
p_{\rm g}/p_b$.

The gas rest-mass flux, specific energy flux, and specific angular momentum
flux are respectively given by
\begin{eqnarray}\label{Dotsmej}
\Mdot &=&  \left|\int\rho \uvec^r dA_{\theta\phi}\right| , \\
\emath \equiv \frac{\dot{E}}{[\Mdot]_t} &=& -\frac{\int (T^r_t+R^r_t) dA_{\theta\phi}}{[\Mdot]_t} , \\
\jmath \equiv \frac{\dot{J}}{[\Mdot]_t} &=& \frac{\int (T^r_\phi+R^r_\phi) dA_{\theta\phi}}{[\Mdot]_t} ,
\end{eqnarray}
where $dA_{\theta\phi} = \gdet dx^{(2)} dx^{(3)}$ for metric
determinant $g={\rm Det}(g_{\mu\nu})$ and uniform code coordinates
with spacing $dx^{(1)}$, $dx^{(2)}$, and $dx^{(3)}$ in the
radial-like, $\theta$-like, and $\phi$-like directions.

The net flow efficiency is given by
\begin{equation}\label{eff}
  \eff = \frac{\dot{E}-\Mdot}{[\Mdot]_t} , \\
\end{equation}
Positive values correspond to an extraction of positive energy from
the system at some radius.  The jet efficiency $\eta_{\rm j}$ includes
all non-radiative terms in $\dot{E}$.

The magnetic flux is given in a normalized form as
\begin{equation}\label{equpsilon}
\Upsilon_{\rm H} \approx 0.7\frac{\int 0.5|B^r| dA_{\theta\phi}}{\sqrt{[\Mdot]_t}} ,
\end{equation}
which accounts for $B^r$ being in Heaviside-Lorentz units
\citep{1999ApJ...522L..57G,pmntsm10}.

\subsubsection{Inflow Equilibrium}

Inflow equilibrium is defined as when the flow is in a complete
quasi-steady-state and the accretion fluxes are constant (apart from
noise) vs. radius and time.  The inflow equilibrium timescale is
\begin{equation}\label{tieofrie}
t_{\rm ie} = N \int_{r_i}^{r_{\rm ie}} dr\left(\frac{-1}{[\langle v_r\rangle_{\rholab}]_t}\right) ,
\end{equation}
where $\langle v_r\rangle_{\rholab}$ is the $\rholab$ weighted radial
velocity, and $N\sim 3$ inflow times from $r=r_{\rm ie}$ and
$r_i=12r_g$.

Viscous theory gives a GR $\alpha$-viscosity estimate for $v_r$ of
$v_{\rm visc}\sim -Q\alpha(H/R)^2 |v_{\rm rot}|$ for rotational
velocity $v_{\rm rot}$, disk thickness $H/R$, and GR correction $Q$
\citep{pt74,pmntsm10}, so we can define an effective $\alpha$
viscosity as
\begin{equation}\label{alphaeq}
\alpha_{\rm eff} \equiv \frac{v_r}{v_{\rm visc}/\alpha} .
\end{equation}
All our models have $\alpha_{\rm eff}\sim 1$ in the quasi-steady
state, as expected for MADs \citep{2012MNRAS.423.3083M}.

\subsubsection{Optical Depth and Radiative Quantities}
\label{opticaldepth}

The scattering optical depth is computed as
\begin{equation}\label{tau}
\tau_{\rm sca}\approx \int \kappa_{\rm s} dl ,
\end{equation}
while the effective optical depth for absorption is computed as
\begin{equation}\label{taueff}
\tau_{\rm eff}\approx \int \sqrt{3\kappa_{\rm a}(\kappa_{\rm s}+\kappa_{\rm a})} dl .
\end{equation}
For the radial direction, $dl=- f_\gamma dr$, $f_\gamma \approx u^t (1
- (v/c)\cos\theta)$, $(v/c)\approx 1-1/(u^t)^2$ (as valid at large
radii), $\theta=0$, and the integral is from $r_0=4000r_g$ to $r$ to
obtain $\tau_r(r)$.  For the angular direction, $dl=f_\gamma r
d\theta$, $\theta=\pi/2$, and the integral is from each polar axis
toward the equator to obtain $\tau_\theta(\theta)$.  Note that the
photospheres in the polar jet region are strongly determined by the
numerical density floors assumed.

To scale $\Mdot c^2$ or a luminosity $L$, one can use the Eddington
luminosity
\begin{equation}\label{Ledddef}
L_{\rm Edd}=\frac{4\pi G \MBH c}{\kappa_{\rm es}} \approx 1.3\times
10^{46} \frac{\MBH}{10^8M_{\odot}}{\erg~{\rm s}^{-1}} ,
\end{equation}
for Thomson electron scattering opacity $\kappa_{\rm es}$. One can
also choose to normalize $\Mdot$ by $\Mdotedd = (1/\eta_{\rm
NT})L_{\rm Edd}/c^2$, where $\eta_{\rm NT}$ is the nominal accretion
efficiency for the Novikov-Thorne thin disk solution
\citep{1973blho.conf..343N} (commonly, a fixed $\eta_{\rm
  NT}=0.1$ is used, but we include the spin dependence).

The radiative luminosity is computed as
\begin{equation}
L_{\rm rad} = -\int dA_{\theta\phi} R^r_t ,
\end{equation}
which is measured just beyond the scattering photosphere to give the
quantity we call $L_{\rm rad,o}$.  The radiative efficiency is
$\eta_{\rm rad, o} = L_{\rm rad,o}/[\Mdot]_t$.  From any cumulative
luminosity $L(\theta)$, we can compute the isotropic equivalent
luminosity
\begin{equation}
L_{\rm iso}(\theta) = \pi \partial_\theta L(\theta)
\end{equation}
and the corresponding beaming factor
\begin{equation}
b = \frac{L_{\rm iso}}{L} .
\end{equation}

The fluid-frame radiation temperature is
\begin{equation}
\hat{T}_\gamma \approx \frac{E}{n \bar{E}_0} ,
\end{equation}
with $\bar{E}_0\approx 2.701\kb$, which is to within 10\% of the full
Bose-Einstein formula given by Eq.~(\ref{tgamma}) that we actually
use. The lab-frame radiation temperature is
\begin{equation}
T_\gamma \approx \frac{-R^t_t}{n_r u_r^t \bar{E}_0} .
\end{equation}
The black-body assumption gives a fluid-frame temperature of
\begin{equation}
\hat{T}_{\rm BB} = \left(\frac{E}{a_{\rm rad}}\right)^{1/4} ,
\end{equation}
for radiation constant $a_{\rm rad}$, while in the lab-frame the
assumption of Planck gives
\begin{equation}
T_{\rm BB} = \left(\frac{-R^t_t}{a_{\rm rad}}\right)^{1/4} .
\end{equation}
From these one can compute the lab-frame photon hardening factor
\begin{equation}
f_{\rm col} = \frac{T_\gamma}{T_{\rm BB}} ,
\end{equation}
and the fluid-frame hardening factor
\begin{equation}
\hat{f}_{\rm col} = \frac{\hat{T}_\gamma}{\hat{T}_{\rm BB}} .
\end{equation}

\subsubsection{Numerical Diagnostics}
\label{sec:numdiagnostics}

The magneto-rotational instability (MRI) is a linear instability
with fastest growing wavelength of
\begin{equation}\label{lambdamri}
\lambda_{x,\rm MRI} \approx  2\pi \frac{|v_{x,\rm A}|}{|\Omega_{\rm rot}|} , \\
\end{equation}
for $x=\theta,\phi$, where $|v_{x,\rm A}|=\sqrt{\bvec_x
  \bvec^x/\epsilon}$ is the $x$-directed~\alf~speed, $\epsilon\equiv
b^2 + \rhorest + u_{\rm g} + p_{\rm g}$, and $r\Omega_{\rm rot} =
v_{\rm rot}$.  $\Omega_{\rm rot},v_{\rm A}$ are separately
angle-volume-averaged at each $r,t$.

The MRI is resolved for grid cells per wavelength
(Eq.~(\ref{lambdamri})),
\begin{equation}\label{q1mri}
\Qx \equiv \frac{\lambda_{x,\rm MRI}}{\Delta_{x}} ,
\end{equation}
of $\Qx\ge 6$, for $x=\theta,\phi$, where $\Delta_{r} \approx
d\xvec^{(1)} (dr/d\xvec^{(1)})$, $\Delta_{\theta} \approx r
d\xvec^{(2)} (d\theta/d\xvec^{(2)})$, and $\Delta_{\phi} \approx r
\sin\theta d\xvec^{(3)} (d\phi/d\xvec^{(3)})$.  Volume-averaging is
done as with $\Qtwo$, except $v_{x,\rm A}/\Delta_{x}$ and
$|\Omega_{\rm rot}|$ are separately $\theta,\phi$-volume-averaged
before forming $\Qx$.  At $t=0$ all our models have $\Qone\sim 40$ and
is constant all along the disk, while in a steady-state all our models
have $\Qone\sim 100$ and $\Qthree\sim 20$ (except model M13 that is
quite thin and has $\Qthree\sim 6$ and so probably has somewhat
underdeveloped MRI turbulence).

The MRI suppression factor corresponds to the number of MRI
wavelengths across the full disk:
\begin{equation}\label{q2mri}
\Qtwo \equiv \frac{2 r (H/R)}{\lambda_{\theta,\rm MRI}} .
\end{equation}
Wavelengths $\lambda<0.5\lambda_{\theta,\rm MRI}$ are stable, so the
linear MRI is suppressed for $\Qtwo<1/2$ when no unstable wavelengths
fit within the full disk \citep{1998RvMP...70....1B,pp05}.  $\Qtwo$
uses averaging weight $w=(b^2\rholab)^{1/2}$, condition $\beta>1$, and
excludes regions where density floors are activated. When computing
the averaged $\Qtwo$, $v_{\rm A}$ and $|\Omega_{\rm rot}|$ are
separately $\theta,\phi$-volume-averaged within $\pm 0.2r$ for each
$t,r$.  The $\Qtwo\sim 0.5$ at $t=0$, while in quasi-steady-state
$\Qtwo\sim 0.1$.  So the field strength has increased considerably due
to magnetic flux accumulation.  All models are MAD with $\Qtwo<1/2$
out to $r\sim 30r_g$.

The flow structure can also be studied by computing the correlation
length scale and then computing how many grid cells cover each
self-correlated piece of turbulence.  We follow our prior works
\citet{2012MNRAS.423.3083M} and \citet{2014MNRAS.441.3177M} and
compute this.  One would desire to have at a minimum $6$ grid cells
per correlation length scale, since otherwise uncorrelated parts of
turbulence are not independently resolved by our piece-wise parabolic
monotonicity-preserving (PPM) scheme that needs $6$ grid cells to
resolve a structure well.  All our models have $\approx 12$ grid cells
per vertical and radial correlation length for density and magnetic
field strength, while our moderate resolution models have $\approx 6$
cells in the $\phi$-direction per correlation length scale.  Our
survey models have only $\approx 3$ cells in the $\phi$-direction per
correlation length.  This means the survey models should be considered
not resolved enough to demonstrate fully resolved turbulence, but they
are still sufficiently interesting to identify what physical effects
(being switch on/off) could be important in fully resolved
simulations.  In addition, we compare some survey models against
moderate resolution versions to confirm the survey models are
reasonable.

\begin{table*}
\caption{Spin, Mass Accretion Rate, Opacity, and Temperature Choices}
\begin{center}
\begin{tabular}[h]{|l||rr|r|r|r|r|}
\hline
Model     &  $a/M$ & $\frac{\dot{M}_{\rm{}H}}{\dot{M}_{\rm Edd}}$ & Opacities            & Radiation Number Density & Radiation Temperature      & Chemical Potential  \\
\hline
M1   &  $0.8$ & 140 & OPAL+Syn+DC    &  Evolved                 & $E/(n \bar{E}_0)$             &  1             \\  %  jonharmrad1
M2   &  $0.8$  & 140 & OPAL           &  Evolved                 & $E/(n \bar{E}_0)$             &  1             \\  %  jonharmrad2
M3   &  $0.8$  & 80 & OPAL           &  Evolved                 & Bose-Einstein                              &  Bose-Einstein \\  %  jonharmrad3
M5   &  $0.8$  & 120 &  OPAL           &  Planck at $\hat{T}_\gamma$    & Planck at $\hat{T}_\gamma$                      &  1             \\  %  jonharmrad5
M6   &  $0$  & 5.9 & OPAL (no TC)   &  Planck at $\hat{T}_\gamma$    & Planck at $\hat{T}_\gamma$                      &  1             \\  %  jonharmrad18
M7   &  $0$  & 4.8 & OPAL           &  Evolved                 & $E/(n \bar{E}_0)$             &  1             \\  %  jonharmrad7
M8   &  $0$  & 5 & OPAL+Syn+DC    &  Evolved                 & $E/(n \bar{E}_0)$             &  1             \\  %  jonharmrad8
M9   &  $0.8$  & 50 & OPAL+Syn+DC    &  Evolved                 & Bose-Einstein                             &  Bose-Einstein \\  %  jonharmrad9
M10  &  $0.8$  & 27 & OPAL+DC        &  Evolved                 & Bose-Einstein                             &  Bose-Einstein \\  %  jonharmrad10
M11  &  $0.8$  & 36 & OPAL+Syn+DC    &  Evolved                 & Bose-Einstein                             &  Bose-Einstein \\  %  jonharmrad11
M13  &  $0.8$  & 1.2 & OPAL+Syn+DC    &  Evolved                 & Bose-Einstein                             &  Bose-Einstein \\  %  jonharmrad13
M14  &  $0.8$  & 3.5 & OPAL+Syn+DC    &  Evolved                 & Bose-Einstein                             &  Bose-Einstein \\  %  jonharmrad14
M14h  &  $0.8$ & 2.4 & OPAL+Syn+DC    &  Evolved                 & Bose-Einstein                             &  Bose-Einstein \\  %  jonharmrad17
M15  &  $0.8$  & 14 & OPAL+Syn+DC    &  Evolved                 & Bose-Einstein                             &  Bose-Einstein \\  %  jonharmrad15
M15h  &  $0.8$  & 31 & OPAL+Syn+DC    &  Evolved                 & Bose-Einstein                             &  Bose-Einstein \\  %  jonharmrad16
\hline \hline
\end{tabular}
\end{center}
\label{tblinit}
\end{table*}

\subsection{Initial Conditions}

The initial disk is Keplerian with a rest-mass density that is
Gaussian in angle with a height-to-radius ratio of $H/R\approx 0.2$
and radially follows a power-law of $\rhorest\propto r^{-0.6}$.  The
solution near and inside the inner-most stable circular orbit (ISCO)
is not an equilibrium, so near the ISCO the solution is tapered to a
smaller density ($\rhorest\to \rhorest (r/15)^7$, within $r=15r_g$)
and a smaller thickness ($H/R\to 0.2 (r/10)^{0.5}$, within $r=10r_g$
-- based upon a low-resolution simulation).  The total internal energy
density $u_{\rm tot}$ is estimated from vertical equilibrium of
$H/R\approx c_s/v_K$ for sound speed $c_s\approx \sqrt{\Gamma_{\rm
    tot} P_{\rm tot}/\rhorest}$ with $\Gamma_{\rm tot}\approx 4/3$ and
Keplerian speed $v_K\approx (r/r_g)/((r/r_g)^{3/2}+a/M)$.  The total
ideal pressure $P_{\rm tot} = (\Gamma_{\rm tot}-1) u_{\rm tot}$ is
randomly perturbed by $10\%$ to seed the MRI.  The disk gas has
$\Gamma_{\rm gas}=5/3$.  The disk has an atmosphere with
$\rhorest=10^{-5} (r/r_g)^{-1.1}$ and gas internal energy density
$\ug=10^{-6} (r/r_g)^{-5/2}$.  The disk's radiation energy density and
flux are set by LTE and flux-limited diffusion
\citep{2014MNRAS.441.3177M}.

The initial magnetic field is large-scale and poloidal.  For
$r<300r_g$, the coordinate basis $\phi$-component of the vector
potential is
\begin{equation}
A_\phi = {\rm MAX}(r^\nu 10^{40} - 0.02,0) (\sin\theta)^{1+h} ,
\end{equation}
with $\nu=0.75$ and $h=4$.  For $r\ge r_0 = 300r_g$, the field
transitions to monopolar using $A_\phi = {\rm MAX}(r_0^\nu 10^{40} -
0.02,0) (\sin\theta)^{1+h(r_0/r)}$.  The field is normalized with
$\sim 1$ MRI wavelength per half-height $H$ giving a ratio of average
gas+radiation pressure to average magnetic pressure of $\beta\approx
40$ for $r<100r_g$.

\subsection{Numerical Setup}
\label{sec:numsetupgrid}

The numerical grid mapping equations and boundary conditions used here
are identical to that given in \citet{2012MNRAS.423.3083M} and
\citet{2014MNRAS.441.3177M}.  The grid focuses on the disk at small
radii and on the jet at large radii.

Models M14h and M15h are moderate resolution models at $N_r\times
N_\theta\times N_\phi=128\times 64\times 64$, while the rest are
survey models at $128\times 64\times 32$ in order to resolve the
lowest order $m$ modes to allow for accretion in MADs.  For MADs such
lower resolution models have shown reasonable convergence
\citep{2012MNRAS.423.3083M,2014MNRAS.441.3177M}.  The grid aspect
ratio is roughly 1:1:1 for radii $5r_g\le r \le 30r_g$ for our
moderate resolution models.

The rest-mass and internal energy densities are driven to zero near
the BH within the jet and near the axis, so we use numerical ceilings
of $b^2/\rhorest = 300$, $b^2/\ug=10^{11}$, and
$\ug/\rhorest=10^{10}$.  The value of $b^2/\rhorest$ is at the code's
robustness limit for the chosen resolution, while the value of
$b^2/\ug$ is chosen to ensure an artificial temperature floor is not
introduced.

At early times in all simulations, we ramp-up the chemical potential
factor from Planck toward its desired target.  This avoids
difficult-to-resolve opacity changes due to synchrotron within the
first $100r_g/c$ in time.  Also, the sharp changes in synchrotron
opacity for high $\phi$ (Eq.~\ref{phisy}) near Planck
(i.e. $\exp{(-\xi)}=1$ with dimensionless chemical potential $\xi=0$)
are difficult to handle.  For synchrotron only, we enforce
$\exp{(-\xi)}<=0.99$ in order to smooth-out these opacity changes
\citep{1993MNRAS.265L..25S}.

\subsection{Models}

Our goal is to consider the physical effect of various choices for the
radiative transfer opacities.  We simulate several low-resolution
``survey'' runs over a time period of about $10000r_g/c$, allowing the
simulations to reach a single inflow time out to $r\sim 50r_g$ and
inflow equilibrium out to $r\sim 20r_g$.  We check these survey runs
with a couple moderate resolution runs, and we consider even lower
resolution runs to see if resolution plays a dominate or sub-dominant
role compared to the opacity effects.

Table~\ref{tblinit} shows the physical choices made for each model,
including the black hole spin, choices for opacities, whether the
radiation photon number density is evolved, how the radiation
temperature is computed, and whether the chemical potential is varied.

The ''OPAL'' opacity refers to the opacity $\kappa_{\rm eff}$ given by
Eq.~(\ref{kappaeff}).  This is physically equivalent to the OPAL-based
opacity used in \citet{2016arXiv160106836J}, except ours is more
approximate, but we also account for radiation temperatures being
different from gas temperatures.  ''DC'' refers to double Compton,
which no simulations have yet accounted for.  ''Syn'' refers to
synchrotron, which has been accounted for in sub-Eddington accretion
cases \citep{fm09,2015ApJ...807...31R,2016arXiv160503184S}.  All
models with photon number evolution include a separately-computed
number mean opacity not yet accounted for in simulations (except only
synchrotron number opacity in \citealt{2016arXiv160503184S}).  The
simplified fluid-frame radiation temperature given by
$\hat{E}/(\hat{n}\bar{E}_0)$ with $\bar{E}_0\approx 2.701\kb$ is the
dominant factor in Eq.~(\ref{tgamma}) \citep{2015arXiv150804980S}.  We
consider both Planck and Bose-Einstein photon distributions.

All models include thermal Comptonization (TC) except model M6 in
order to study how turning that off affects gas temperatures.  Model
M6 would be like many existing radiative transfer simulations
\citep{2012ApJS..201....9F,2014ApJ...796..106J,2014ApJ...796...22F,2016ApJ...826...23T}
except those by \citet{fm09,2009PASJ...61..769K,2014ApJ...784..169J}
and ourselves \citep{2014MNRAS.439..503S,2014MNRAS.441.3177M}.

Models M9, M11, M12, M13, M14, M14h, M15, M15h are used to explore how
varying $\dot{M}$ affects the results while using our full opacity
physics, with M9 and M11 not behaving much differently due to having
similar $\dot{M}$.

Models M4 and M12 are very low resolution ($128\times 64\times 16$)
test models that otherwise match models M3 and M11, respectively.
These models and others at different resolutions exhibit all of the
same radiative properties we will discuss, which shows that resolution
is unlikely a dominant factor in the controlling our results.  We do
not discuss models M4 and M12 further.

Throughout our discussion of model results, we focus on specific
models, often showing more details for model M15 because it has an
intermediate mass accretion rate and demonstrates properties of all
our models.

\subsection{Initial and Final State}

\begin{figure}
  \centering
  \includegraphics[width=0.99\columnwidth]{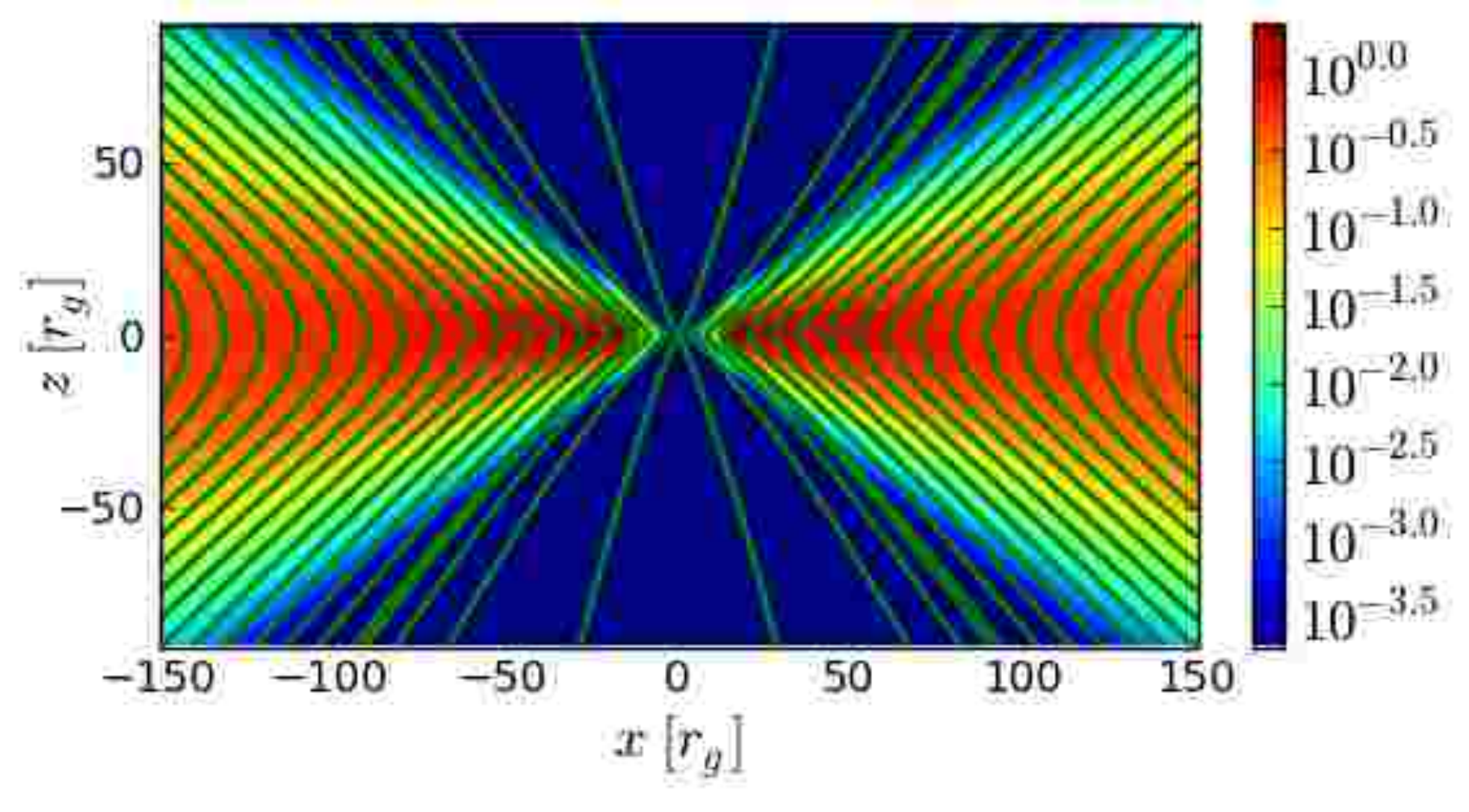}\hfill
  \includegraphics[width=0.99\columnwidth,trim={1.5cm 0.0cm 0.0cm 0.0cm},clip]{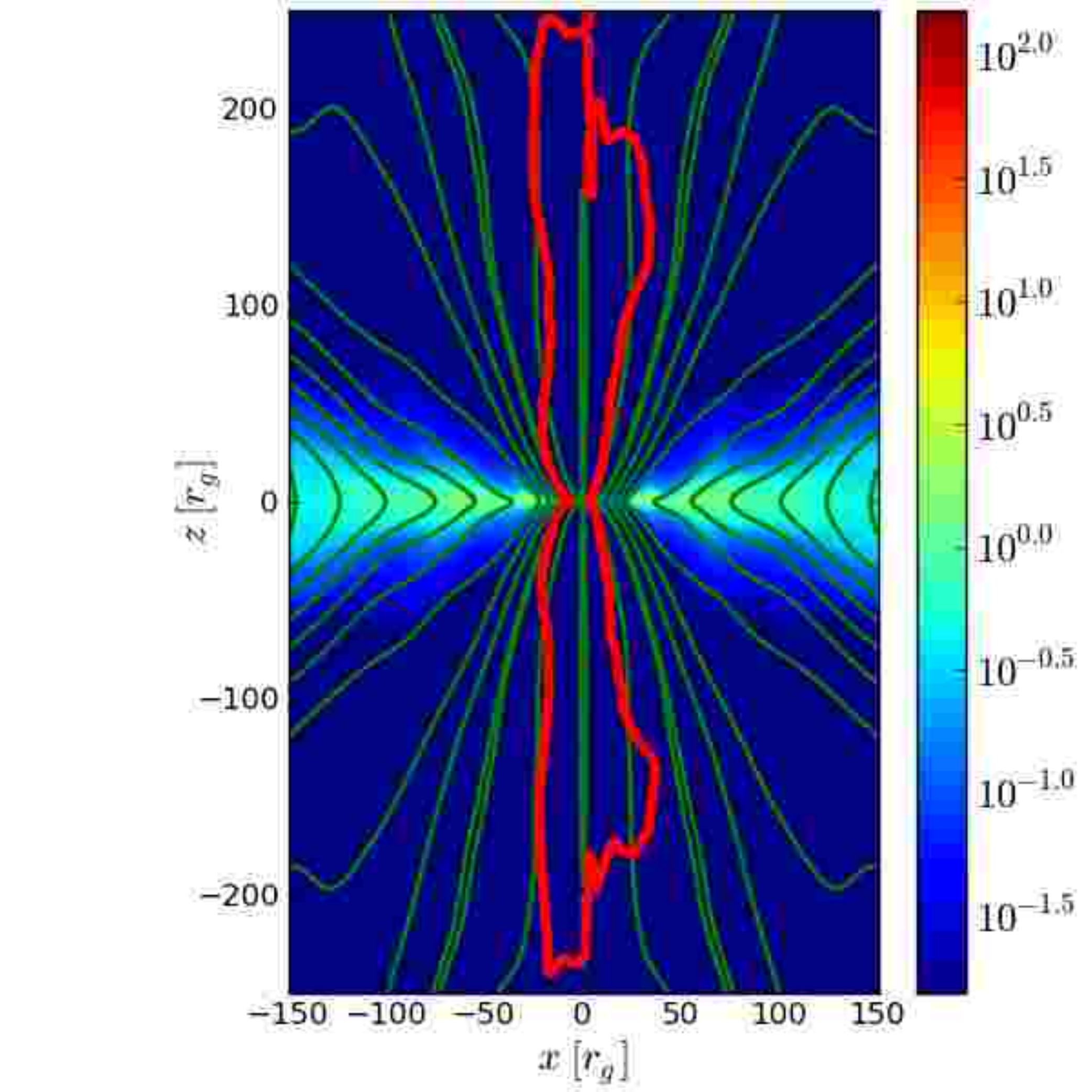}\hfill
  \caption{Model M15, showing the initial condition (top panel) and
    final state (bottom panel).  The initial disk is shown as
    rest-mass density (scaled by an Eddington density value inferred
    from $\dot{M}_{\rm Edd}$, $r_g$, and $c$, color with legend)
    threaded by magnetic field lines (green lines).  The final disk
    has become much higher density near the black hole and the
    rotating black hole and disk have launched a
    magnetically-dominated jet due to the accumulation of magnetic
    flux into the MAD state.  The BH-spin-driven jet emerges and has a
    boundary that can be seen where $b^2/\rho=1$ (red line).  At large
    radii at late time, the jet has become kinetically-dominated with
    Lorentz factor $\gamma\sim 2$ by $r=500r_g$, so that $b^2/\rho<1$
    there.}
  \label{fig:initfinal}
\end{figure}

Fig.~\ref{fig:initfinal} shows the initial and final state of the
accretion flow for model M15, which is typical of all models with
rotating black holes.  The initial magnetic field structure is
large-scale and poloidal and threads the disk that is initially a
Gaussian disk with $H/R\approx 0.2$ and a power-law radial behavior.
The rotating black hole and disk have launched a jet that starts out
magnetically-dominated but converts its energy into kinetic energy at
large radii.

Table~\ref{tblfinal} shows results for dynamical quantities (like
fluxes and efficiencies) for all our models as time-averaged from
$4000r_g/c$ till the end of the simulation.  This table can be used to
compare results for models with different opacity physics and
different $\dot{M}$.  The models vary in mass accretion rate through
the horizon with $\dot{M}/\dot{M}_{\rm Edd}=1$--$140$, and the
radiative luminosity is given by $L_{\rm rad,o}$.  Other quantities
are measured on the horizon (e.g. $\eta_{\rm H}$), at an inner radius
of $10r_g$ for inner ('i') quantities, or at large radii for outer
('o') quantities.

The efficiency $\eta_{\rm H}$ is the total efficiency of the system as
measured at the horizon, which is constant to within $30\%$ out to the
scattering photosphere where outer quantities are measured.  The total
and radiative efficiency at large radii ($\eta_{\rm RAD,o}$) are
comparable to the NT standard thin disk efficiencies ($\eta_{\rm
  NT}$).  The total non-radiative jet efficiency $\eta_{\rm j,in}$ is
the efficiency at $r=10r_g$ in the jet with $b^2/\rho>1$.  This tracks
each model's total efficiency, but the jet energy is progressively
lost to the surrounding material that heats-up and radiates
\citep{2015MNRAS.454L...6M}, leading to relatively low gas jet
efficiencies by $r=1000r_g$.  The total radiative efficiency is
therefore dependent upon the mass-loading physics, which is controlled
partially by numerical floor injection in our simulations.  For
non-rotating BH models, the jet efficiency is low at $1\%$, except for
the model without thermal Comptonization with jet efficiency at $4\%$
due to the thermal energy content of the jet (this also leads to a
slightly higher radiative efficiency).

The normalized magnetic flux on the horizon $\Upsilon_{\rm H}\sim 10$
for relatively thick disks at higher super-Eddington rates, while
lower $\dot{M}$ lead to down to $\Upsilon_{\rm H}\sim 4$ as seen in
thin MAD simulations \citep{2015arXiv150805323A}.  A few models (M9,
M10, and M11) show up to $\Upsilon_{\rm H}\sim 20$ and have quite high
efficiencies as apparently due to radiative suppression of the
magnetic Rayleigh-Taylor modes due to opacity effects at their
intermediate $\dot{M}/\dot{M}_{\rm Edd}\sim 30$--$50$, but one
suspects that higher resolutions would show no such effect (which M15h
approaches and does show more moderate $\Upsilon_{\rm H}$ and
efficiencies).

\begin{table}
\caption{Accretion Rates, Luminosities, Efficiencies [$\%$], and Magnetic Fluxes}
\begin{center}
\begin{tabular}[h]{|l|r|r|r|r|r|r|r|r|r|r|r|r|r|r|r|}
 \hline    	  	  	    	  	   
 Model & $\frac{\dot{M}_{\rm{}H}}{\dot{M}_{\rm Edd}}$ & $\frac{L_{\rm{}rad,o}}{L_{\rm Edd}}$	 & $\eta_{\rm{}H}$	 & $\eta_{\rm{}j,in}$	 & $\eta^{\rm{}RAD}_{\rm{}o}$ & $\eta_{\rm{}NT}$	 & $\Upsilon_{\rm{}H}$	 \\  
 \hline    	  	  	    	  	   
 M1 & 140 & 67	 & 54.7	 & 30.3	 & 5.68 & 12.2	 & 9.8	 \\ % jonharmrad1
 M2 & 140 & 110	 & 58.3	 & 40.7	 & 9.76 & 12.2	 & 9	 \\ % jonharmrad2
 M3 & 80 & 81	 & 114	 & 84.6	 & 12.3 & 12.2	 & 13	 \\ % jonharmrad3
 M5 & 120 & 81	 & 59.2	 & 36.3	 & 8.39 & 12.2	 & 9.5	 \\ % jonharmrad5
 M6 & 5.9 & 6	 & 9.83	 & 4.24	 & 5.86 & 5.72	 & 3.9	 \\ % jonharmrad18
 M7 & 4.8 & 2.4	 & 7.98	 & 0.926	 & 2.84 & 5.72	 & 3.9	 \\ % jonharmrad7
 M8 & 5 & 1.9	 & 8.39	 & 0.979	 & 2.21 & 5.72	 & 5	 \\ % jonharmrad8
 M9 & 50 & 240	 & 124	 & 71.6	 & 59.4 & 12.2	 & 14	 \\ % jonharmrad9
 M10 & 27 & 160	 & 267	 & 184	 & 71.8 & 12.2	 & 20	 \\ % jonharmrad10
 M11 & 36 & 290	 & 185	 & 110	 & 95.5 & 12.2	 & 18	 \\ % jonharmrad11
 M13 & 1.2 & 1.1	 & 18.5	 & 8.2	 & 10.7 & 12.2	 & 3.6	 \\ % jonharmrad13
 M14 & 3.5 & 2.8	 & 23.7	 & 11.4	 & 9.79 & 12.2	 & 5.1	 \\ % jonharmrad14
 M14h & 2.4 & 2.5	 & 21.8	 & 11.3	 & 12.7 & 12.2	 & 4.7	 \\ % jonharmrad17
 M15 & 14 & 18	 & 32.3	 & 14.2	 & 15.6 & 12.2	 & 6.7	 \\ % jonharmrad15
 M15h & 31 & 11	 & 30.9	 & 17.7	 & 4.57 & 12.2	 & 7.1	 \\ % jonharmrad16
\hline
\hline
\end{tabular}
\end{center}
\label{tblfinal}
\end{table}

Fig.~\ref{fig:multipanel} shows a snapshot from the simulation and
shows various fluxes and efficiencies vs. time, whose constancy
indicates that the flow has reached a quasi-steady state in which the
total efficiency is $\eta_{\rm H}\sim 30\%$, almost three times the NT
thin disk efficiency and the radiative efficiency is $\eta_{\rm
rad,o}\sim 5\%$.  The disk is in a MAD state out to about $r\sim
50r_g$ with evident magnetic Rayleigh-Taylor instabilities in the
$y-x$ plane. The dimensionless magnetic flux $\Upsilon\approx 7$,
comparable with non-radiative disks with $H/R\approx 0.3$
\citep{2012MNRAS.423.3083M}. The effective photosphere reaches close
to the disk, except where the jet has relatively high densities.

\begin{figure*}
  \centering
  \includegraphics[width=1.99\columnwidth]{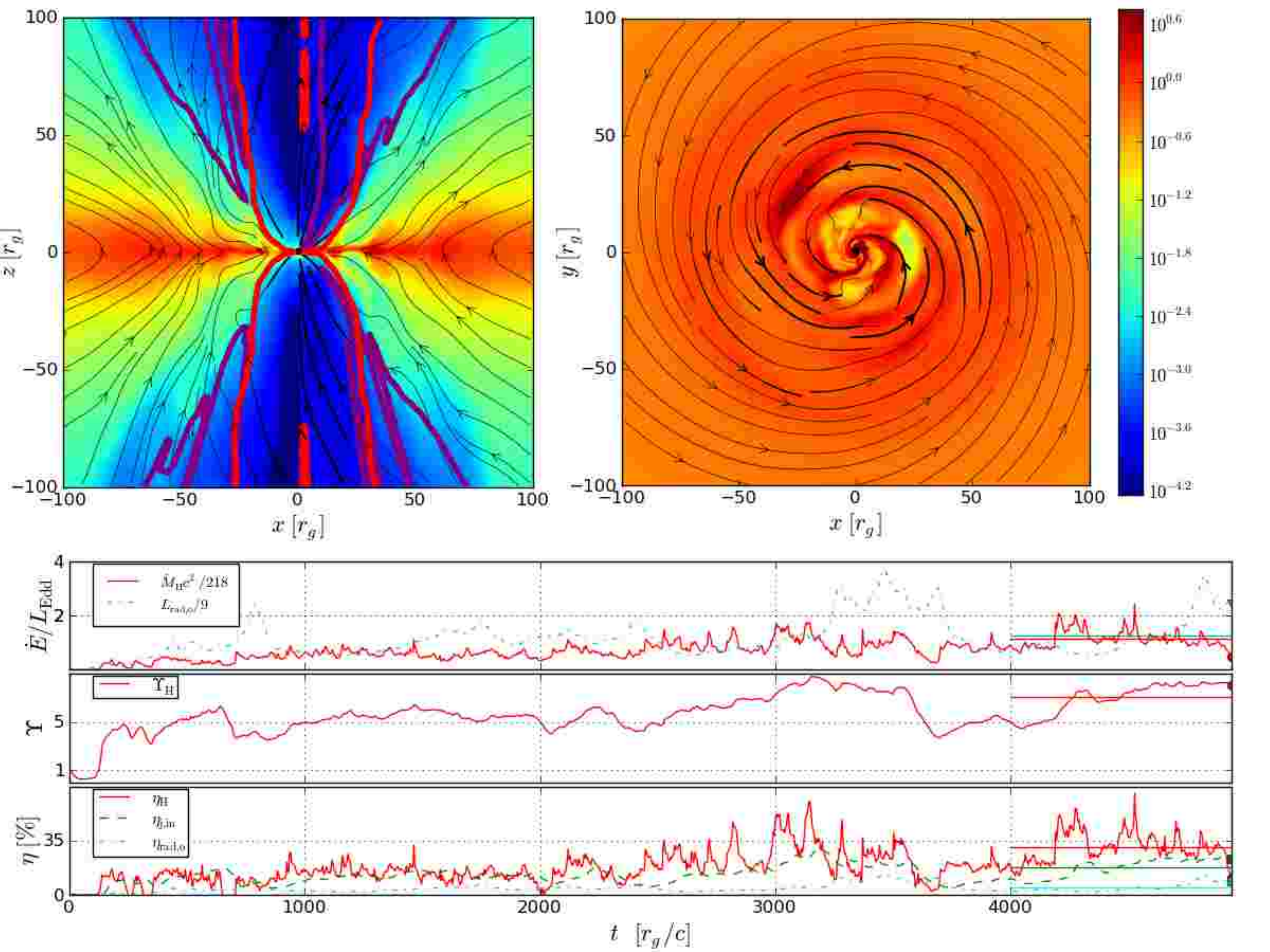}
  \caption{Evolved snapshot of model M15h at $t\approx 5000r_g/c$
    showing log of rest-mass density (scaled by an Eddington density
    value inferred from $\dot{M}_{\rm Edd}$, $r_g$, and $c$, shown in
    color with legend on right) in both the $z-x$ plane at $y=0$
    (top-left panel) and $y-x$ plane at $z=0$ (top-right panel).
    Black lines trace field lines.  In the top-left panel, the thick
    red line corresponds to where $b^2/\rhorest=1$ and the purple line
    corresponds to the effective photosphere (computed radially
    inward). The bottom panel has 3 subpanels.  The top subpanel shows
    $\Mdot$ through the BH ($\dot{M}_{\rm H}$) and radiative
    luminosity ($L_{\rm rad,o}$). The middle subpanel shows the
    magnetic flux passing through the horizon ($\Upsilon_{\rm H}$).
    The bottom subpanel shows the total efficiency ($\eta_{\rm H}$),
    inner jet efficiency ($\eta_{\rm j,in}$), and radiative efficiency
    ($\eta_{\rm rad, o}$).  Horizontal solid lines of the same colors
    show the averages over the averaging period.  For super-Eddington
    accretion at $\dot{M}/\dot{M}_{\rm Edd}\approx 31$, the total BH
    efficiency is moderate at $\eta_{\rm H}\approx 31\%$ with
    radiative efficiency of $\eta_{\rm rad,o}\approx 5\%$.}
  \label{fig:multipanel}
\end{figure*}

Table~\ref{tbl19} shows results for our radiative diagnostics, which
can be used to deduce how the radiative properties are affected by
different opacity choices and different $\dot{M}$.  This table
includes the fluid-frame radiation temperature per unit gas
temperature at $r=10r_g$ in the disk, fluid-frame radiation
temperature in the disk, lab-frame radiation temperature in the
radiation beam at $r=100r_g$, fluid-frame hardening factor
$\hat{f}_{\rm col}$ in the disk, lab-frame hardening factor $f_{\rm
  col}$ in the radiation beam at $r=100r_g$, chemical potential
factors in the disk and radiation beam, radiation beam half-opening
angle ($\theta_{\rm r}$) at $r=1000r_g$, and jet
(kinetic+enthalpy+electromagnetic) half-opening angle ($\theta_{\rm
  j}$) at $r=1000r_g$.  Disk quantities are computed as weighted by
volume of the grid cell times square of density, while radiation or
electromagnetic beam quantities are measured at the peak in the
luminosity per unit angle ($\partial_\theta L(\theta)$).

The low $\dot{M}\sim \dot{M}_{\rm Edd}$ models have higher gas
temperatures in the disk, but gas temperatures are at most about $10$
times the radiation temperatures with thermal Comptonization.  Only
model M6 without thermal Comptonization shows very low
$\hat{T}_\gamma/T_{\rm gas}$. The radiation beam and disk have
hardening with a Wien spectrum in models without double Compton or in
models with $\dot{M}\sim \dot{M}_{\rm Edd}$.

The photon distributions tend to be somewhat Wien in the coronae for
models with chemical potential evolution.  Models without double
Compton and synchrotron (like M2, M3, and M7) show significant photon
hardening, which becomes much more limited when including these
opacities.  Radiation beam lab-frame temperatures are comparable to
the disk core, except for models without double Compton and
synchrotron.  Models without double Compton and synchrotron (e.g. M7)
have much higher radiation temperatures than otherwise identical
models (e.g. M8) with double Compton and synchrotron.  This shows that
double Compton and synchrotron are crucial to include in order to
obtain accurate observer-frame radiation temperatures for flows with
$\dot{M}\gtrsim \dot{M}_{\rm Edd}$.

The half-opening angles in radians identify the maximum in $L_{\rm
  iso}$ within the radiation beam or gas jet.  We also computed (not
in table) the beaming factor ($b=L_{\rm iso}/L$, i.e. isotropic
equivalent luminosity per unit total luminosity) measured at
$r=1000r_g$.  The electromagnetic jet is beamed by factors up to
$b=10$ for rotating black hole models and up to $b=5$ for non-rotating
black hole models.  The radiation beaming factor is up to $b=8$ for
rotating black hole models and $b=3$ for non-rotating black hole
models (similar to seen in \citealt{2015arXiv150300654S}).  This
corresponds to an enhanced radiative flux at specific viewing angles,
with higher beaming for higher $\dot{M}$ and higher $a/M$. The M1
closure slightly overestimates the beaming factors
\citep{2016MNRAS.457..608N}.

\begin{table*}
\caption{Non-LTE and non-Planck Radiative Properties and Radiation/Jet Opening Angles in Radians}
\begin{center}
\begin{tabular}[h]{|l|r|r|r|r|r|r|r|r|r|}
\hline
Model   &  $\left(\frac{\hat{T}_\gamma}{T_{\rm{}gas}}\right)_{r=10r_g}$  &  $\hat{T}_{\gamma,r=10r_g}{\rm{}[K]}$  &  $T_{\gamma,r=100r_g,\rm{rad.beam}}{\rm{}[K]}$  &  $\hat{f}_{\rm{col}}^{r=10r_g,\rm{disk}}$  &  $f_{\rm{col}}^{r=100r_g,\rm{rad.beam}}$  &  ${\rm{}e}^{-\xi}_{r=10r_g,\rm{disk}}$  &  ${\rm{}e}^{-\xi}_{r=100r_g,\rm{rad.beam}}$  &  $\theta_{\rm{}r}^{r=1000r_g}$  &  $\theta_{\rm{}j}^{r=1000r_g}$  \\
\hline
M1      &  0.96                                                          &  4.2e7                                 &  1.7e7                                          &  1                                         &  1.4                                      &  1                                      &  1                                           &  0.12                           &  0.067                          \\  %  jonharmrad1
M2      &  0.93                                                          &  1.3e8                                 &  7.7e8                                          &  3.9                                       &  55                                       &  1                                      &  1                                           &  0.089                          &  0.0038                         \\  %  jonharmrad2
M3      &  1                                                             &  1.1e8                                 &  6e8                                            &  3.2                                       &  42                                       &  0.88                                   &  0.36                                        &  0.084                          &  0.021                          \\  %  jonharmrad3
M5      &  0.97                                                          &  3.8e7                                 &  9.1e6                                          &  1                                         &  0.77                                     &  1                                      &  1                                           &  0.092                          &  0.025                          \\  %  jonharmrad5
M6      &  9.8e-5                                                        &  1.5e7                                 &  3.3e6                                          &  1                                         &  0.58                                     &  1                                      &  1                                           &  0.6                           &  0.12                           \\  %  jonharmrad18
M7      &  0.87                                                          &  1.4e8                                 &  2.5e7                                          &  8.9                                       &  5.4                                      &  1                                      &  1                                           &  0.6                            &  0.19                           \\  %  jonharmrad7
M8      &  0.63                                                          &  1.9e7                                 &  6.7e6                                          &  1.1                                       &  1.6                                      &  1                                      &  1                                           &  0.6                            &  0.15                           \\  %  jonharmrad8
M9      &  0.98                                                          &  4.1e7                                 &  1.5e7                                          &  1                                         &  1.1                                      &  0.97                                   &  0.96                                        &  0.13                           &  0.16                           \\  %  jonharmrad9
M10     &  0.99                                                          &  4.9e7                                 &  2.1e7                                          &  1.3                                       &  1.6                                      &  0.87                                   &  0.69                                        &  0.13                           &  0.041                          \\  %  jonharmrad10
M11     &  0.99                                                          &  4.3e7                                 &  1.5e7                                          &  1                                         &  1                                        &  0.99                                   &  0.97                                        &  0.098                          &  0.21                           \\  %  jonharmrad11
M13     &  0.12                                                          &  3e7                                   &  9.2e6                                          &  4.5                                       &  2.6                                      &  0.43                                   &  0.45                                        &  0.28                           &  0.0091                         \\  %  jonharmrad13
M14     &  0.23                                                          &  1.5e7                                 &  2.1e6                                          &  1.3                                       &  0.48                                     &  0.79                                   &  0.94                                        &  0.26                           &  0.011                          \\  %  jonharmrad14
M14h    &  0.2                                                           &  1.3e7                                 &  2e6                                            &  1.2                                       &  0.45                                     &  0.76                                   &  0.99                                        &  0.47                           &  0.055                          \\  %  jonharmrad17
M15     &  0.85                                                          &  2.4e7                                 &  7.3e6                                          &  1.1                                       &  1.2                                      &  0.86                                   &  0.76                                        &  0.16                           &  0.028                          \\  %  jonharmrad15
M15h    &  0.97                                                          &  2.5e7                                 &  7.2e6                                          &  1                                         &  1.1                                      &  0.99                                   &  0.92                                        &  0.37                           &  0.097                          \\  %  jonharmrad16
\hline
\hline
\end{tabular}
\end{center}
\label{tbl19}
\end{table*}

\subsection{Magnetic and Radiative Fluxes}

Fig.~\ref{fig:emradflux} shows the magnetic flux lines (with
electromagnetic efficiency) and lab-frame radiation flux stream lines
(with radiative efficiency).  The radiation is broadly distributed,
but has an enhanced beamed region that sits in angle at about twice
larger angle compared to the electromagnetic jet.  Models with zero
black hole spin (not plotted) show the peak EM luminosity per unit
angle emerging from cylindrical radius near the ISCO
\citep{2012JPhCS.372a2040T}, instead of the rotating black hole models
where the peak power per unit angle emerges from near the equatorial
region of the black hole horizon
\citep{2004ApJ...611..977M,2006MNRAS.368.1561M,2009MNRAS.394L.126M}.

\begin{figure}
  \centering
  \includegraphics[width=0.99\columnwidth]{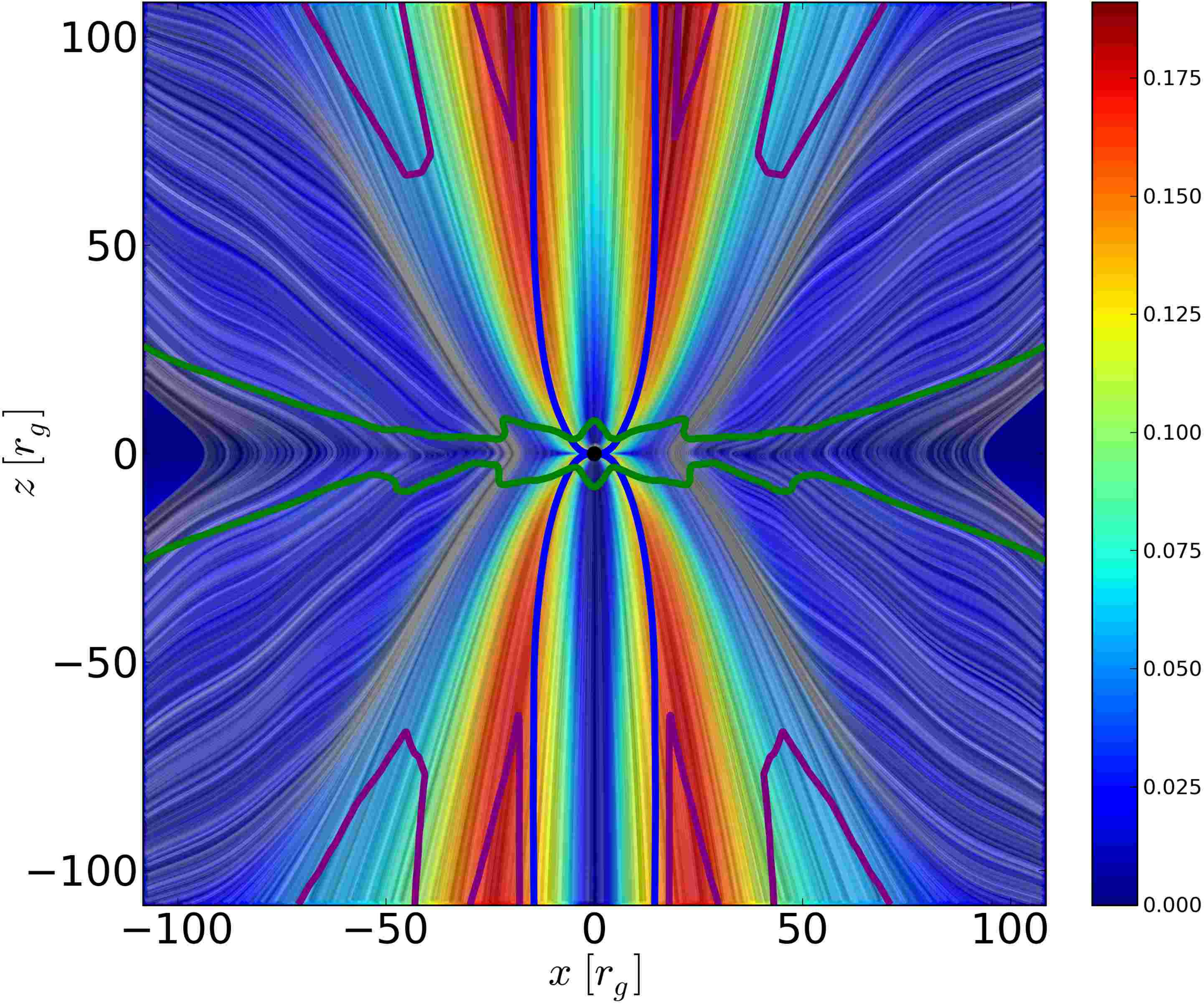}\hfill
  \includegraphics[width=0.99\columnwidth]{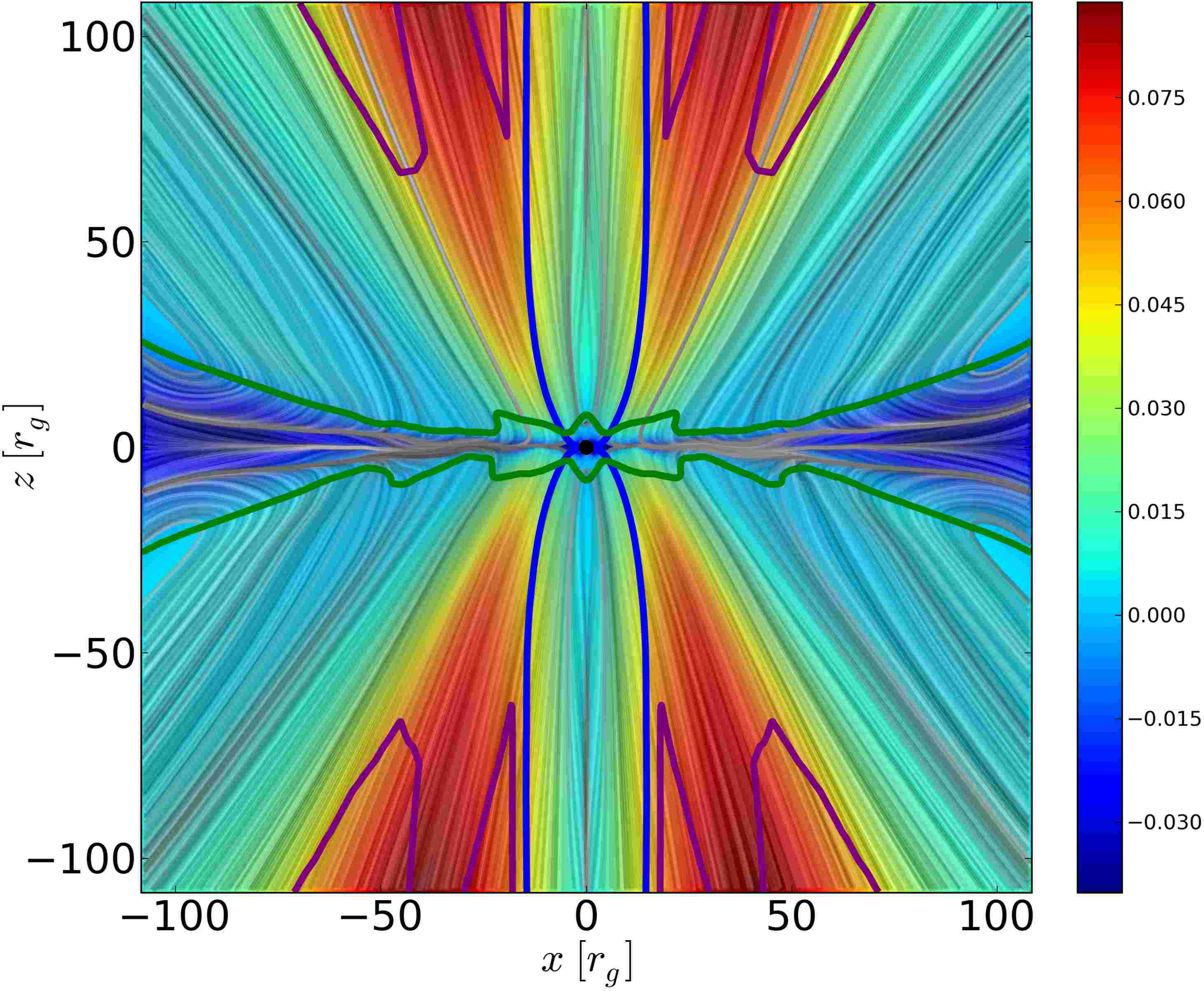}\hfill
  \caption{Model M15, with top panel showing time-$\phi$-averaged
    magnetic flux lines (translucent gray lines) with electromagnetic
    luminosity per unit angle $(\partial_\theta L_{\rm EM})/(\dot{M}_{\rm
      H}c^2)$ (color with legend), with blue line showing where
    $b^2/\rho=1$, green line showing where $u^r=0$, and purple line
    showing total effective photosphere.  Image is duplicated across
    the $x=0$ line.  The bottom panel shows same things for lab-frame
    radiation flux lines and radiation luminosity.  The
    electromagnetic energy flux acts as a funnel-wall jet by following
    (and sitting just outside) the boundary where $b^2/\rho=1$.  Most
    of the electromagnetic energy is released from the black hole's
    spin energy.  The MAD equatorial region is quite hot and dynamic,
    leading to radiation emerging from quite close to the BH (and in a
    time-averaged sense, some appears to emerge from above the BH due
    to transient but powerful polar magnetic fields that jump between
    the BH and disk). Radiation moves inward within the disk that
    accumulates more radiative energy, which somewhat follows the path
    of the wind and ultimately becomes more radially-directed at
    larger distances.}
  \label{fig:emradflux}
\end{figure}

\subsection{Effective Energy Photospheres}

Fig.~\ref{fig:taueff} shows the effective photospheres for model M15.
At large radii, the total effective photosphere sits above the disk
and disk wind.  Sitting inside the total effective photosphere is the
free-free photosphere, the double Compton photosphere, and the
synchrotron photosphere.  The free-free, double Compton, and
synchrotron opacities all merge within some radius, showing they
become comparably important.  While free-free and DC are clearly
important in metal-free plasmas, bound-free and bound-bound
contributions are important with solar abundances, which leads to an
effective photosphere far beyond the free-free photosphere. For higher
$\dot{M}$ models, these different opacities become more comparable at
larger radii than in this lower $\dot{M}$ model.  The scattering
photosphere is at $r\sim 800r_g$ in this model.

\begin{figure}
  \centering
  \includegraphics[width=0.99\columnwidth]{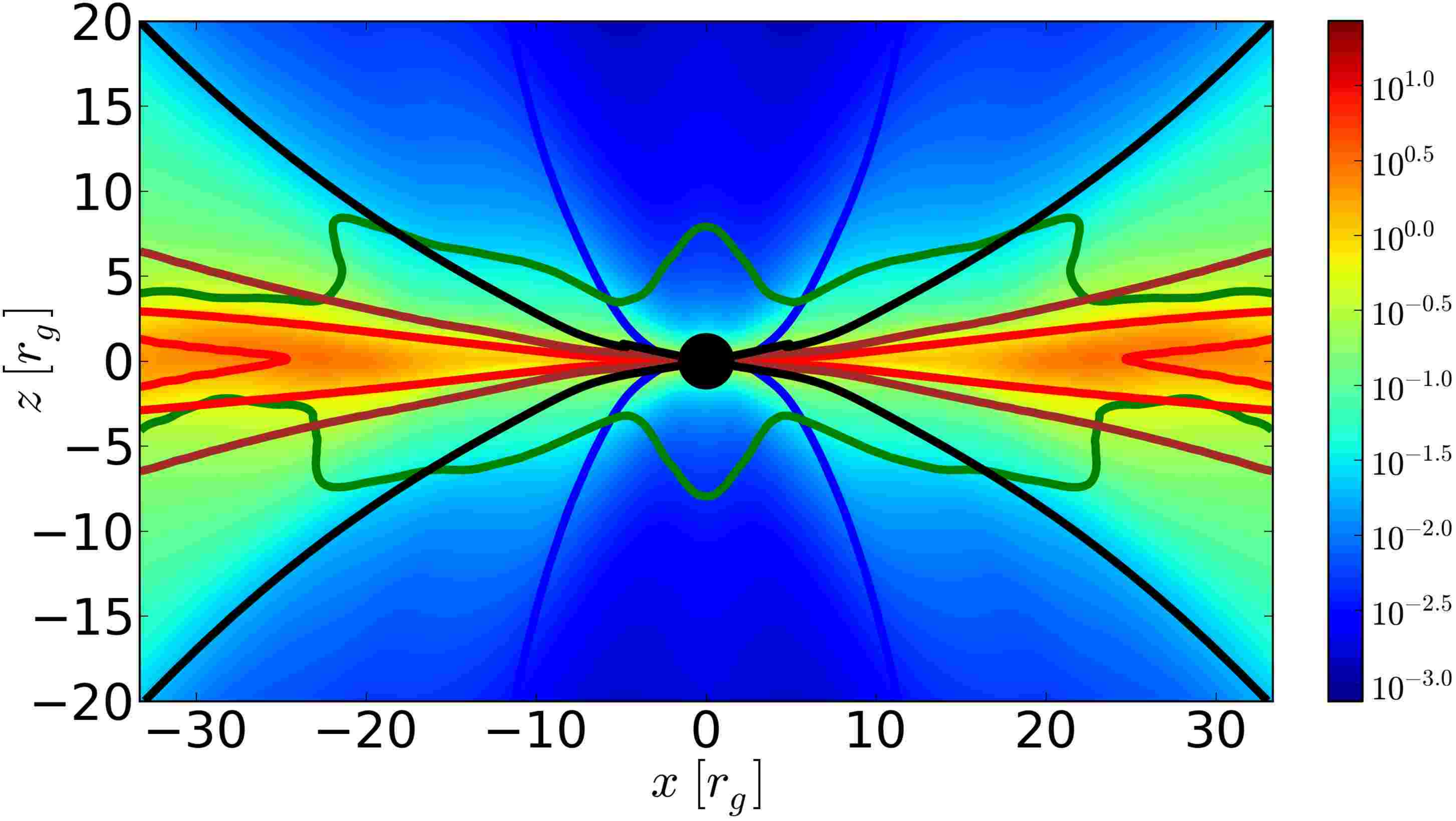}\hfill
  \includegraphics[width=0.99\columnwidth]{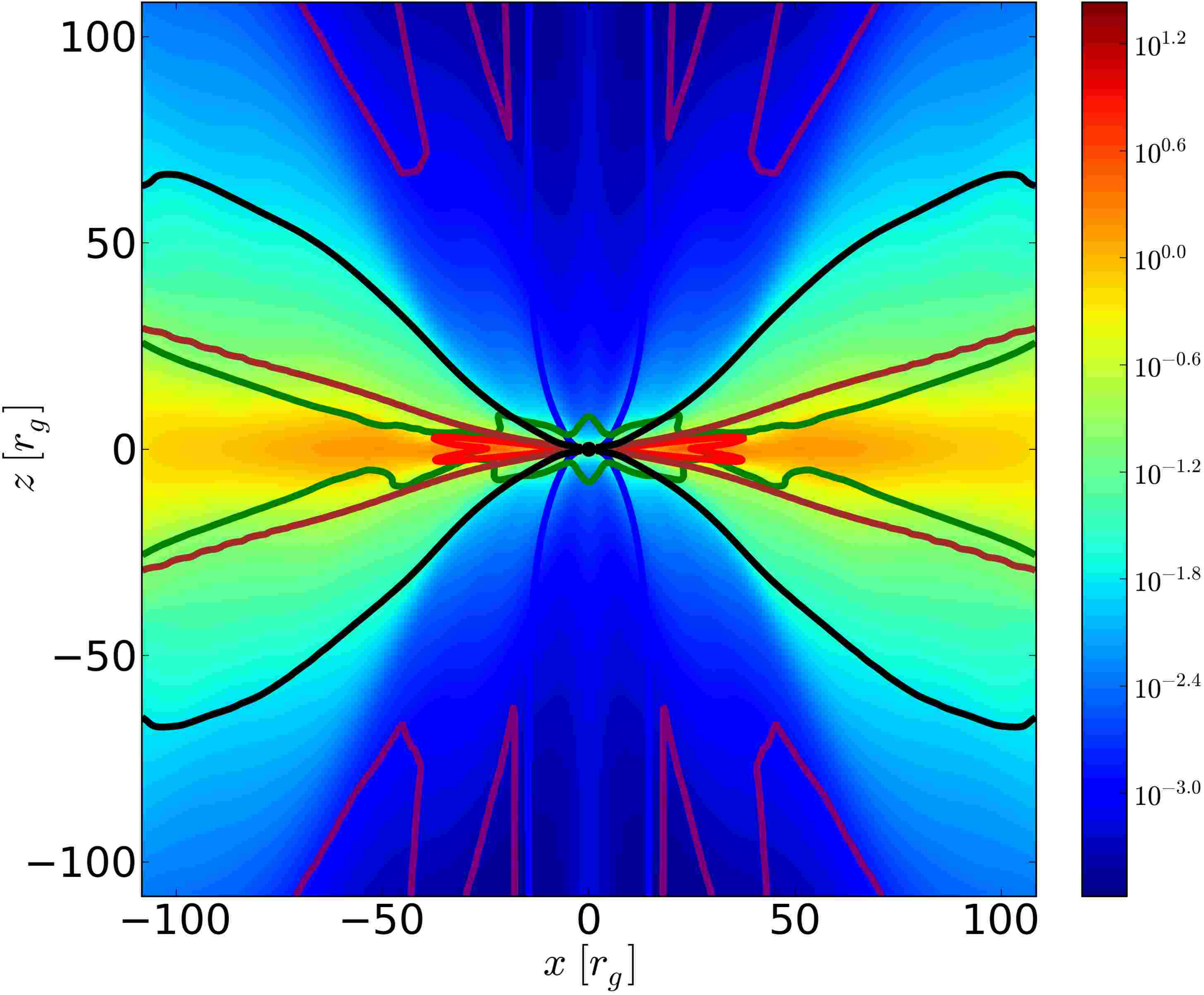}\hfill
  \includegraphics[width=0.99\columnwidth]{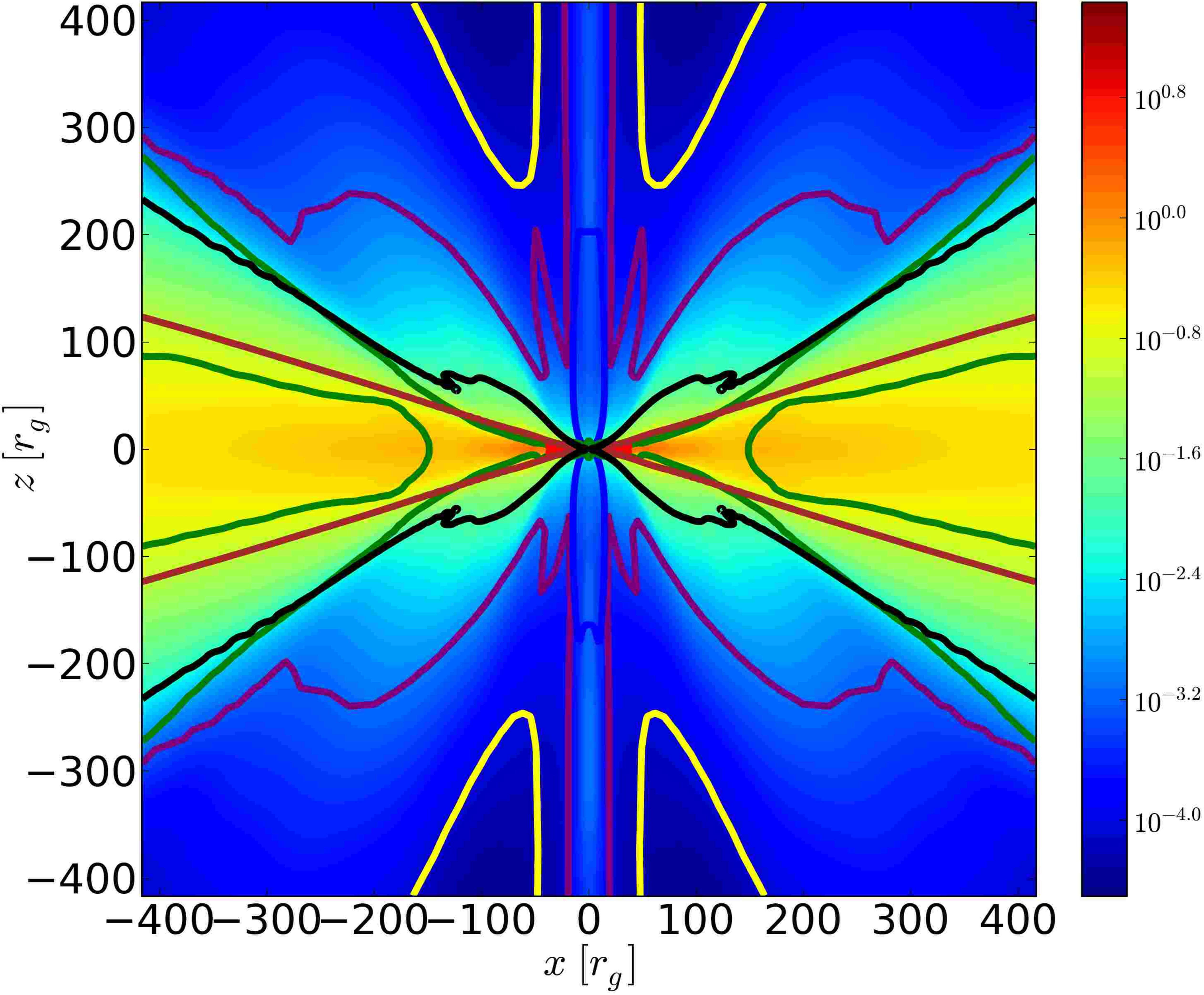}\hfill
  \caption{Model M15, showing effective photospheres for different
    processes, including OPAL+DC+synchrotron (purple line),
    synchrotron (red line), DC (brown line), free-free (black line),
    scattering only (yellow line) on different size regions (top,
    middle, bottom panels).  The photospheres are computed radially
    inward from $r=4000r_g$.  Rest-mass density shown in color (with
    legend), scaled by an Eddington density value inferred from
    $\dot{M}_{\rm Edd}$, $r_g$, and $c$.  Green, blue, yellow, and
    purple lines are, if present, as in Fig.~\ref{fig:emradflux}.  For
    $r\lesssim 5r_g$, DC and free-free become comparably important
    processes in the disk.  The effective photosphere somewhat follows
    the transition between disk inflow and wind outflow.  The polar
    axis regions contains a high-density portion of the jet (launched
    by mass injection near the black hole) which leads to higher
    opacities there.}
  \label{fig:taueff}
\end{figure}

\subsection{Gas Over-Heated Regions}

Fig.~\ref{fig:TradandTgas} shows the lab-frame radiation temperature
and fluid-frame gas temperature for model M15.  The radiation
temperature reaches up to $T_\gamma\sim 10^8$K, while the gas
temperature reaches up to $T_{\rm gas}\sim 10^9$K in the jet region.
Given our discussion in the introduction, this suggests that double
Compton should be important through-out the flow, while synchrotron is
likely important in the jet region.  Notice that as $\dot{M}$ drops
that the disk becomes thinner, although such MAD type disks are also
magnetically-compressed by the large-scale poloidal and toroidal
fields threading the black hole and disk \citep{2012MNRAS.423.3083M}..

\begin{figure}
  \centering
  \includegraphics[width=0.99\columnwidth]{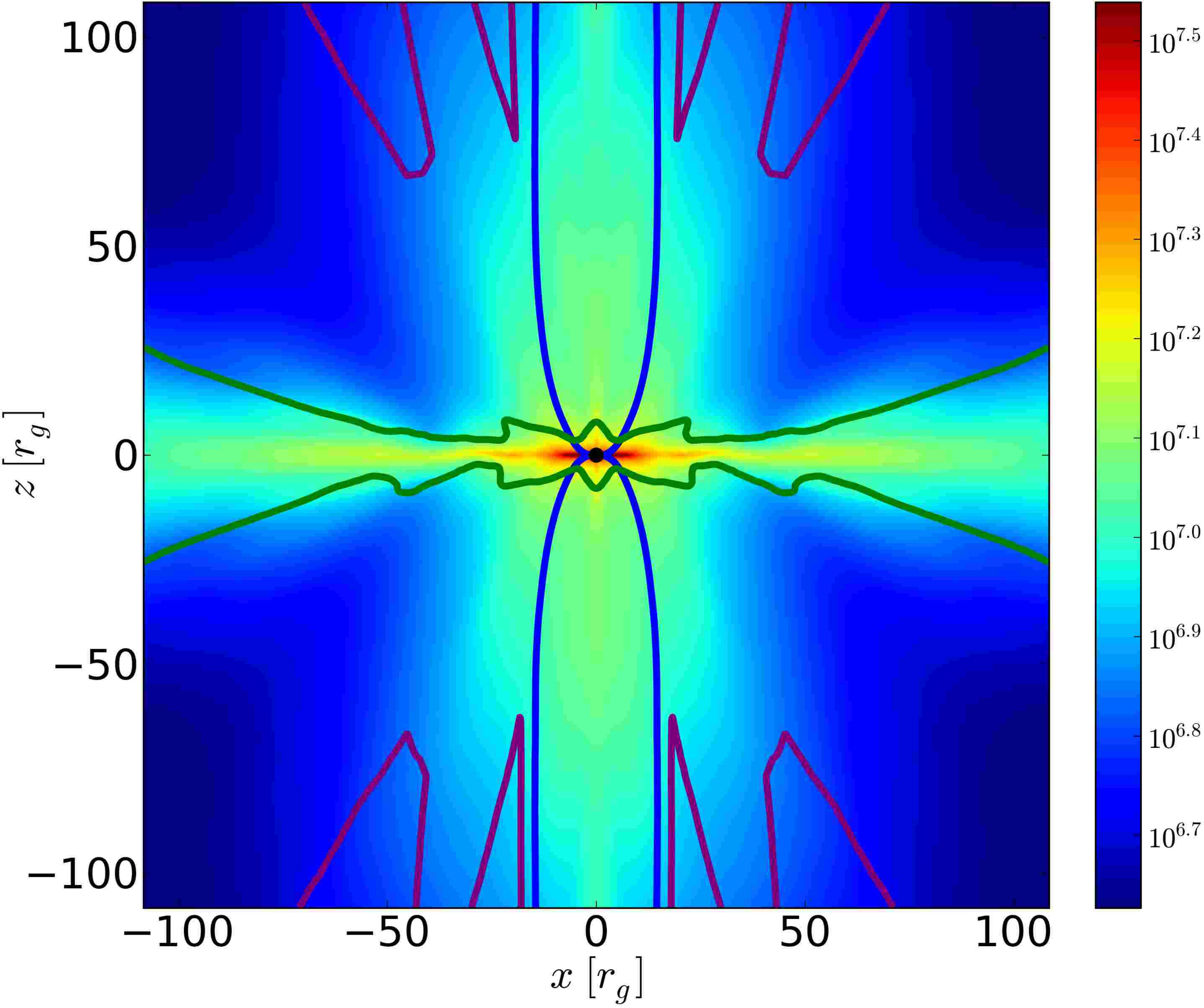}\hfill
  \includegraphics[width=0.99\columnwidth]{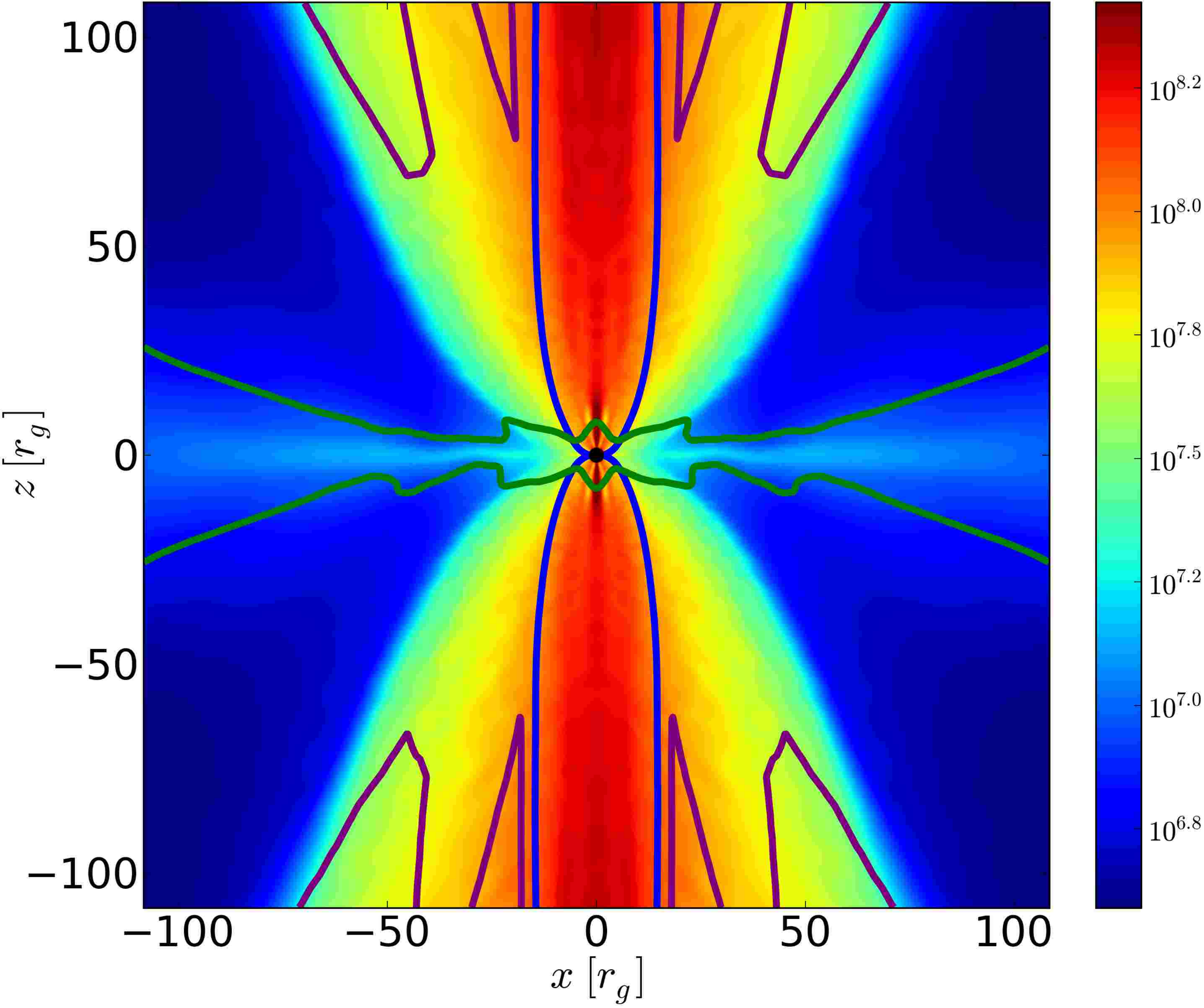}\hfill
  \caption{Model M15, showing lab-frame radiation temperature in
    Kelvin (top panel) and fluid-frame gas temperature in Kelvin
    (bottom panel).  Green, blue, yellow, and purple lines are, if
    present, as in Fig.~\ref{fig:emradflux}.  Radiation temperatures
    are high in the equatorial and polar regions, while gas
    temperatures are high in the jet region due to insufficient
    Comptonization.}
  \label{fig:TradandTgas}
\end{figure}

\begin{figure}
  \centering
  \includegraphics[width=0.99\columnwidth]{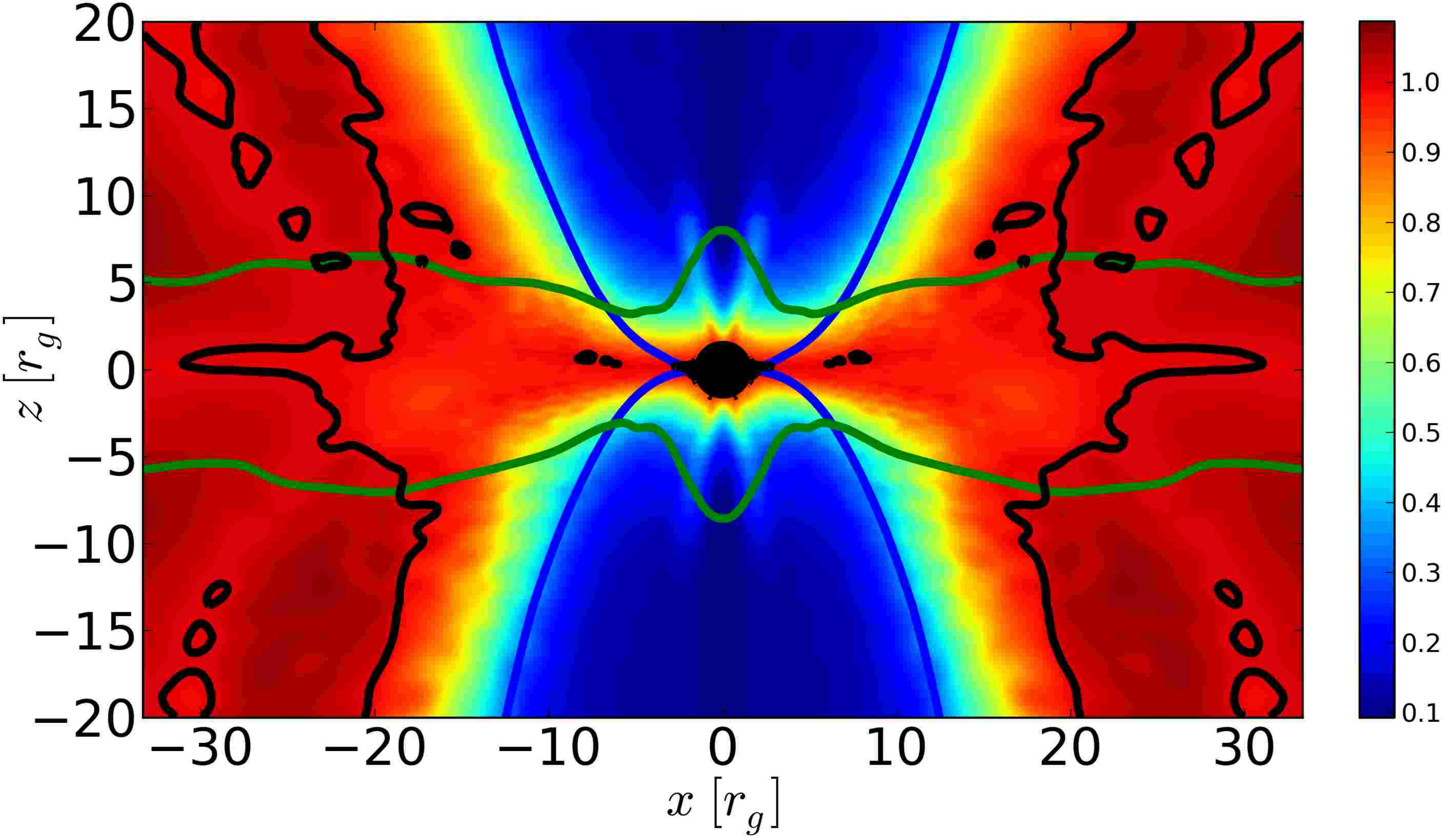}\hfill
  \includegraphics[width=0.99\columnwidth]{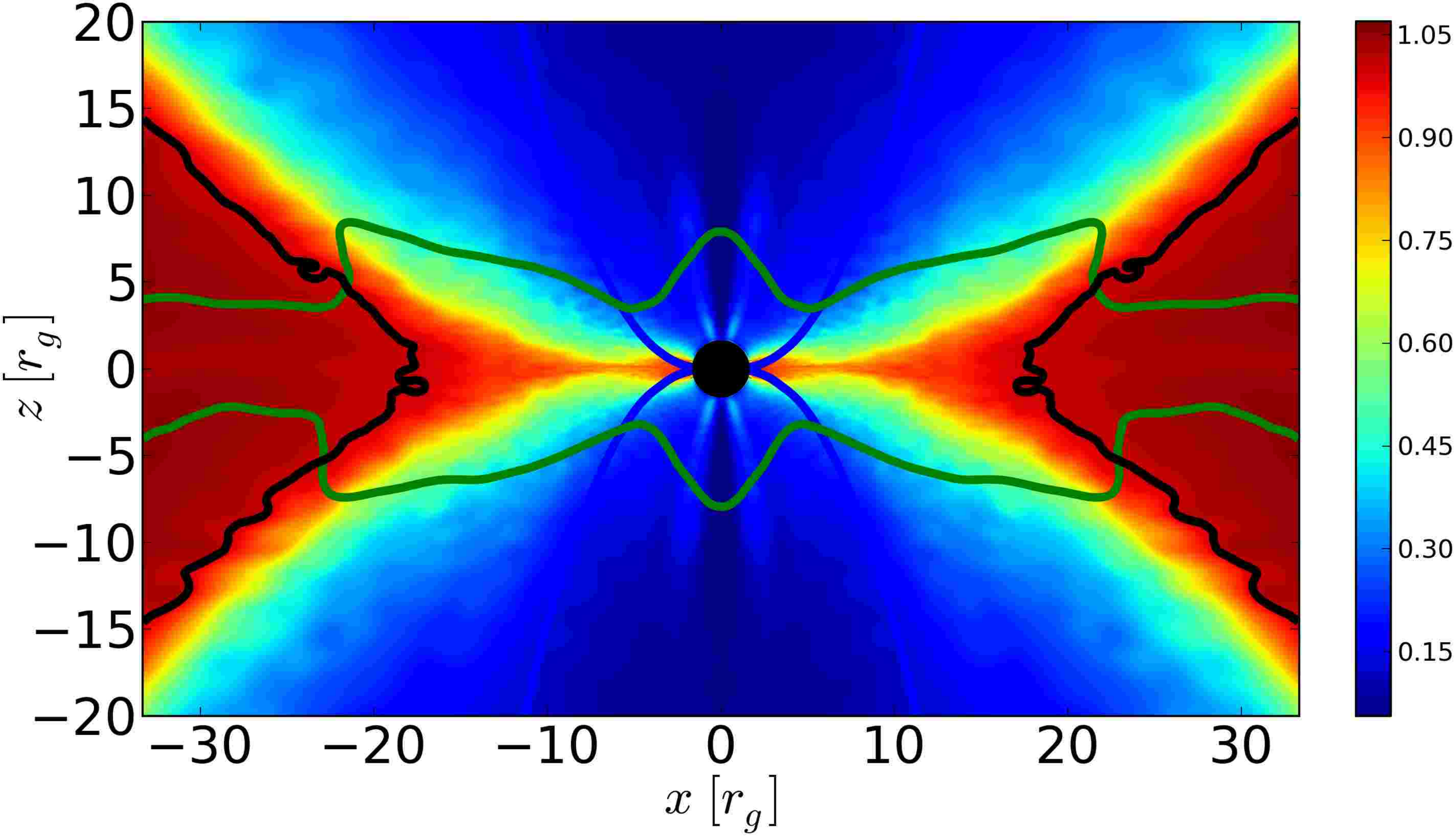}\hfill
  \includegraphics[width=0.99\columnwidth]{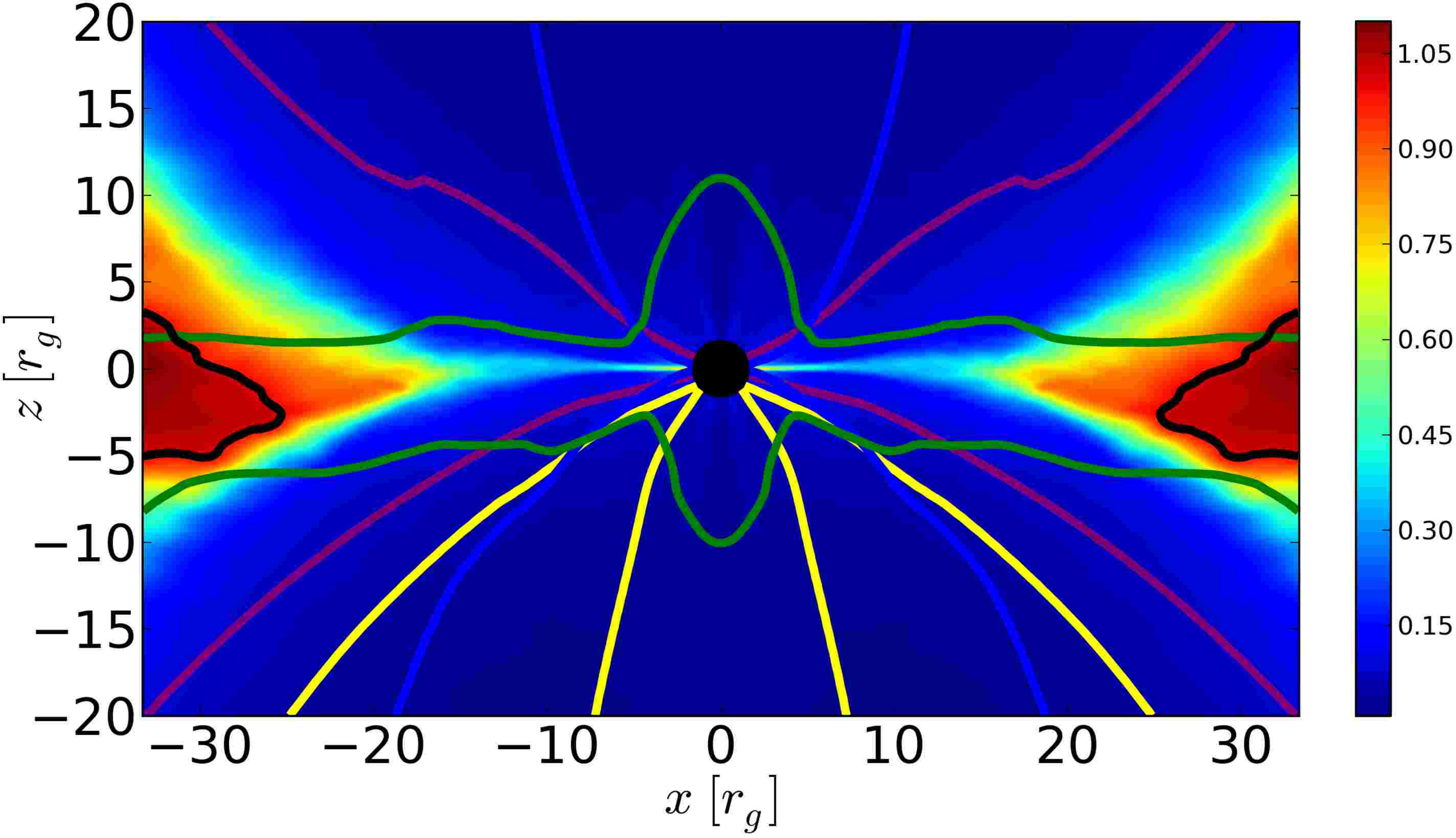}\hfill
  \includegraphics[width=0.99\columnwidth]{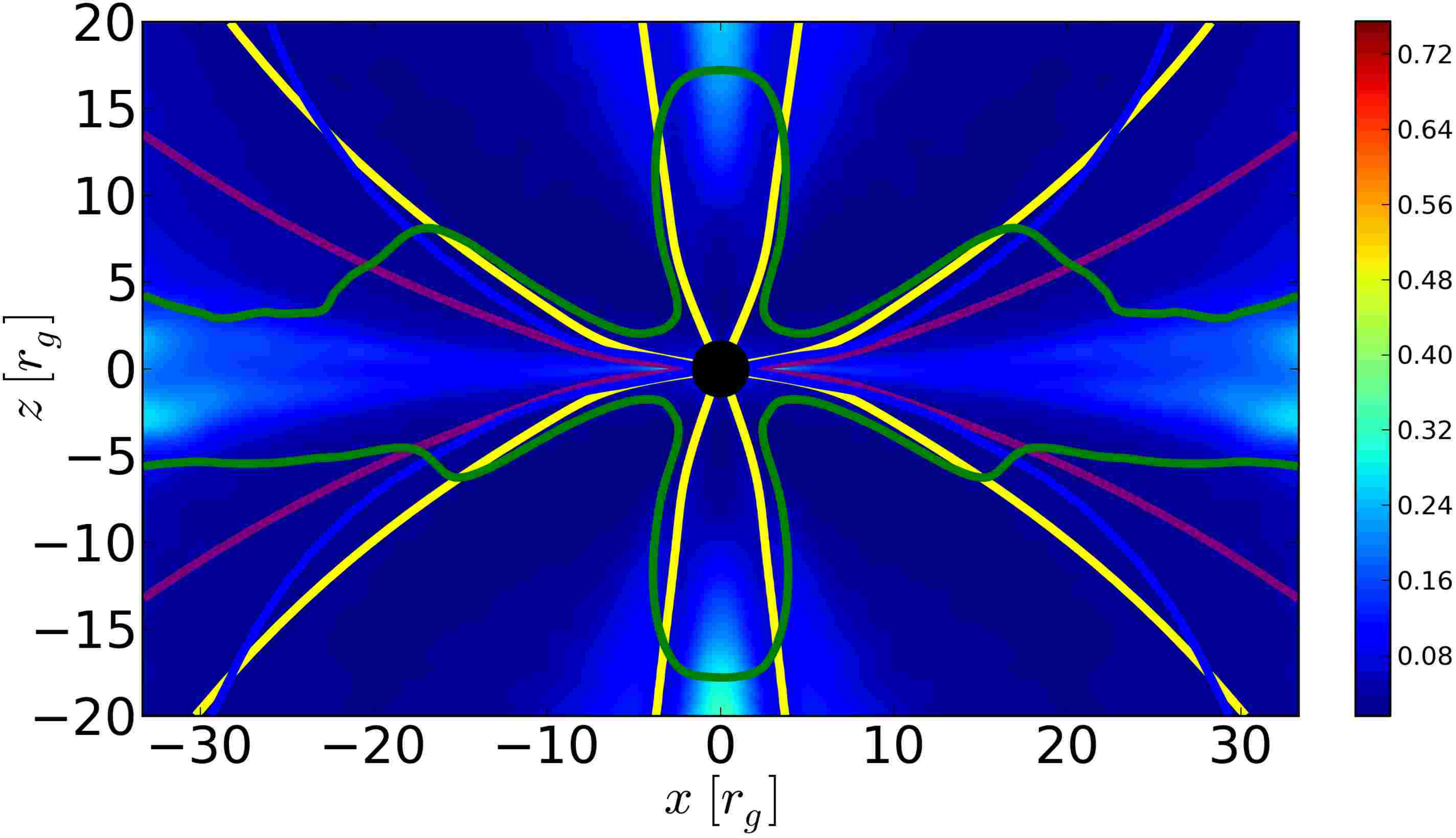}\hfill
  \caption{Model M1 at high $\dot{M}\sim 100\dot{M}_{\rm Edd}$ (top
    panel), model M15 at lower $\dot{M}\sim 10\dot{M}_{\rm Edd}$
    (middle panel), model M14 at low $\dot{M}\sim 3\dot{M}_{\rm Edd}$
    (next panel), and model M13 at low $\dot{M}\sim 1\dot{M}_{\rm
      Edd}$ (bottom panel), showing fluid-frame $\hat{T}_\gamma/T_{\rm
      gas}$.  Black line has $\hat{T}_\gamma/T_{\rm gas}=1$.  Green,
    blue, yellow, and purple lines are, if present, as in
    Fig.~\ref{fig:emradflux}.  Models M13 and M14 have a photosphere
    near the disk, except the jet that is launched which keeps the
    density high to large radii.  At high or low $\dot{M}$, the value
    of $\hat{T}_\gamma/T_{\rm gas}$ is order unity due to thermal
    Comptonization.  The disk is evidently thinner and cooler at lower
    mass accretion rates of $\dot{M}\sim 10\dot{M}_{\rm Edd}$ even
    though the inflow is still quite super-Eddington.  Only for the
    lowest $\dot{M}\sim \dot{M}_{\rm Edd}$ model M13 does
    $\hat{T}_\gamma/T_{\rm gas}\sim 0.1$ in the central disk, but
    progressively more coronal material has higher gas temperatures as
    $\dot{M}$ drops.}
  \label{fig:TradoTgas}
\end{figure}

Fig.~\ref{fig:TradoTgas} shows the fluid-frame radiation temperature
per unit gas temperature.  Thermal Comptonization acts to regulate gas
temperatures toward the radiation temperature in radiation-dominated
plasmas. We show several model's poloidal plane temperature ratio in
order to present how this effect works at high $\dot{M}\sim
100\dot{M}_{\rm Edd}$ to low $\dot{M}\sim 1\dot{M}_{\rm Edd}$.  Notice
that model M14 has a long-lived hemispherical asymmetry, leading to
scattering photosphere far away on the upper hemisphere.  The figure
shows fluid-frame $\hat{T}_\gamma/T_{\rm gas}$ (numerator and
denominator separately time-$\phi$-averaged).  While lower $\dot{M}$
models have a slightly over-heated gas region within some radius, the
core disk region has at most 10\% higher gas temperatures than
radiation temperatures for $\dot{M}\sim 10\dot{M}_{\rm Edd}$ models.

Model M13 with $\dot{M}\approx 1\dot{M}_{\rm Edd}$ shows an optically
thin corona with gas temperatures about $100$ times larger than the
disk's black body temperature and about $20$ times larger than the
disk's radiation temperature that is hardened by $\hat{f}_{\rm
  col}\approx 4.5$ (see related data from Fig.~\ref{fig:fluidvstheta}).
The gas pressure is up to a tenth of the radiation pressure and the
disk thickness $H/R\sim 0.1$ in this model.

Table~\ref{tbl19} includes a sequence of $a/M=0$ models M6, M7, and M8
which have no thermal Comptonization and no photon hardening (M6),
have thermal Comptonization and photon hardening but without double
Compton or synchrotron (M7), and have both along with all our
opacities (M8).  This shows that the lack of thermal Comptonization
leads to unphysically high gas temperatures similar to seen in
\citet{2016ApJ...826...23T}, who did not include thermal
Comptonization.  Hence, thermal Comptonization is required for
Eddington to super-Eddington accretion models.

\subsection{Photon Hardening}

Fig.~\ref{fig:fcol0} shows the lab-frame photon hardening factor
($f_{\rm col} = T_\gamma/T_{\rm BB}$) for the $a/M=0$ models M6, M7,
and M8 that we discussed above.  For an observer in the lab-frame,
this would correspond to the color correction factor assuming the
observers fit spectra with Planck or BE distributions.  We find that
double Compton and synchrotron play a crucial role in limiting photon
hardening.  These opacities generate much more radiation at high
temperatures, unlike free-free, and result in the primary radiation
beam changing from $f_{\rm col}=5$ down to $f_{\rm col}=1.5$.  Our
model M7 is similar to the model in \citet{2015arXiv150804980S}, who
find up to $f_{\rm col}=7$ for comparable, but slightly higher,
accretion rates.  This shows that photon hardening in Eddington to
super-Eddington flows must account for these additional opacities.

Table~\ref{tbl19} shows $\hat{f}_{\rm col}$ in the disk and $f_{\rm
  col}$ in the jet for other models.  These other models have a
similar distribution for $f_{\rm col}$ vs. radius and angle as model
M8 for models with double Compton and synchrotron or as model M7 for
models without these processes.  As $\dot{M}\to 1\dot{M}_{\rm Edd}$,
the photon hardening becomes stronger, a trend seen in AGN
\citep{2006AJ....131.2826S,2007ApJ...665.1004J}.

\begin{figure}
  \centering
  \includegraphics[width=0.99\columnwidth]{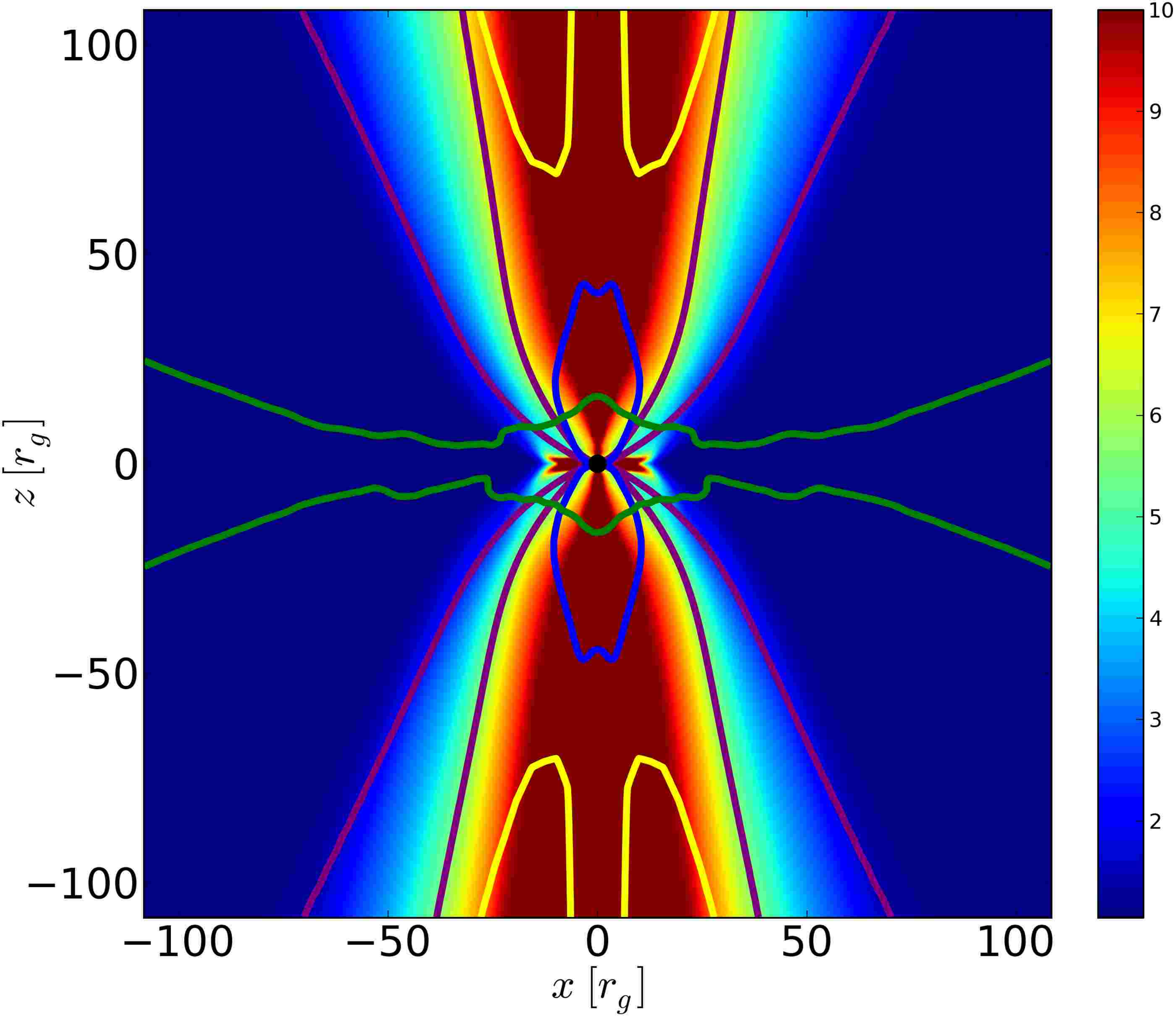}\hfill
  \includegraphics[width=0.99\columnwidth]{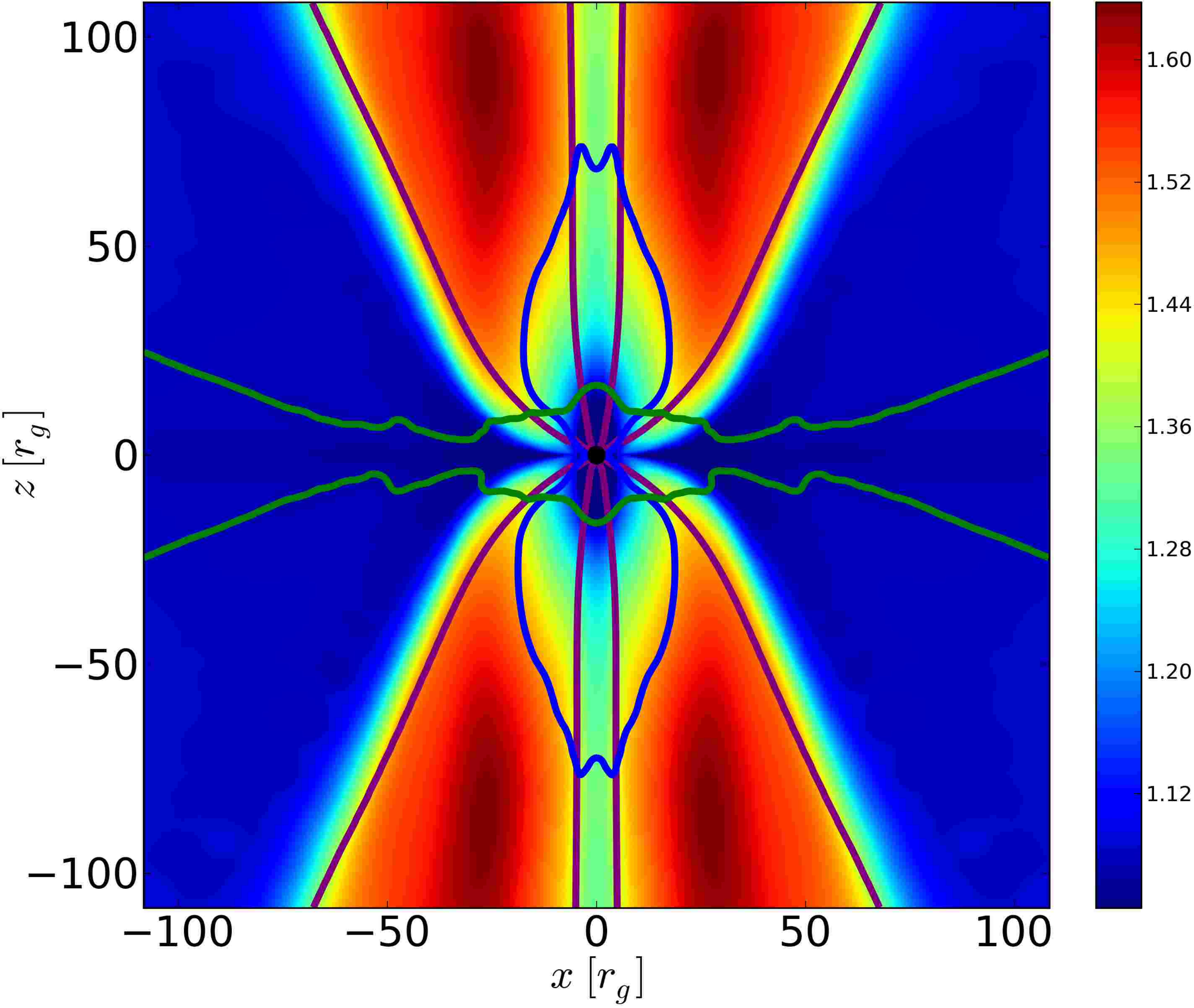}\hfill
  \caption{Models M7 and M8 with $a/M=0$, which have no BH-spin-driven
    jet, showing lab-frame $f_{\rm col}=T_\gamma/T_{\rm BB}$
    (numerator and denominator separately time-$\phi$-averaged).
    Green, blue, yellow, and purple lines are, if present, as in
    Fig.~\ref{fig:emradflux}.  Without double Compton or synchrotron,
    the radiation beam has $f_{\rm col}\sim 5$ (with overly high
    values in the funnel), while with these processes $f_{\rm col}\sim
    1.5$ -- a significant reduction due to the high opacity of double
    Compton at high radiation temperatures and the plentiful soft
    photons produced by synchrotron in regions where $T_{\rm
      gas}\gtrsim 10^8$K.}
  \label{fig:fcol0}
\end{figure}

Table~\ref{tbl19} includes two models, M10 and M11, where M10 has no
synchrotron while M11 has all our opacities.  These models have
comparable $\dot{M}$ and fluxes as shown in Table~\ref{tblfinal}.  The
goal here is to see if double Compton by itself regulates photon
hardening and to what extent synchrotron goes beyond double Compton.
In model M10, the radiation beam is only moderately hardened due to
the sensitivity of double Compton to radiation temperature, and
synchrotron in M11 provides soft photons that lead to essentially no
hardening of the radiation beam.  This means double Compton by itself
regulates photon hardening away from large values seen in other
simulations with no double Compton (e.g. M7).

Table~\ref{tbl19} includes two models, M3 and M5, where M3 has a
Bose-Einstein chemical potential, while M5 has a Planck chemical
potential.  The goal here is to see if chemical potential evolution
has an effect in the case where neither double Compton or synchrotron
are included.  These models are dynamically quite similar as shown in
Table~\ref{tblfinal}.  Table~\ref{tbl19} shows significantly more
photon hardening in the disk and radiation beam when using the
Bose-Einstein chemical potential.  This shows that the underlying
assumption about the photon distribution has a strong effect on the
photon hardening.

Fig.~\ref{fig:varexpf} shows how Wien the spectrum is, with the
primary radiation beam in model M15 being down from Planck
($\exp{(-\xi)}=1$) to $\exp{(-\xi)}\approx 0.85$.  This distribution
with radius and angle is typical for models with double Compton and
synchrotron.

\begin{figure}
  \centering
  \includegraphics[width=0.99\columnwidth]{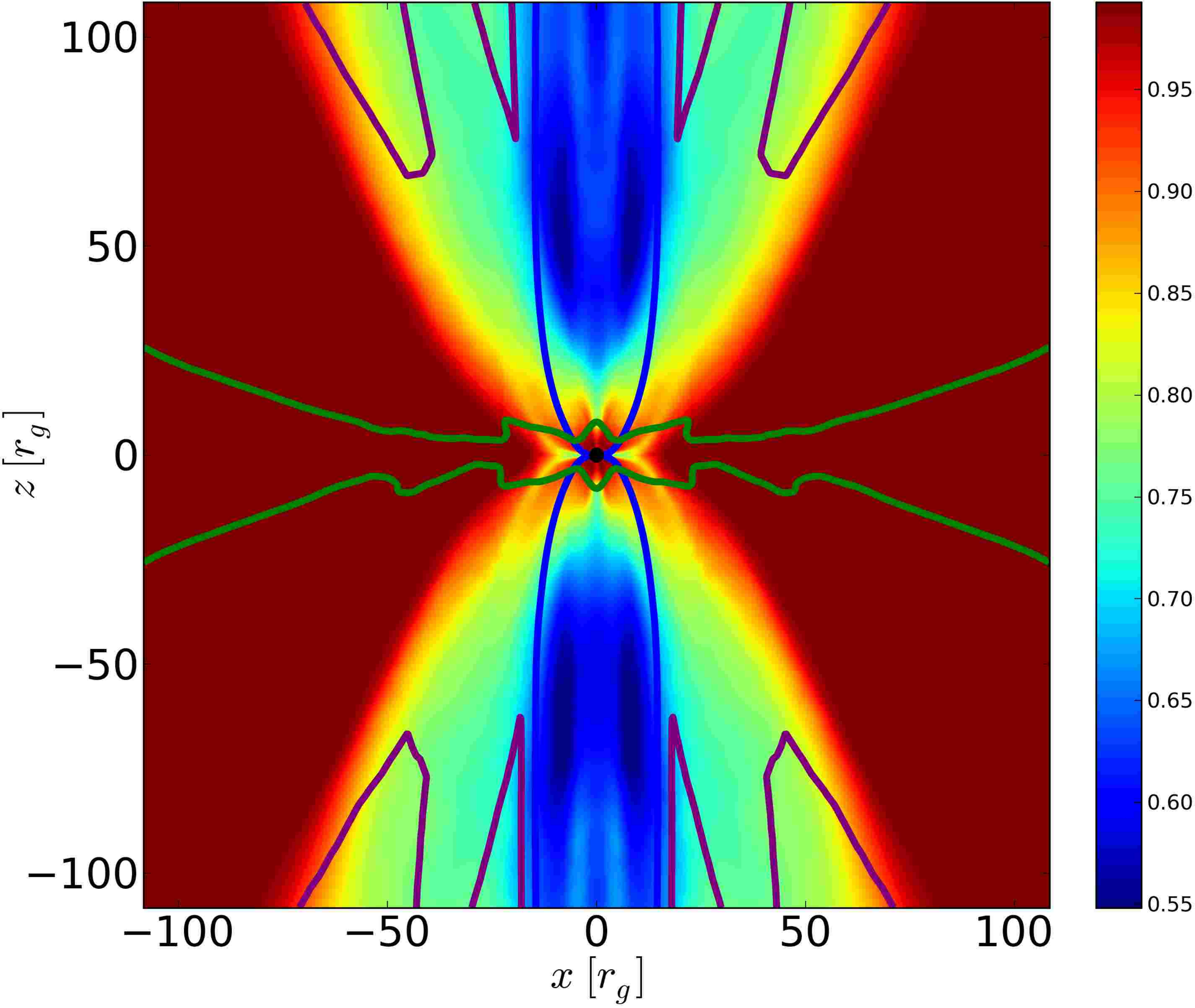}
  \caption{Model M15, showing fluid-Frame time-$\phi$-averaged
    dimensionless chemical potential factor ($\exp{(-\xi)}$ given by
    Eq.~\ref{expmu}, as color with legend) for the Bose-Einstein
    distribution.  Green, blue, yellow, and purple lines are, if
    present, as in Fig.~\ref{fig:emradflux}.  The radiation beam is
    slightly Wien with $\exp{(-\xi)}\approx 0.85$, and the spatial
    distribution is similar for other models.}
  \label{fig:varexpf}
\end{figure}

\subsection{Radial and Angular Dependencies for All Models}

Here we consider how radiative quantities are affected by photon
conservation, opacity choices, etc. Quantities like rest-mass density,
internal energy density, radiation-energy density, velocity, magnetic
field, etc. behave like power-laws vs. radius within $r\sim 20r_g$
within the inflow equilibrium region.  Such radial power-laws are
typical of MADs \citep{2012MNRAS.423.3083M}, so we do not discuss the
radial behavior here.

Fig.~\ref{fig:fluidvstheta} shows rest-mass density, gas temperature,
and radiation to gas temperature ratio vs. $\theta$.  The gas
temperatures are artificially high in models without thermal
Comptonization or in models with photon number evolution that do not
include double Compton and synchrotron.  For example, models M2 and
M10 do not include double Compton and synchrotron, while comparable
models M1 and M9 do.  Models with double Compton and synchrotron have
much lower gas and radiation temperatures in the corona and jet.  So,
the thermodynamical and radiative properties of the disk, corona, and
jet are only accurate with double Compton and synchrotron.

\begin{figure}
  \centering

  \includegraphics[width=0.99\columnwidth,trim={0.0cm 2.0cm 0.0cm 0.9cm},clip]{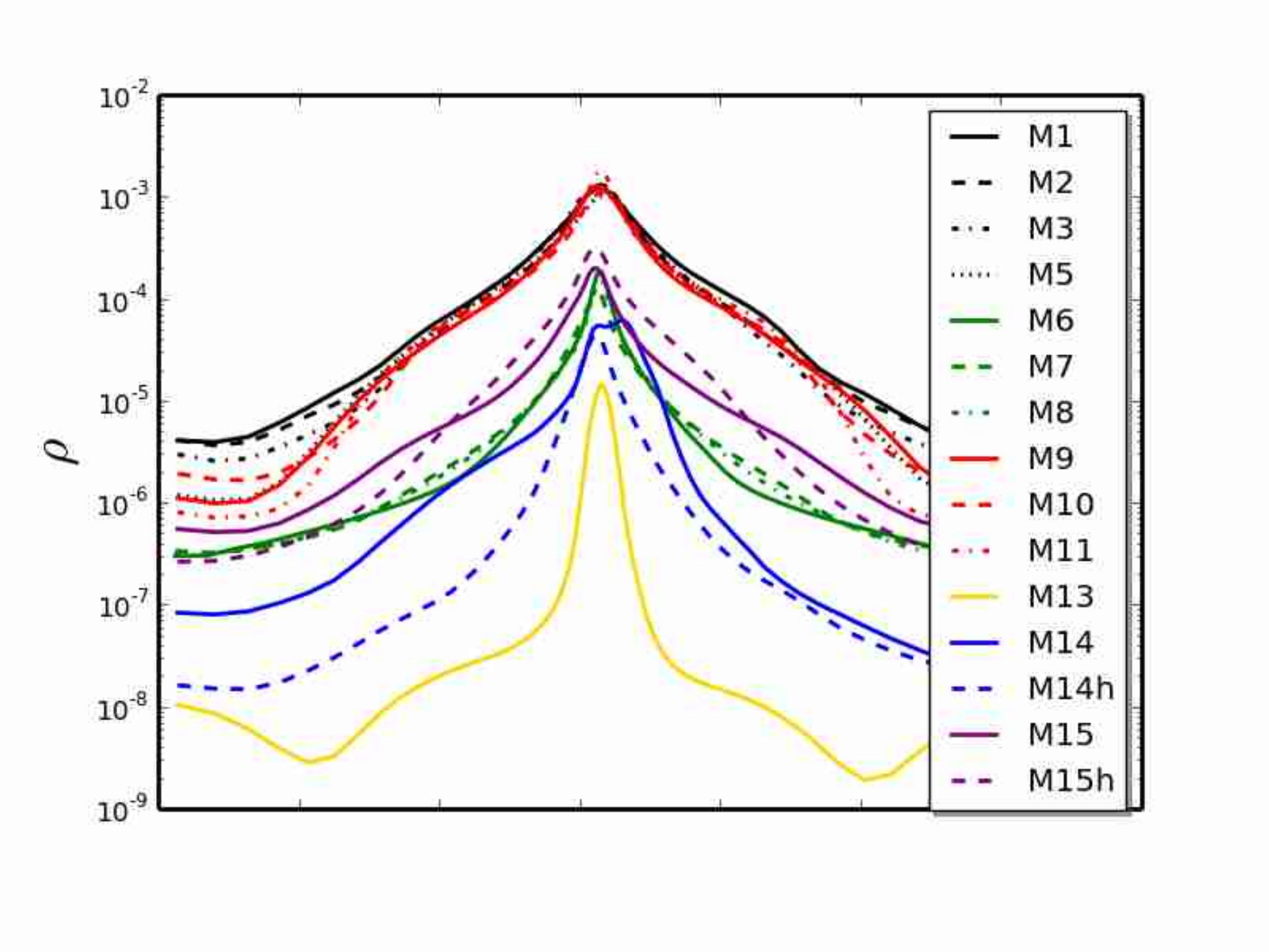}\hfill
  \includegraphics[width=0.99\columnwidth,trim={0.0cm 2.0cm 0.0cm 0.9cm},clip]{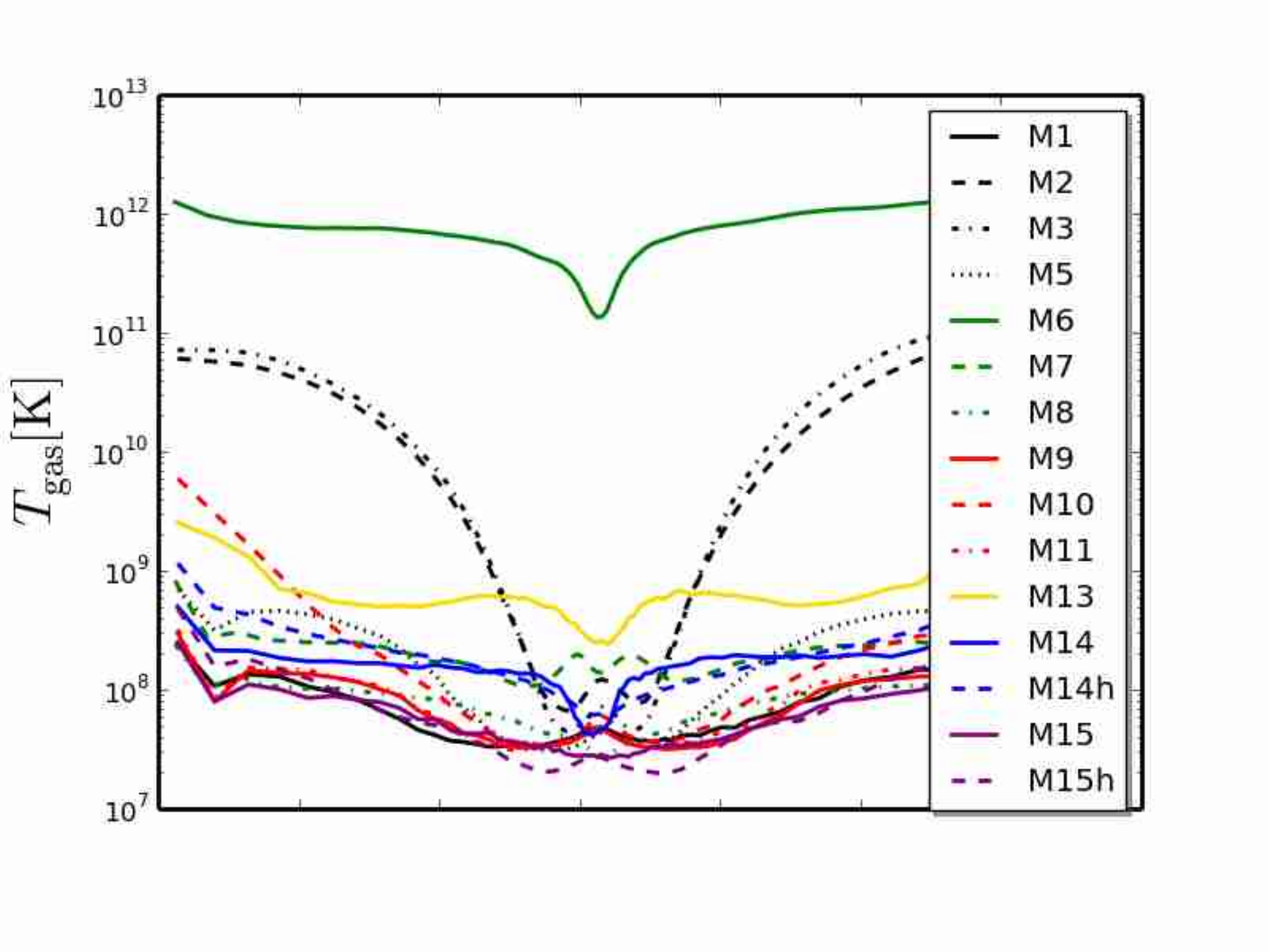}\hfill

  \includegraphics[width=0.99\columnwidth,trim={0.0cm 0.00cm 0.0cm 0.9cm},clip]{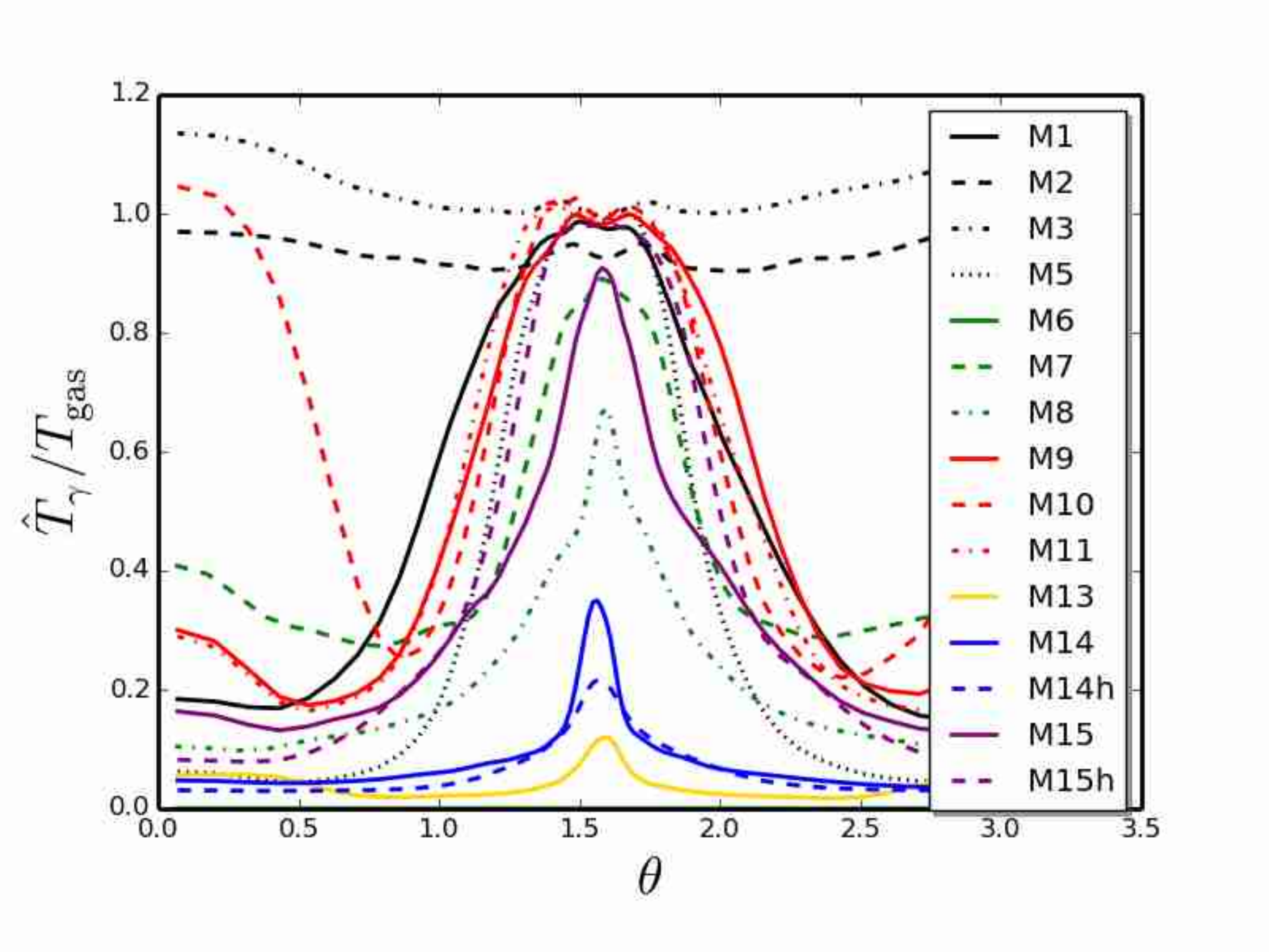}\hfill

  \caption{All models, showing fluid-frame density ($\rho$ in ${\rm
      g}~{\rm cm}^{-3}$, top panel), fluid-frame gas temperature
    ($T_{\rm gas}$, second panel), and ratio of fluid-frame radiation
    to gas temperatures ($\hat{T}_\gamma/T_{\rm gas}$, last panel)
    measured at $r=10r_g$.  Legend shows model types, where models are
    grouped by color for similar $\dot{M}$, $a/M$, or resolution
    changes for otherwise same initial density.  M1-M5 are high
    $\dot{M}\sim 100\dot{M}_{\rm Edd}$ models, M6-M8 are $a/M=0$
    models, M9-M11 are intermediate $\dot{M}\sim 50\dot{M}_{\rm Edd}$
    models, M13 is our lowest $\dot{M}\approx \dot{M}_{\rm Edd}$
    model, M14-M14h are our slightly higher $\dot{M}\approx
    3\dot{M}_{\rm Edd}$ models at different resolutions, and M15-M15h
    are intended to be $\dot{M}\approx 10\dot{M}_{\rm Edd}$ models
    (but M15h ended up with much higher $\dot{M}\approx 30\dot{M}_{\rm
      Edd}$).  Lines are drawn in order from M1-M15h, so latter lines
    may overlap earlier lines. High $\dot{M}$ models have a broader
    density profile compared to the more sharply-peaked low $\dot{M}$
    models.  Models with high $\dot{M}\sim 100\dot{M}_{\rm Edd}$
    without double Compton and synchrotron (M2,M3) or models without
    thermal Comptonization (M6) have unphysically hot coronae and jets
    by three orders of magnitude in temperature, showing that these
    processes are crucial to include in order to obtain accurate
    thermodynamical and radiative properties in flows with
    $\dot{M}\gtrsim \dot{M}_{\rm Edd}$.}
  \label{fig:fluidvstheta}
\end{figure}

Fig.~\ref{fig:TradoTgasvsr} shows the fluid-frame
$\hat{T}_\gamma/T_{\rm gas}$ (numerator and denominator separately
time-$\phi$-averaged) as weighted by rest-mass density times each grid
cell volume size.  This focuses the measurement of temperature on the
core of the disk at the highest densities.  Models at high $\dot{M}$
have $\hat{T}_\gamma\sim T_{\rm gas}$, and progressively lower
$\dot{M}$ down to $\dot{M}\sim \dot{M}_{\rm Edd}$ have gas
temperatures as high as ten times the radiation
temperatures. Resolution should modify the results the most for the
lowest $\dot{M}$ models, but the model like M14 and M14h (at twice the
$\phi$ resolution of M14) only show up to 30\% relative differences.
So, the corona region's temperatures are roughly converged.

Fig.~\ref{fig:TradoTgasvsr} also shows the fluid-frame
$\hat{T}_\gamma$ (weighted like $\hat{T}_\gamma/T_{\rm gas}$
above). Lower $\dot{M}\sim \dot{M}_{\rm Edd}$ are radiatively cooler
than higher $\dot{M}$ -- even for super-Eddington rates.  Compared to
the temperatures in \citet{2015arXiv150804980S} that were for
weakly-magnetized models at $\dot{M}\sim 10\dot{M}_{\rm Edd}$, our
results are comparable.  However, our model M7 without double Compton
and synchrotron demonstrates that our more magnetized disks would be
much hotter if it were not for double Compton and synchrotron like
included in model M8.  So double Compton and synchrotron are required
to regulate both gas and radiation temperatures in magnetized flows in
general.  Model M6 without thermal Comptonization shows extremely high
gas temperatures, similar to seen in \citet{2016ApJ...826...23T}, but
thermal Comptonization severely limits the difference between gas and
radiation temperatures by driving them toward equilibrium.

\begin{figure}
  \centering
  \includegraphics[width=0.99\columnwidth,trim={0.0cm 2.0cm 0.0cm 0.9cm},clip]{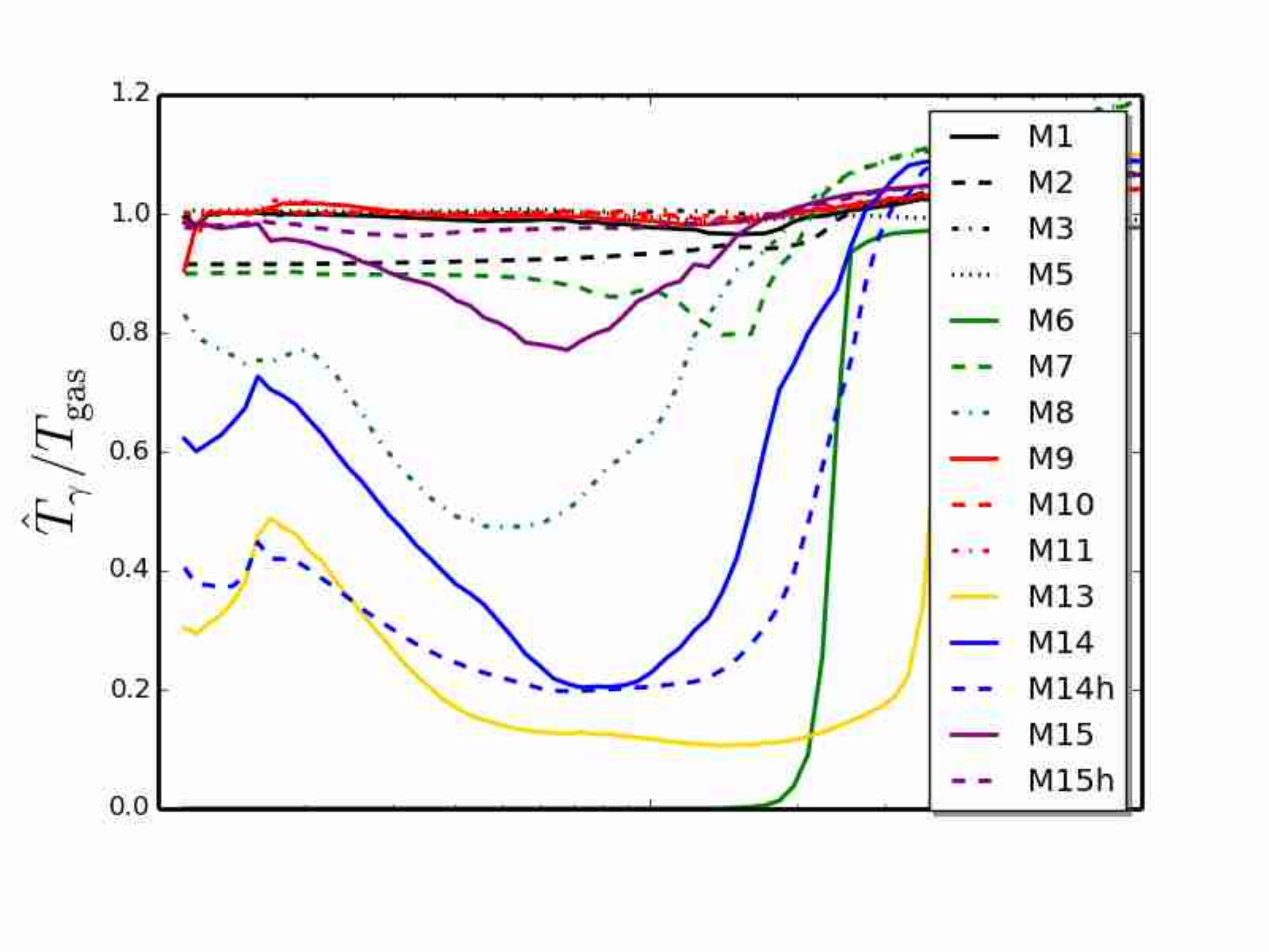}
  \includegraphics[width=0.99\columnwidth,trim={0.0cm 0.0cm 0.0cm 0.9cm},clip]{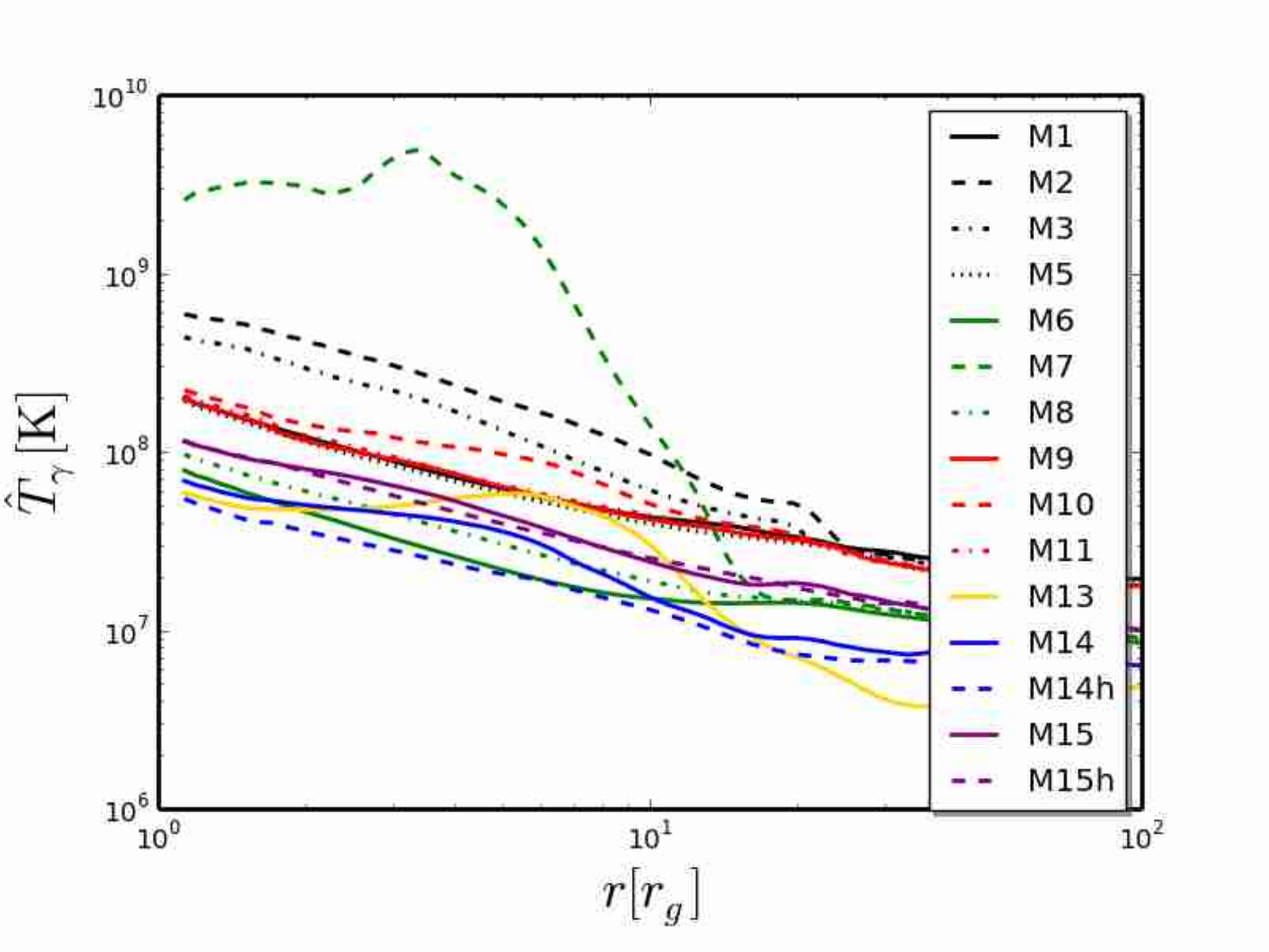}
  \caption{All models, showing fluid-frame $\hat{T}_\gamma/T_{\rm
      gas}$ (top panel) and fluid-frame $\hat{T}_\gamma$ (bottom
    panel) in the disk.  Even for $\dot{M}\sim \dot{M}_{\rm Edd}$, gas
    temperatures reach no more than ten times radiation temperatures
    at $r\sim 10r_g$ due to efficient thermal Comptonization.  Higher
    $\dot{M}$ tend to have higher radiation temperatures.  Model M7
    with photon number density evolution but without double Compton
    has much higher radiation temperatures than model M8 that has
    double Compton and is otherwise identical.  Model M6 is like M7
    but also has no thermal Comptonization, and as a consequence it
    has extremely high gas temperatures.  These results demonstrate
    the importance of thermal Comptonization and double Compton as
    thermostats that, respectively, lower gas and radiation
    temperatures in the disk.}
  \label{fig:TradoTgasvsr}
\end{figure}

Fig.~\ref{fig:nfcolvsr} shows the fluid-frame $\hat{f}_{\rm
  col}=\hat{T}_\gamma/T_{\rm BB}$, weighted like
$\hat{T}_\gamma/T_{\rm gas}$ above.  The photon distribution within
the disk is essentially Planckian at large radii but for low
$\dot{M}\sim \dot{M}_{\rm Edd}$ the distribution becomes harder at
intermediate radii of $r\sim 5r_g$.  Model M7 (without double Compton
and synchrotron, as compared to otherwise identical model M8 with
double Compton and synchrotron) has excessive unphysical hardening
even in the disk due photon conserving Comptonization driving the
photon distribution toward Wien.

Fig.~\ref{fig:nfcolvsr} also shows the fluid-frame chemical potential
factor $\exp{(-\xi)}$ given by Eq.~(\ref{expmu}), weighted like
$\hat{T}_\gamma/T_{\rm gas}$ above.  The photon distribution within
the disk is essentially Planckian at large radii, but for low
$\dot{M}\sim \dot{M}_{\rm Edd}$ (e.g. model M13) the distribution
becomes more Wien as densities become small enough that there is
inefficient photon production.

\begin{figure}
  \centering
  \includegraphics[width=0.99\columnwidth,trim={0.0cm 2.0cm 0.0cm 0.9cm},clip]{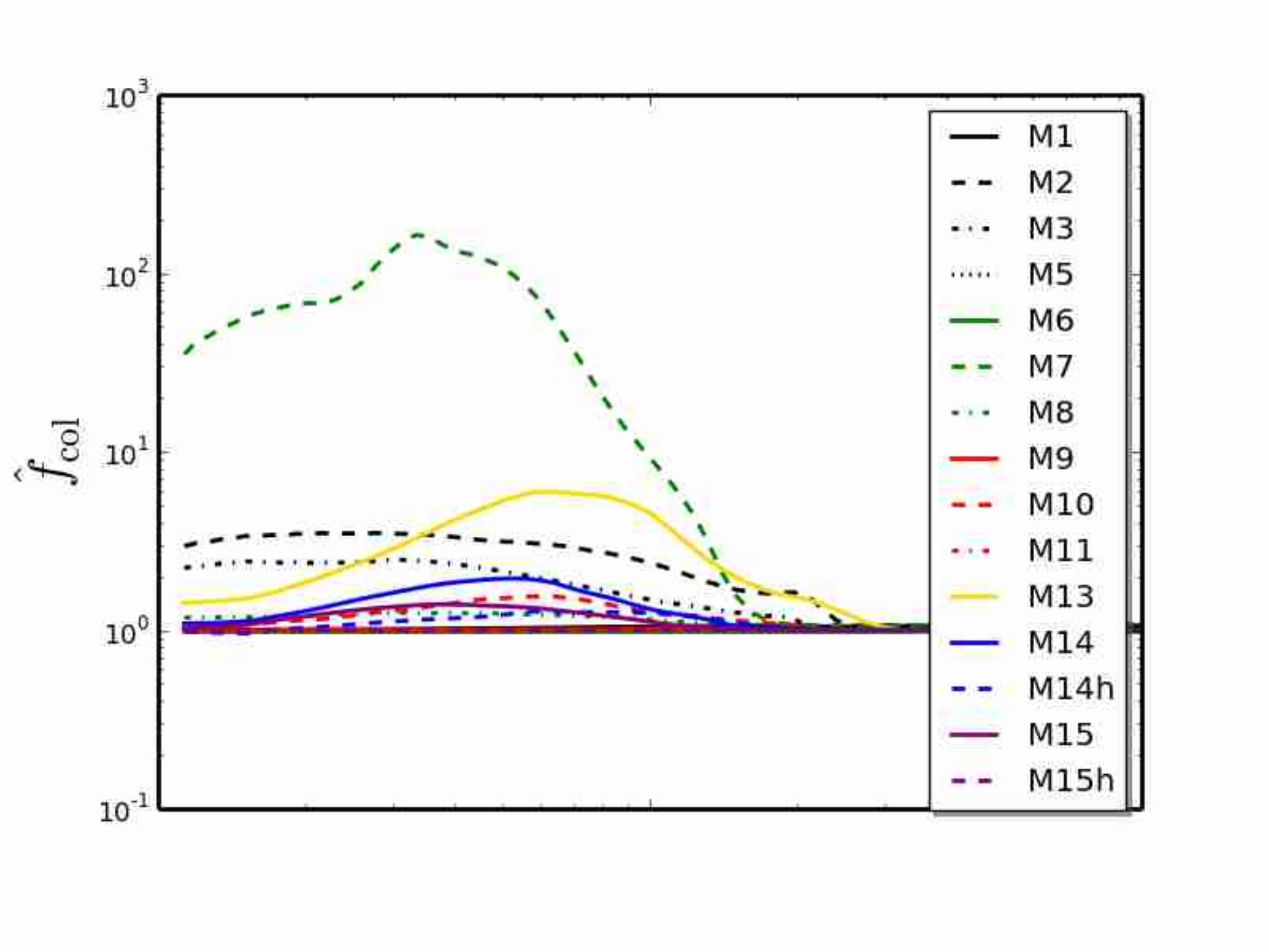}\hfill

  \includegraphics[width=0.99\columnwidth,trim={0.0cm 0.0cm 0.0cm 0.9cm},clip]{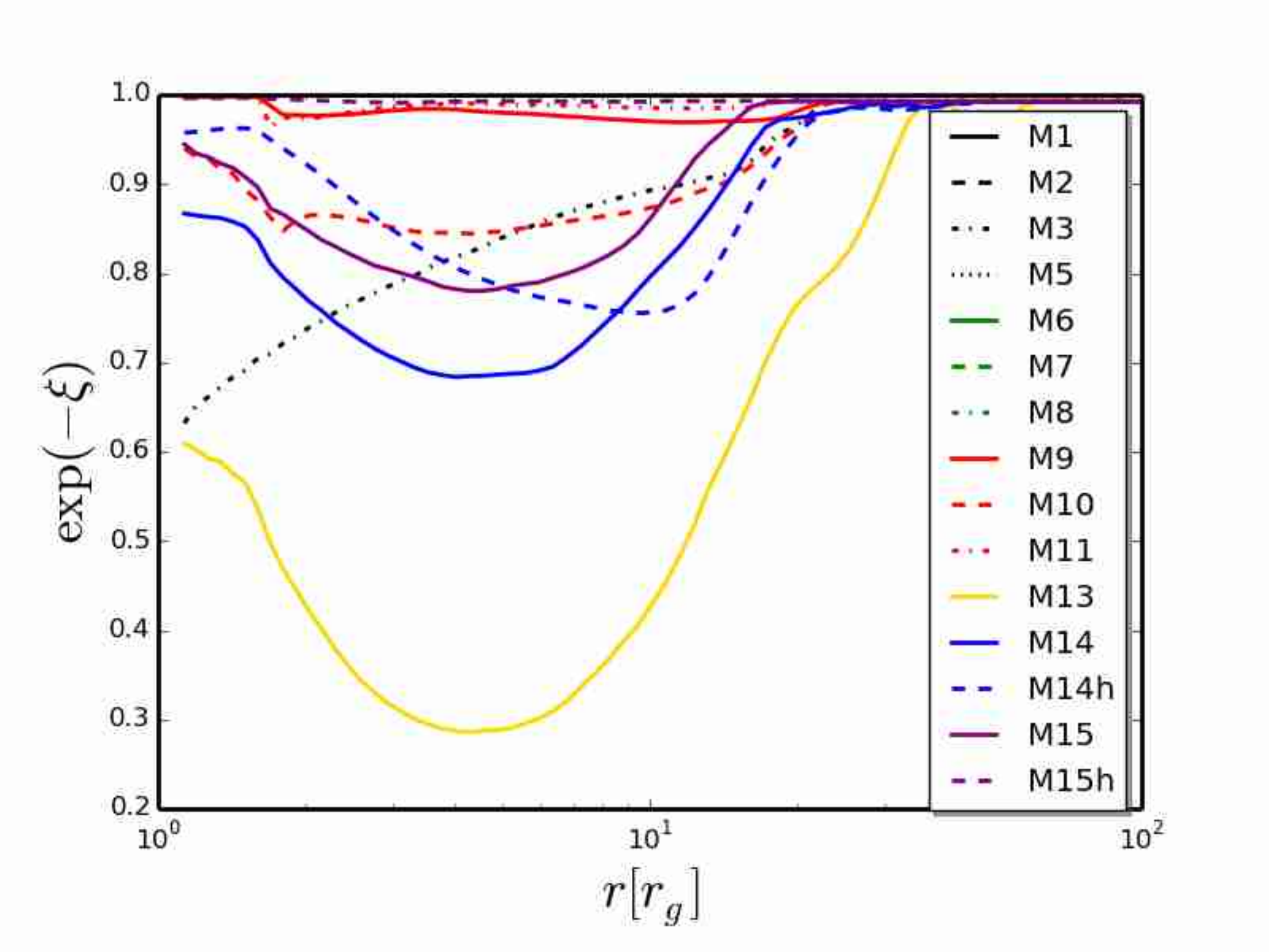}
  \caption{All models, showing fluid-frame $\hat{f}_{\rm col}$ (top
    panel) and chemical potential factor $\exp{(-\xi)}$ (bottom panel)
    in the disk.  Lower $\dot{M}$ (e.g. M13) tend to have regions in
    the disk around $r\sim 5r_g$ where the disk photon distribution is
    hardened, while at high $\dot{M}$ the disk photons are essentially
    Planckian as long as double Compton and synchrotron are included.
    Model M2, M3, and M7 have no double Compton or synchrotron,
    leading to unphysically large photon hardening in the disk.  This
    shows that double Compton and synchrotron are required in order to
    accurately track photon hardening.}
  \label{fig:nfcolvsr}
\end{figure}

Fig.~\ref{fig:kappadcvsr} shows the ratio of double Compton to
free-free opacities (units ${\rm cm}^{-1}$, which includes a density
scale) with no additional weighting, as averaged over a density scale
height for each simulation at each radius.  Higher $\dot{M}$ lead to
progressively more importance of double Compton relative to free-free
within the equatorial disk region.  For $\dot{M}\sim 10\dot{M}_{\rm
  Edd}$, double Compton dominates free-free within $r\sim 8r_g$,
within which significant radiation is generated.

Table~\ref{tbl19} includes models M1 and M2 that primarily differ in
that M1 includes double Compton and synchrotron while M2 does not.
Double Compton and synchrotron provide plentiful soft photons that
lead to much lower photon hardening than seen in M2, so this shows
that double Compton and synchrotron are crucial to include in
highly-magnetized MAD-type super-Eddington flows with $\dot{M}\sim
100\dot{M}_{\rm Edd}$.

\begin{figure}
  \centering
  \includegraphics[width=0.99\columnwidth,trim={0.0cm 0.0cm 0.0cm 0.9cm},clip]{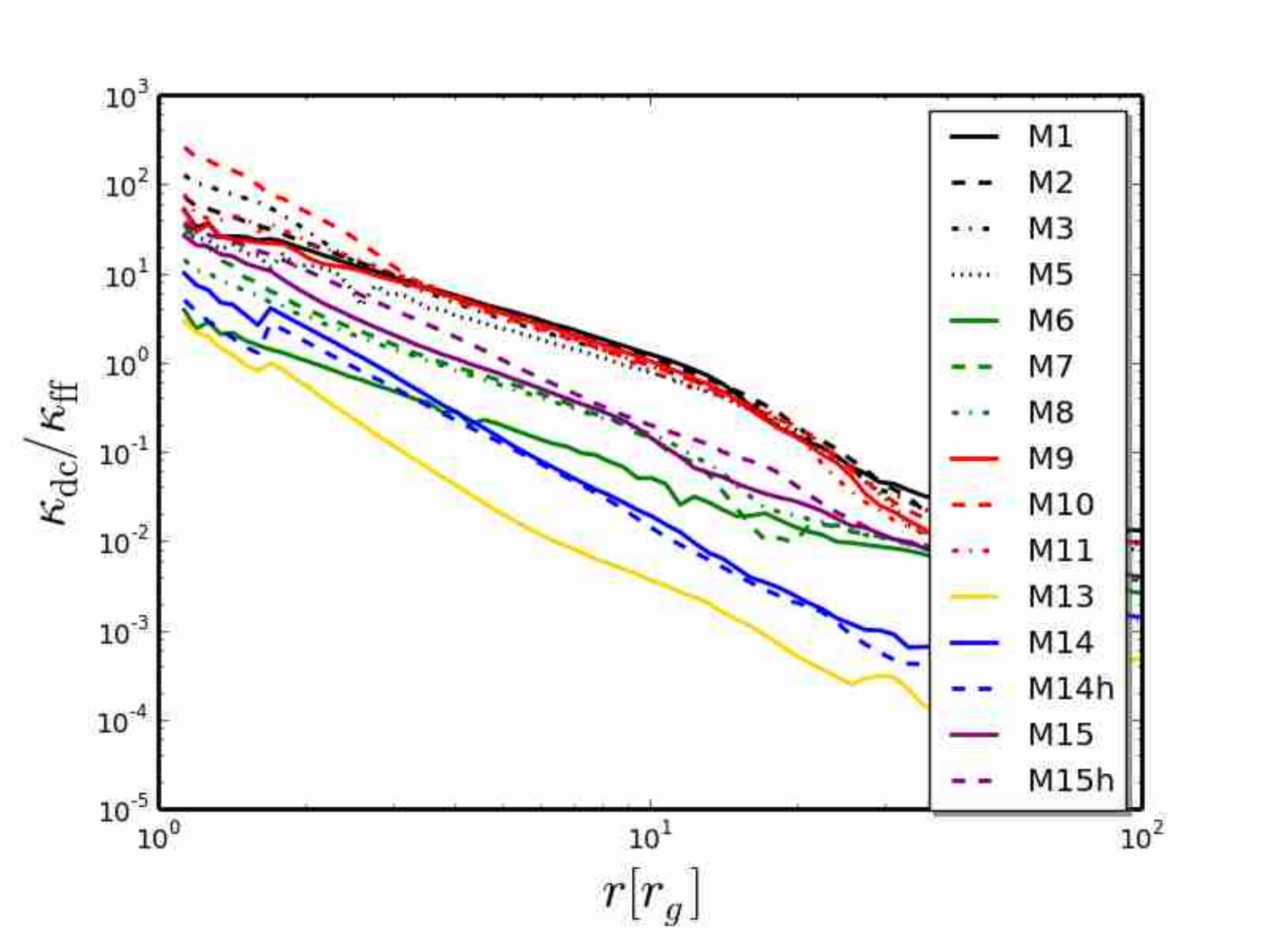}
  \includegraphics[width=0.99\columnwidth,trim={0.0cm 0.0cm 0.0cm 0.9cm},clip]{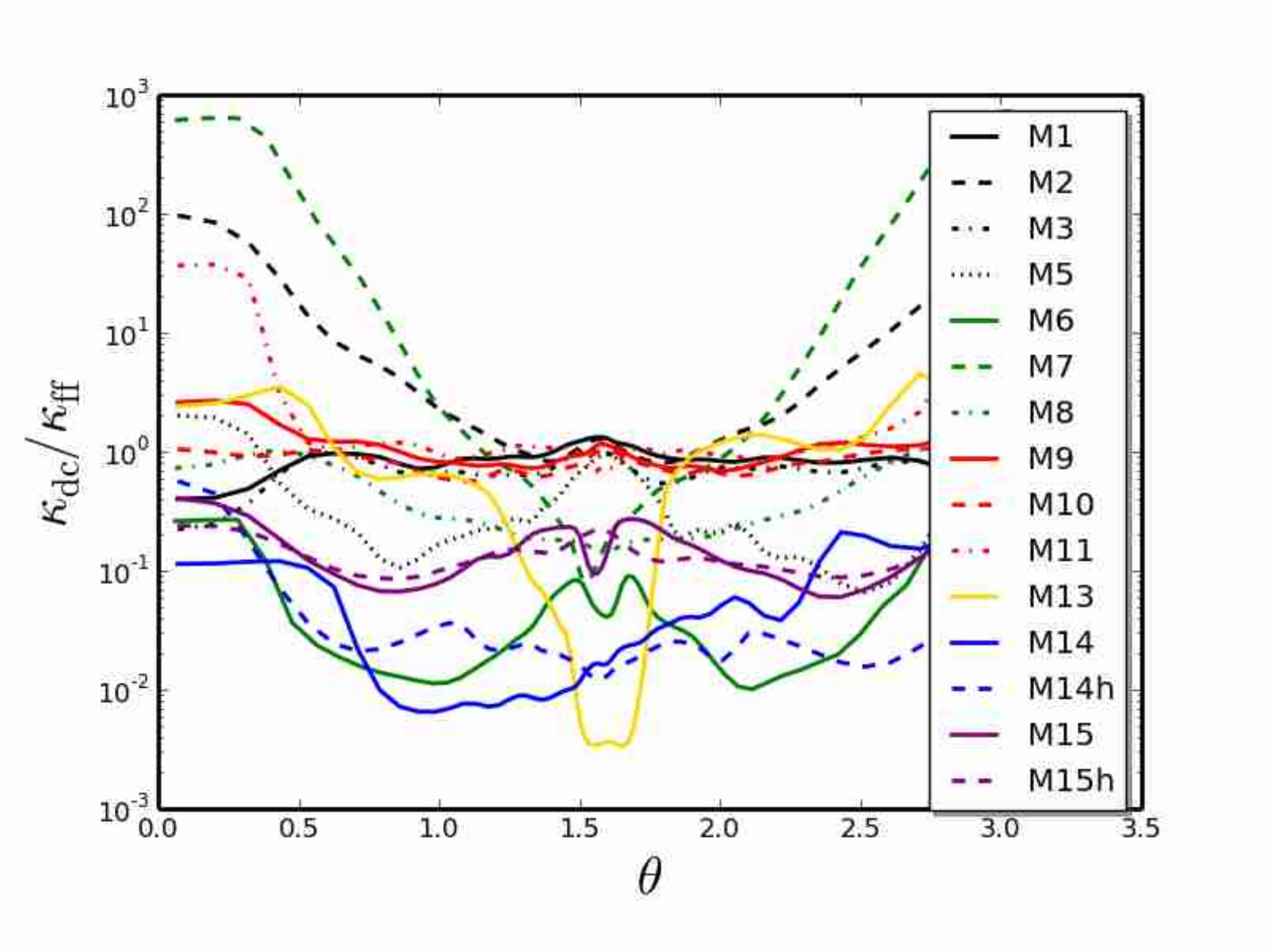}
  \caption{All models, showing fluid-frame double Compton opacity
    divided by free-free opacity in the disk (top panel) and across
    the disk at $r=10r_g$ (bottom panel).  Some models, like
    M2,M3,M5,M7, and M10, have no double Compton and show excessive
    double Compton opacity because it was not present to regulate the
    radiation temperature to lower values.  For $\dot{M}\sim
    \dot{M}_{\rm Edd}$ double Compton dominates free-free inside the
    disk within $r<3r_g$, for $\dot{M}\sim 10\dot{M}_{\rm Edd}$ double
    Compton dominates free-free inside the disk within $r<8r_g$, and
    for $\dot{M}\sim 100\dot{M}_{\rm Edd}$ double Compton dominates
    free-free inside the disk within $r<15r_g$ (and potentially
    further if simulation duration was extended, as inflow equilibrium
    is only out to $r\sim 20r_g$).  Comparing non-rotating black hole
    models M7 (no double Compton) and M8 (with double Compton), double
    Compton acts as a thermostat that keeps the plasma in a state
    where free-free and double Compton are comparable in the corona.
    Since significant radiation emerges from within $r=10r_g$ and the
    temperature of the corona affects the radiation that reaches large
    radii, double Compton is an important process to include for
    highly-magnetized super-Eddington accretion flows.}
  \label{fig:kappadcvsr}
\end{figure}

Fig.~\ref{fig:kappasyvsr} shows the synchrotron opacity divided by the
free-free opacity in the disk as well as $b^2/\rho$ (each $b^2$ and
$\rho$ weighted with density like done for $\hat{T}_\gamma/T_{\rm
  gas}$ above) in the disk.  At small radii, where magnetic energy
density approaches rest-mass energy density, synchrotron dominates
free-free in the disk for rapidly rotating black hole models.  This is
due to the jet and its interaction with the disk in a MAD state, where
at $r=10r_g$, the orthonormal $B^\phi$ is ten times higher at
$a/M=0.8$ than at $a/M=0$ outside the disk.

Fig.~\ref{fig:kappasyvstheta} shows the synchrotron energy opacity to
free-free energy opacity across the disk as well as $b^2/\rho$ (with
$\rho$ evaluated at the equator) across the disk at $r=10r_g$.
Synchrotron is not important as an energy opacity in the disk at such
radii, but in the jet is the dominant process due to the large
magnetic energy density.

By contrast, the synchrotron number opacities in the disk, corona, and
jet dominate the free-free (and even OPAL) opacities out to $r\sim
20r_g$ for $\dot{M}\gtrsim 100\dot{M}_{\rm Edd}$, out to $r\sim 5r_g$
for $\dot{M}\gtrsim 3\dot{M}_{\rm Edd}$, and only in the jet for
nearly sub-Eddington accretion rates.  This is due to the relatively
large $\phi\gtrsim 10^3$ (Eq.~\ref{phisy}) while also having
relatively stronger magnetic fields at higher accretion rates.

Model M6 without thermal Comptonization shows an artificial dominance
of synchrotron due to the artificially high gas temperatures, which
are suppressed by thermal Comptonization.

\begin{figure}
  \centering
  \includegraphics[width=0.99\columnwidth,trim={0.0cm 2.0cm 0.0cm 0.9cm},clip]{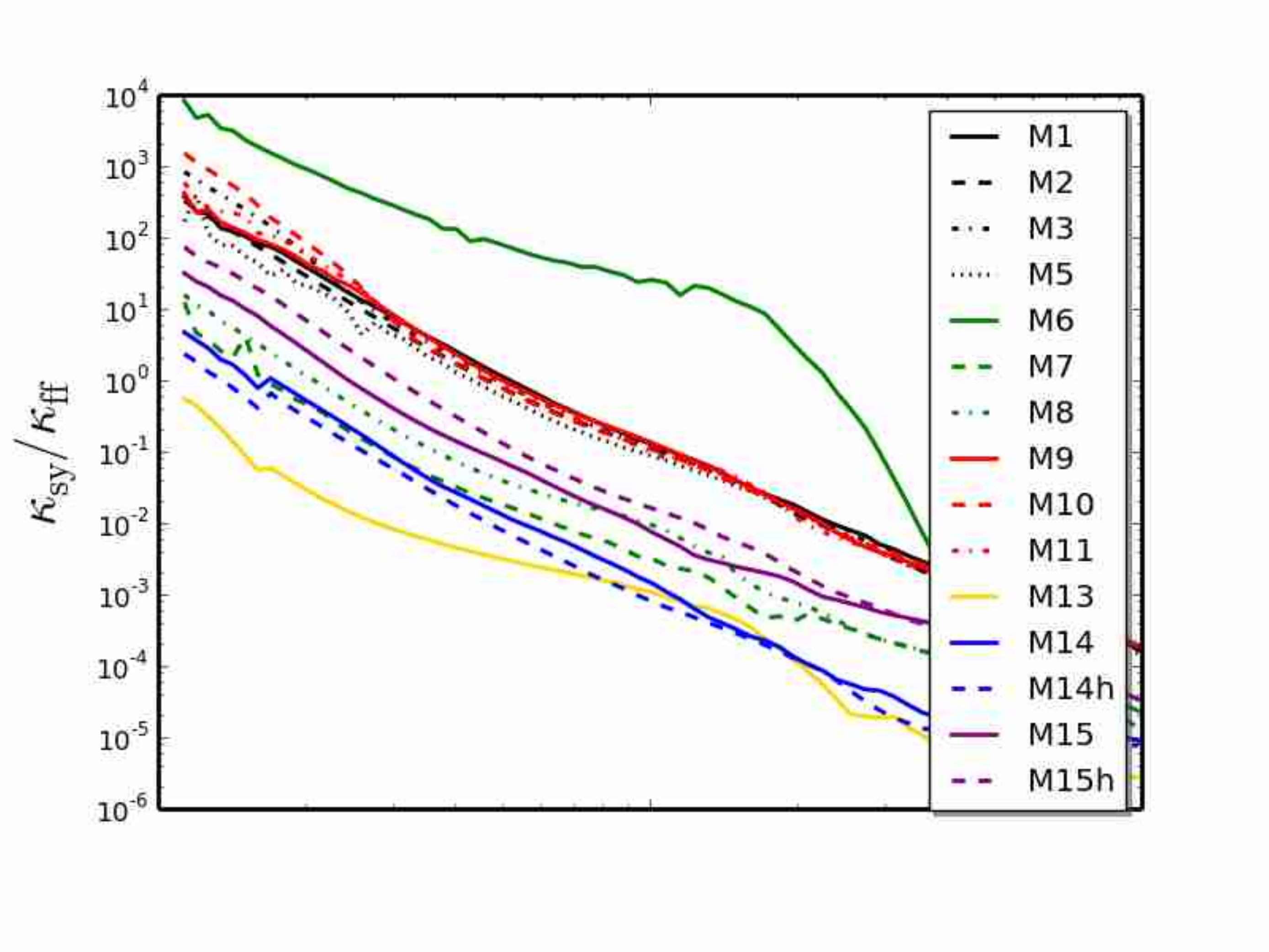}\hfill
  \includegraphics[width=0.99\columnwidth,trim={0.0cm 0.0cm 0.0cm 0.9cm},clip]{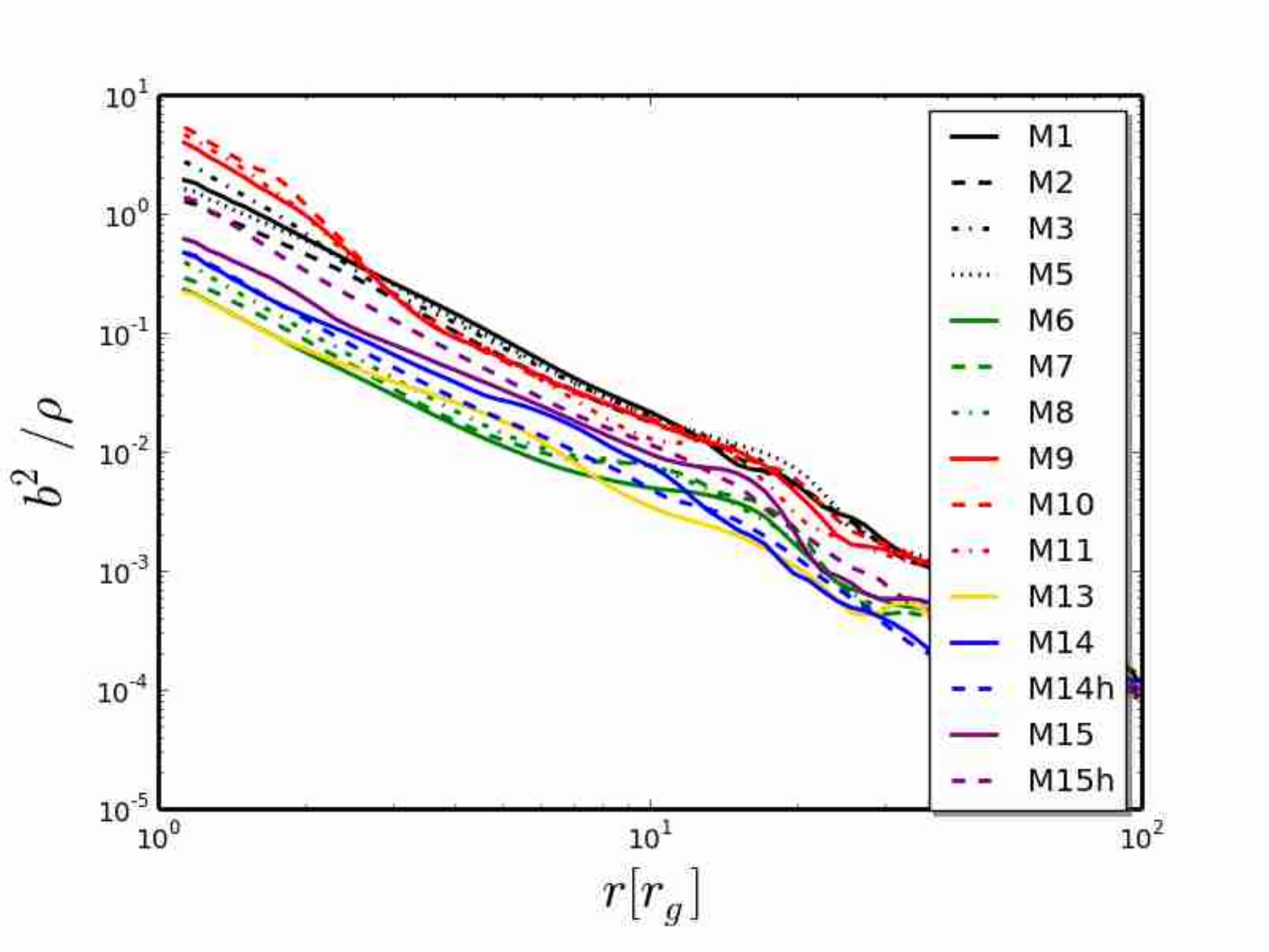}\hfill
  \caption{All models, showing fluid-frame synchrotron opacity divided
    by free-free opacity in the disk (top panel) and $b^2/\rho$ in the
    disk (bottom panel).  Synchrotron dominates free-free in the disk
    at small radii once temperatures rise to $T\sim 10^8$K and
    $b^2/\rho\sim 1$.  Low $\dot{M}\sim \dot{M}_{\rm Edd}$ models only
    have comparable synchrotron and free-free quite close to the
    horizon, but higher $\dot{M}$ have progressively important
    synchrotron that dominates free-free to larger radii -- out to
    $r\sim 5r_g$ for $\dot{M}\sim 100\dot{M}_{\rm Edd}$.  Model M6 has
    no thermal Comptonization or synchrotron, so gas temperatures and
    synchrotron emission opacities are unphysically high.  For models
    with thermal Comptonization, synchrotron is not important in the
    central disk except for quite high $\dot{M}$.}
  \label{fig:kappasyvsr}
\end{figure}

\begin{figure}
  \centering
  \includegraphics[width=0.99\columnwidth,trim={0.0cm 2.0cm 0.0cm 0.9cm},clip]{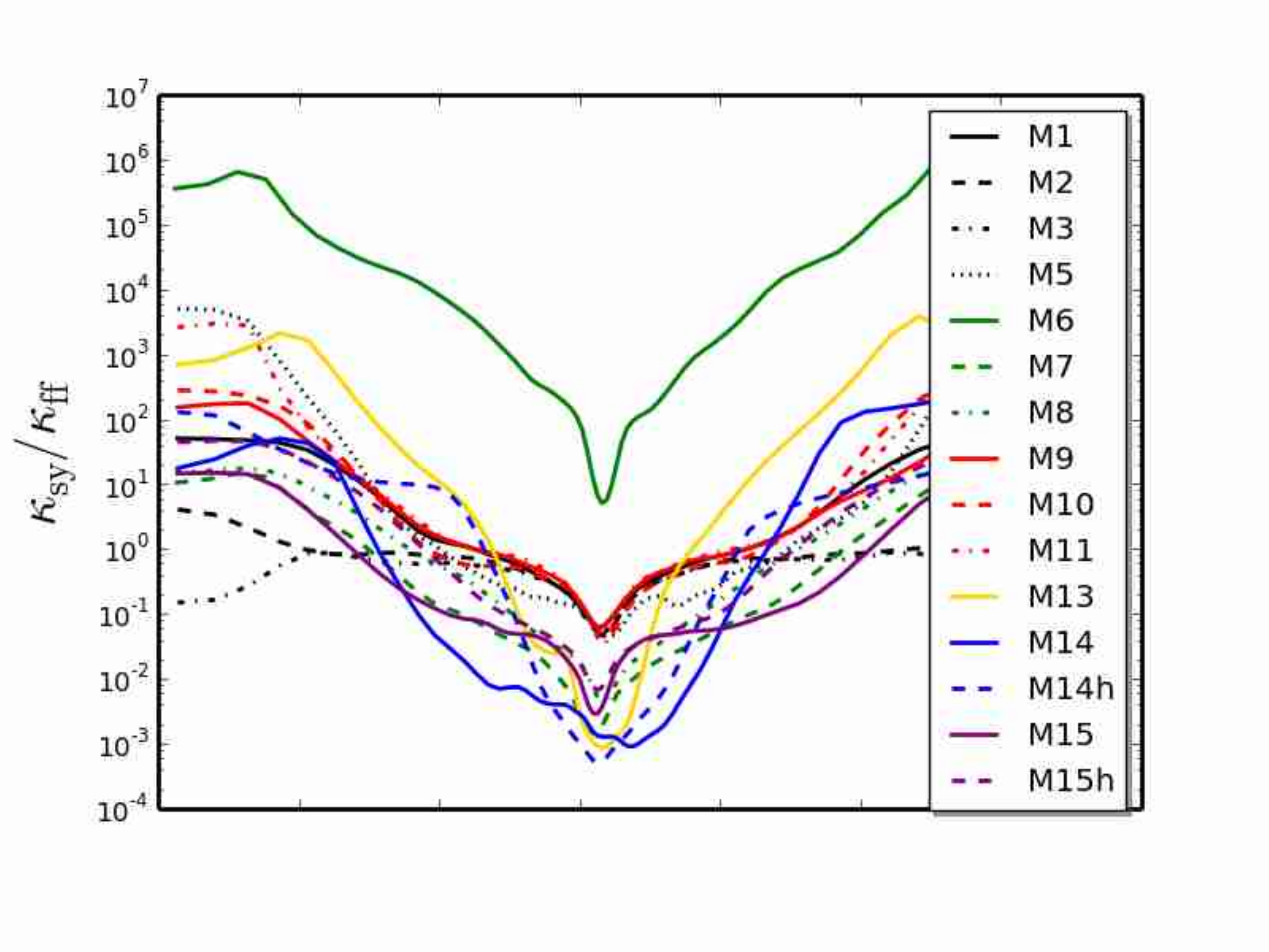}\hfill
  \includegraphics[width=0.99\columnwidth,trim={0.0cm 0.0cm 0.0cm 0.9cm},clip]{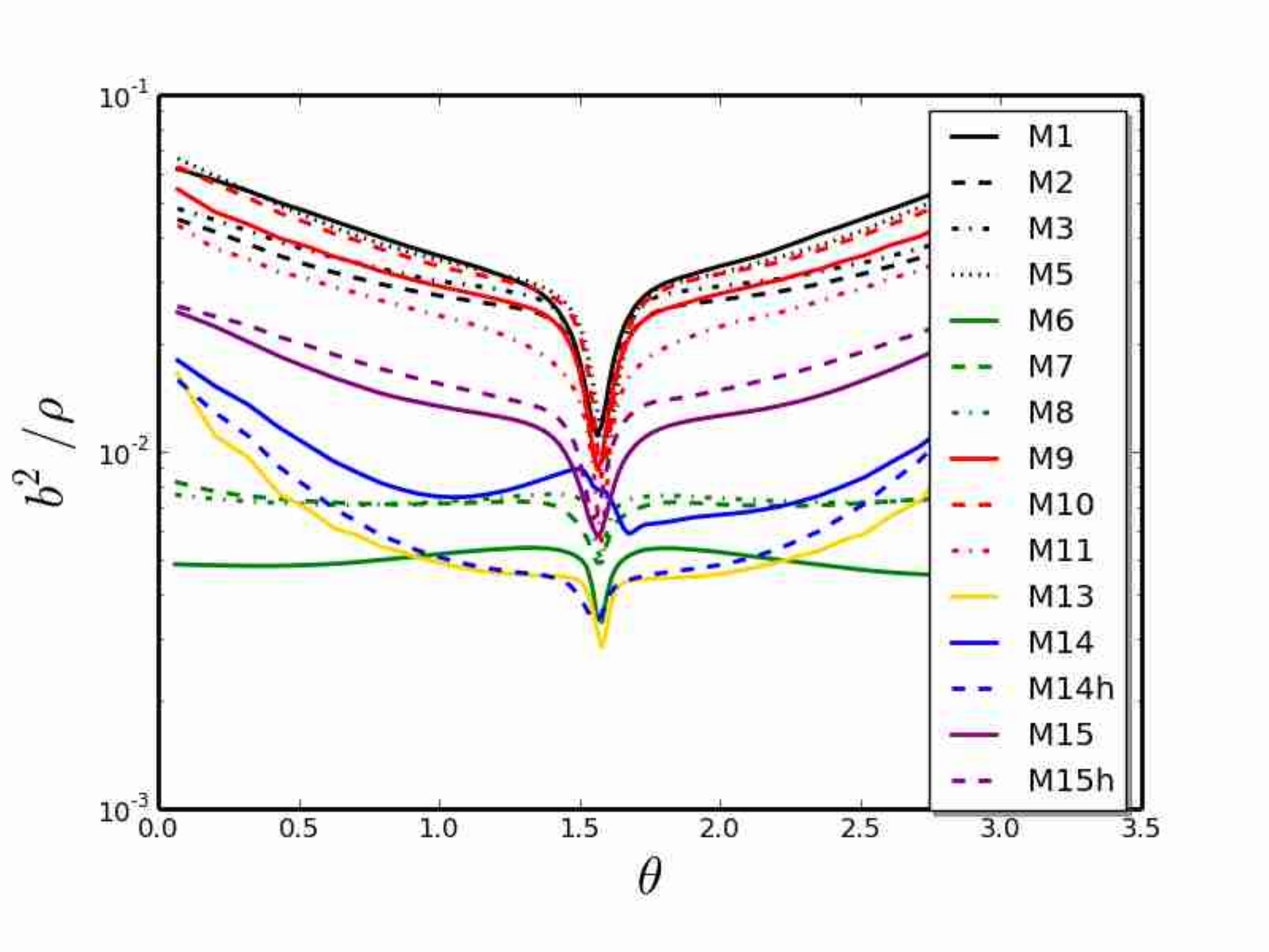}\hfill
  \caption{All models, showing fluid-frame synchrotron opacity divided
    by free-free opacity as well as $b^2/\rho$ (using density
    at equator for $b^2/\rho$) across the disk at $r=10r_g$.
    Synchrotron becomes as important as
    free-free in the corona in models with $\dot{M}\sim
    100\dot{M}_{\rm Edd}$ and becomes much more important than
    free-free in the jet for most models.  So synchrotron is crucial
    to include for the corona and jet thermodynamics in
    highly-magnetized MAD models with $\dot{M}\gtrsim \dot{M}_{\rm
      Edd}$.}
  \label{fig:kappasyvstheta}
\end{figure}

\section{Summary}\label{sec:summary}

We have incorporated several new opacity effects within HARMRAD as
applicable to black hole accretion flows accreting at Eddington to
super-Eddington rates.  We investigated a range of accretion rates
from $\dot{M}\sim 1\dot{M}_{\rm Edd}$ up to $\dot{M}\sim
100\dot{M}_{\rm Edd}$ in order to consider how thermal Comptonization,
double Compton, synchrotron, and other opacity effects control the
thermodynamic and radiative properties of highly-magnetized MAD type
accretion flows.

We found that double Compton dominates free-free in the central core
of the disk even out to fairly large radii of $r\sim 15r_g$ (and
potentially much further if we considered longer duration simulations)
in highly super-Eddington accretion flows with $\dot{M}\sim
100\dot{M}_{\rm Edd}$.  We also found for these MAD models that
synchrotron dominates free-free within $r\sim 5r_g$ for highly
super-Eddington accretion flows, and synchrotron provides plentiful
soft photons to regulate radiation temperatures throughout the flow
for $\dot{M}\gtrsim 10\dot{M}_{\rm Edd}$.  Progressively lower
accretion rates are dominated by double Compton and synchrotron within
progressively smaller radii until for sub-Eddington accretion these
processes do not dominate in the central core of the disk.

For our MAD models, we found that the coronal gas and radiation
temperatures are regulated by double Compton and synchrotron, while
jet temperatures are regulated by synchrotron.  Of our more realistic
models with all our opacity physics, our model with $\dot{M}\approx
\dot{M}_{\rm Edd}$ shows the highest disk gas temperatures compared to
higher accretion rate models.  The surrounding coronal gas has
temperatures of $T_{\rm e}\approx 7\times 10^8$K corresponding to
$60$keV, which is $\approx 100$ times higher than the disk's black
body temperature and $\approx 20$ times higher than the disk's
radiation temperature (hardened by $f_{\rm col}\approx 4.5$).  This
compares to other recent local shearing box simulations that studied
coronae and saw only up to $T\sim 10^7$K in the corona and a factor of
$10$ cooler disk black body temperatures \citep{2014ApJ...784..169J}.
The polar jet region near the black hole has $T_e\sim 4\times 10^9$K
corresponding to $350$keV, about ten times higher than models with
$\dot{M}\approx 30\dot{M}_{\rm Edd}$, which should also contribute to
Comptonized emission \citep{2016ApJ...819...95O,2016arXiv160701060O}.
The gas heating in the corona is plausibly driven by the strong
highly-dynamic magnetic fields threading the black hole and MAD type
disk, but more detailed analysis of this state will be presented in
future work.

Thermal Comptonization with double Compton and synchrotron were
crucial to include, otherwise gas temperatures were excessively high
by factors of ten (with thermal Comptonization, but without double
Compton or synchrotron) and up to factors of a thousand (when no
thermal Comptonization is included) as seen in
\citet{2016ApJ...826...23T}.  However, their suggestion that hot gas
regions could lead to the very high state \citep{2004MNRAS.353..980K}
might still apply to our model with $\dot{M}\sim \dot{M}_{\rm Edd}$,
which has disk gas temperatures ten times higher than disk radiation
temperatures.

We also found that for rotating black hole models, the isotropic
equivalent luminosity is enhanced compared to the total luminosity by
a beaming factor of up to $b=12$ for the jet and up to $b=8$ for the
radiation.  Non-rotating models show only moderate beaming of the jet
by $b=5$ and radiation by factor $b\sim 2$--$3$, which agrees with
\citet{2015arXiv150300654S}.  The rotating black hole models with
significant beaming of radiation could help explain ultra-luminous
X-ray sources (ULXs).

None of our models show signs of thermal instability, as consistent
with our prior studies of super-Eddington flows
\citep{2014MNRAS.441.3177M,2015MNRAS.454L...6M}.  The stability of
radiation-dominated disks is uncertain, but affected by numerical
method details, opacities used, the degree of magnetization, and how
these lead to enhanced advection
\citep{2009ApJ...691...16H,2013ApJ...778...65J,2016MNRAS.459.4397S,2016arXiv160106836J,2016arXiv160304082M}. Free-free
does not limit the thermal instability for increasing temperatures due
to its opacity dropping, but processes like double Compton increase
with temperature and so might ultimately limit such runaways.

The primary limitation of our study is the use of the M1 closure
approximation because rays cannot intersect in the optically thin
regime.  This limits applications to cases where most of the radiation
is in the scattering-dominated regime (true for all of our models in
the disk, and true for many of our models out to large radii) or where
radiation emerges primarily from a single region (as common in
accretion disks, for which radiation comes from near the BH)
\citep{2015MNRAS.453.1108T}.  Multi-frequency transport could also be
important \citep{2016arXiv160407848R}.

Our study is most relevant to observations related to tidal disruption
events (TDEs) \citep{2015ApJ...812L..39D}, super-Eddington black hole
X-ray binary systems like GRS1915+105 and SS433
\citep{2011MNRAS.414.1183K}, ULXs, and numerous quasars that accrete
near or above Eddington.  Our results show that radiation beaming and
photon hardening effects are strong, and that Eddington to
super-Eddington models require double Compton and synchrotron in order
to obtain accurate temperatures in the corona and jet.  Double Compton
may be an important thermostat (that enforces slightly lower
temperatures than pair creation seen in \citealt{2015MNRAS.451.4375F})
in quasars and other radiatively efficient super-massive BHs.  Either
improvements to our radiative transfer scheme or multi-frequency
radiative transfer post-processing \citep{2016MNRAS.457..608N} could
allow these simulations to be compared directly with spectra and
timing from these systems \citep{2016MNRAS.458.1839C}.

\section*{Acknowledgments}

JCM thanks Jane Dai for useful discussions and acknowledges support by
NASA/NSF/TCAN (NNX14AB46G) as well as computing time from
NSF/XSEDE/TACC (TG-PHY120005) and NASA/Pleiades (SMD-14-5451).  JC
acknowledges support by the Royal Society as a Royal Society
University Research Fellow at the University of Manchester, UK.  MW
acknowledges support of the Foundation for Polish Science within the
START programme and the Polish NCN grant UMO-2013/08/A/ST9/00795 RN
was supported in part by NSF grant AST1312651 and NASA grant
NNX14AB16G.  AS acknowledges support by NASA through the Einstein
Postdoctoral Fellowship number PF4-150126.

\appendix

\section{Bose-Einstein Photon Distribution}\label{sec:be}

The fluid-frame quantities $E$ and $n$ described in \S\ref{sec:eom}
provide two parameters that describe the photon distribution function.
All quantities in these appendices are in the fluid frame.  A
reasonable assumption for general optical depths is a Bose-Einstein
(BE) distribution \citep{2011fxts.confE..24M}, which correctly
captures the behavior of the photon distribution in a
scattering-dominated atmosphere and how it goes from Planck to Wien as
radiation and gas temperatures equilibrate.  A diluted Planck might be
reasonable in some cases \citep{2016arXiv160503184S}. We assume BE
holds in the fluid frame, which will be valid in the optically thick
regime or when the radiation is fairly isotropic in the fluid frame
and energy exchange with the electrons drives the photon distribution
towards kinetic equilibrium.  This neglects Doppler effects, because
the radiation should be isotropic in its own frame according to the M1
closure.

Let us define $x=h\nu/(k_b T_\gamma)$ and $\xi=\mu/(k_b T_\gamma)$.  Then the
Bose-Einstein distribution function is
\begin{equation}
n_x = \frac{{\rm e}^{-\xi}}{{\rm e}^x - {\rm e}^{-\xi}} .
\end{equation}
The angle-integrated distribution of photons is given by the number
distribution
\begin{equation}
{BN}dx = C (k_b T_\gamma)^3 {\rm e}^{-\xi} \frac{x^2 dx}{{\rm e}^x -
  {\rm e}^{-\xi}} = C (k_b T_\gamma)^3 {\rm e}^{-\xi} dI_n ,
\end{equation}
with $C = (8\pi)/(h^3 c^3)$.  The corresponding energy distribution is
\begin{equation}
{BE}dx = C (k_b T_\gamma)^4 {\rm e}^{-\xi} \frac{x^3 dx}{{\rm e}^x - {\rm
e}^{-\xi}} = C (k_b T_\gamma)^4 {\rm e}^{-\xi} dI_E .
\end{equation}
Then, the radiation number and energy densities are
\bsub
\beal
n(\xi)&= C (k_b T_\gamma)^3 {\rm e}^{-\xi} I_n(\xi) ,
\\
E(\xi)&= C (k_b T_\gamma)^4 {\rm e}^{-\xi} I_E(\xi) ,
\end{align}
\esub
respectively. This corresponds to two equations and two unknowns for $T_\gamma$ and $\xi$
(or equivalently ${\rm e}^{-\xi}$).

Solving these equations, one obtains ${\rm e}^{-\xi} = n^4 I_E^3/(C E
I_n^4)$ and $k_b T_\gamma = E I_n/(I_E n)$. Noting the behavior of
$T_\gamma$ and ${\rm e}^{-\xi}$ is like exponentials, and searching
for a simple fitting function to avoid an inversion of a Polylog, we
find a fit for radiation temperature of
\begin{equation}\label{tgamma}
k_b T_\gamma = \frac{E/n}{0.33333 + 0.060725/(0.646756 + 0.121982 C E^3/n^4)}
\end{equation}
with a corresponding equation for the chemical potential ($\mu$)
written in terms of a dimensionless $\xi=\mu/(k_b T_\gamma)$ as
\begin{equation}\label{expmu}
{\rm e}^{-\xi} = \frac{1.64676}{0.646756 + 0.121982 C E^3/n^4} .
\end{equation}
These fits agree with the full solution to $<2\%$ from Planck
($\mu=0$) to Wien ($\mu\to\infty$) limits. To match the Bose-Einstein
condensate limit $\mu=0$ when $T_\gamma\to 0$, we restrict ${\rm
e}^{-\xi}<1$, while $T_\gamma$ naturally tends to zero as
$n\to \infty$ for fixed $E$. Now given the equations of motion value
for $E$ and $n$, we can obtain $T_\gamma$ and ${\rm e}^{-\xi}$ that
are used to compute any opacity or emission rate.

Note that for Planck distributions, the integration over the $BE_\nu$
distribution function gives $u_0 = B (4\pi/c)$ with Planck photon
energy density of $u_0=aT^4$ ($a$ is the radiation constant) and
integration over the $BN_\nu$ distribution gives $n_0=(30 {\rm
  Zeta}(3)/\pi^4) (a/k_b) T^3\approx (1/2.7) (u_{0}/(k_b T))$ where $N
= n_{0}c/(4\pi)$.

More generally, a BE distribution can be considered as a rough fit to
the frequency position of the peak and the shape of the soft photon
portion for the general distribution $n_x$. This fit would break-down
when the emission is optically thin and absorption occurs far from the
peak in emission.

\section{Mean Emission and Absorption}\label{meanopacities}

Here we discuss how the mean opacity should be computed for Boltzmann
moment methods like the M1 closure method. Here, the emitted mean
intensity is $j_\nu = e_\nu/(4\pi)$ and $e_\nu$ is the emissivity
(energy loss per unit volume per unit time per unit frequency).  The
energy loss rate per unit volume is then $\lambda_e=\int e_\nu d\nu$.

\subsection{Integrated evolution equations}\label{integratedeqs}

We obtain the form of Kirchhoff's law and the correct mean opacity
from the Boltzmann equation in the fluid frame in a homogeneous plasma
without Compton scattering (not required to obtain our result). The
Boltzmann equation for the photon occupation number $n_x$ with
$x\equiv h\nu/(k_b T_\gamma)$ and $\xe\equiv h\nu/(k_b \Te)$, caused
by bremsstrahlung (BR) and double Compton (DC) emission without
Kompaneets, reads

\begin{eqnarray}
\label{eq:B_BR_DC}
\partial_\tau n_x &=& \frac{\Lambda_{\rm BR}(x, \Tg, \Te)}{x^3}\left[1-n_x (\expfun{\xe}-1)\right]\\
 &+& \frac{\Lambda_{\rm DC}(x, \Tg, \Te)}{x^3}\left[1-n_x (\expfun{x}-1)\right]\nonumber,
\end{eqnarray}

in the fluid-frame for a fluid element with proper time $\tau$, where
$\Lambda_{\rm X}(x, \Tg, \Te)/x^3 = dn_x/dt$ of emission, describes
the respective emission rates (excluding stimulated emission). The BR
process drives the photon occupation number towards a black-body at
temperature $\Te$, i.e., $n_x=1/(\expfun{\xe}-1)$, while DC pushes
towards a black-body at temperature $\Tg$, i.e.,
$n_x=1/(\expfun{x}-1)$ (see \S\ref{subsec:dc}).  Neither of these two
solutions make the Boltzmann collision term vanish for $\Te\neq \Tg$
and the general equilibrium solution with respect to the BR and DC
processes is given by

\begin{eqnarray}
n^{\rm eq}_x= \left(\frac{\Lambda_{\rm BR} \expfun{\xe}
+ \Lambda_{\rm DC} \expfun{x}}{\Lambda_{\rm BR}+\Lambda_{\rm DC}}-1\right)^{-1} .
\end{eqnarray}

As expected, for $\Te=\Tg$ this reduces to $n^{\rm
eq}_x=1/(\expfun{x}-1)$.  So assuming a BE-type distribution is only
accurate in the equilibrium limit when a single process dominates at
some point in time and space or if $T_\gamma\sim T_{\rm e}$, the
latter being true in equilibrium because we include thermal
Comptonization.

From Eq.~\eqref{eq:B_BR_DC}, we wish to obtain evolution equations for
the moments $N_\gamma=\int x^2 n_x \id x$ and $\rho_\gamma=\int x^3
n_x \id x$ of the photon distribution. We furthermore want to express
the equations in a form $\partial_\tau N_\gamma = \Lambda_N - \kappa_N
N_\gamma$ and $\partial_\tau \rho_\gamma = \Lambda_\rho
- \kappa_\rho \rho_\gamma$, where the coefficients $\Lambda_i$ and
$\kappa_i$ are functions of the two temperatures and possible
dependent parameters describing the underlying photon distributions
function. To obtain integrated equations one has to make an ansatz for
$n_x$, which is given by only two {\it free} parameters, since only
two moments of the Boltzmann equation are being used ($N_\gamma$ and
$\rho_\gamma$).

Let us start with $N_\gamma$. Inserting $n_x=n_{\rm
BE}=1/(\expfun{x+\mu}-1)$ into Eq.~\eqref{eq:B_BR_DC} with a cutoff
$\xc$ yields

\begin{eqnarray}
\label{eq:B_BR_DC_N}
\partial_\tau N_\gamma(\Tg, \mu) &=& \int_{\xc}^\infty \frac{\Lambda_{\rm BR+DC}(x, \Tg, \Te)}{x} \id x \\
&-& \int_{\xc}^\infty \frac{\Lambda_{\rm BR}(x, \Tg, \Te)}{x} \frac{\expfun{\xe}-1}{\expfun{x+\mu}-1} \id x\nonumber\\
&-& \int_{\xc}^\infty \frac{\Lambda_{\rm DC}(x, \Tg, \Te)}{x} \frac{\expfun{x}-1}{\expfun{x+\mu}-1} \id x \nonumber,
\end{eqnarray}

so that one has the emission and absorption coefficients

\bsub
\beal
\label{eq:Lambda}
\Lambda_N &= \int_{\xc}^\infty \frac{\Lambda_{\rm BR+DC}}{x} \id x
\\
\kappa_N&= \frac{1}{N_\gamma(\Tg, \mu)}\left[
\int_{\xc}^\infty \frac{\Lambda_{\rm BR}(\expfun{\xe}-1)+\Lambda_{\rm DC}(\expfun{x}-1)}{x \,(\expfun{x+\mu}-1)} \id x\right]
\end{align}
\esub

Similarly, for $\rho_\gamma$ we find

\bsub
\beal
\label{eq:Lambda2}
\Lambda_\rho &= \int_{\xc}^\infty \Lambda_{\rm BR+DC} \id x
\\
\kappa_\rho&= \frac{1}{\rho_\gamma(\Tg, \mu)}\left[
\int_{\xc}^\infty \frac{\Lambda_{\rm BR}(\expfun{\xe}-1)+\Lambda_{\rm DC}(\expfun{x}-1)}{\expfun{x+\mu}-1} \id x\right].
\end{align}
\esub

One can also obtain the moment equation associated with momentum
conservation, which would introduce a radiation flux (and so
radiation frame velocity in the M1 closure) as a weight.  We do not
try to model the angular distribution and frequency dependence on
the M1 radiation velocity, and in general the M1 flux mean behaves
in a more complicated way as a function of optical depth
\citep{1997AnPhy.257..111S,2014JPhD...47J3001C}.

These equations shows how one must construct $\alpha_\nu$ via the
correct Kirchhoff's law (i.e. which temperature enters) and the
related mean opacity (i.e. how to weight the emission rate), where
$\lambda = C (k_b T_\gamma)^4\Lambda_\rho$, $\lambda_n = C (k_b
T_\gamma)^3\Lambda_N$, $\kappa_a = \kappa_\rho/c$, and $\kappa_{an}
= \kappa_N/c$.

\subsection{Kirchhoff's Law}

The opacity for each process, $\alpha_\nu$ (in ${\rm cm}^{-1}$, where
$\kappa$, in the same units, is reserved for the mean opacity), for a
thermal electron distribution is given by Kirchhoff's law
\begin{equation}\label{kir4}
\alpha_\nu = \frac{j_\nu}{B_\nu(T_{\rm eq},\mu_{\rm eq})} ,
\end{equation}
where the last section shows that $T_{\rm eq}=T_{\rm e}$ for a process like
BR and $T_{\rm eq}=T_\gamma$ for a process like DC, where $\mu_{\rm
  eq}=0$.

\subsection{Mean Opacities}

We use an approximation to the absorption mean, flux mean, and
emission mean that enter the 4-force.  For a more general discussion,
see section 82 of \citet{1984oup..book.....M}, section 2.3 of
\citet{huebner2014opacity}, page 38 in \citet{1986psun....1.....S},
page 174 in \citet{2004rahy.book.....C}, or section 15.2 in
\citet{modest2013}.

The Planck or emission mean is
\begin{equation}
\kappa_P = \frac{\int_\nu \alpha_\nu B_\nu}{\int_\nu B_\nu} =
\frac{\int_\nu j_\nu}{\int_\nu B_\nu} = j/B = e/(4\pi B) ,
\end{equation}
for optically thin emission rate $e$.  The Planck mean would be what
is required if the radiation were LTE and one wanted to ensure the
optically thin emission rate was correct if using our 4-force in the
comoving frame with an energy loss rate of
\begin{equation}
du_g/d\tau = \kappa_a (E - 4\pi B) ,
\end{equation}
such that in the limit that $E\to 0$ as $\tau=\kappa_a L\to 0$ for
some length $L$ of the region, then the gas would lose energy at a
rate of $4\pi \kappa_a B = e \equiv \lambda_e$ as required.

In the LTE diffusion limit, the Rosseland mean opacity
$\kappa_R$ with
\begin{equation}
\kappa_R^{-1} = \frac{\int_\nu \alpha_\nu^{-1} dB_\nu/dT}{\int_\nu
dB_\nu/dT}
\end{equation}
is a good approximation to the flux mean \citep{2014ApJ...796..107D}.
The Rosseland mean should not be generally used in the energy equation
as it would fail to give the correct emission rate in the optically
thin limit.  Another approach is the use the Planck or absorption mean
for the energy equation while using the Rosseland mean to approximate
the flux mean for the momentum equation \citep{2016arXiv160503184S},
but this only gives a correct momentum exchange in the diffusive limit
for each grid cell.

Section~\ref{integratedeqs} shows that the opacity which enters the energy
and number equations is the absorption mean given by
\begin{equation}
\kappa_A = \frac{\int_\nu \alpha_\nu J_\nu}{\int_\nu J_\nu}
\end{equation}
for a mean intensity $J_\nu=\int_{4\pi} I_\nu d\Omega /(4\pi)$ of
radiation.  Nominally, one might not know or it may be hard to model
$J_\nu$, but we model the distribution function as BE. The absorption
mean has also been called the two-temperature Planck mean, and it has
been used to handle irradiated planetary
atmospheres \citep{2003ApJ...594.1011H,2012MNRAS.420...20H} and
circumstellar atmospheres \citep{2014A&A...568A..91M}.

For the momentum equation and its flux mean, the mean absorption
opacity should vary from the full Rosseland mean (including every
emission process and scattering in a single integration) in the
diffusion limit to the above absorption mean in the streaming
limit. This could be approximately achieved by interpolation
\citep{1965JQSRT...5..211S,1967JQSRT...7..611P,1994A&A...284..105L,2004A&A...421..741V},
or by splitting the trapped and streaming radiation (e.g.,
\citealt{2015MNRAS.449.4380R}) with different opacities for each
component.  For simplicity, we use the same way of computing the
absorption mean opacity for the energy, number, and momentum
equations.  This may be reasonable for accretion disks, whose global
photospheres are often optically thin across each grid cell where
source terms are applied.  In the deep optically thick limit within
the disk, the diffusion times are often slower than an inflow or
outflow time.  Then, radiation is trapped in the flow, so that the
opacity just needs to be high enough to maintain trapping.

The energy absorption energy-mean opacity is given by
\begin{equation}\label{kappaafinal}
\kappa_a = \frac{\int_\nu \alpha_\nu {BE}_\nu(T_{\rm abs},\mu_{\rm abs})}{\int_\nu
{BE}_\nu(T_{\rm abs},\mu_{\rm abs})} .
\end{equation}
for a Bose-Einstein energy distribution ${BE}$. The corresponding
number absorption number-mean opacity is given by
\begin{equation}\label{kappaanfinal}
\kappa_{an} = \frac{\int_\nu \alpha_\nu {BN}_\nu(T_{\rm abs},\mu_{\rm abs})}{\int_\nu {BN}_\nu(T_{\rm abs},\mu_{\rm abs})} .
\end{equation}
for a Bose-Einstein number distribution ${BN}$.  For the distribution
${BE}$ and ${BN}$, one must choose $T_{\rm abs}\to T_\gamma$ and
$\mu_{\rm abs}\to \mu$ for an ambient BE radiation field at
temperature $T_\gamma$ and chemical potential $\mu$ as consistent with
Eq.~(\ref{integratedeqs}).  The lower integration range over frequency
is assumed to be set by the Razin effect for each process.

Note that if one used $\kappa_a$ instead of $\kappa_{an}$ in the
number evolution equation, it would not give back a consistent
optically thick thermal equilibrium density of photons.  From
Eq.~(\ref{dotn}), one would have obtained $c\kappa_a n = \lambda_n$.
However, the denominator of $\kappa_a$ integrates to $B=u_0c/(4\pi)$
while the numerator of $\kappa_a$ integrates to the energy density
loss rate per solid angle $\lambda_e/(4\pi)$.  So one would have
obtained an equilibrium number density of $n = \lambda_n/(c\kappa_a) =
u_0 \lambda_n/\lambda_e$, which while dimensionally correct is not the
required Planck answer of $n_0$.  However, by using the number
absorption number-mean opacity, one obtains an equilibrium number
density from $c\kappa_{an} n = \lambda_n$ giving $n =
\lambda_n/(c\kappa_{an}) = n_0$ as required.

\subsection{Modifying the BE distribution with inverse Compton and self-absorption}\label{regcompt}

An ambient photon distribution as BE with a frequency independent
$\mu_{\rm abs}\neq 0$ is only a reasonable assumption if the
absorption timescale is longer than a typical flow timescale $t_{\rm
  flow}$ and longer than the timescale for inverse Comptonization to
upscatter photons (and so avoid absorption).  Otherwise, if absorption
is effective, then $\mu_{\rm abs}\to 0$
\citep{1970Ap&SS...7....3S,1981ApJ...244..392L,1991A&A...246...49B,1995A&A...303..323B}. Because
DC and bremsstrahlung emission become increasingly effective at low
frequencies, this condition is always fulfilled at very low
frequencies.

When $k_b T_{\rm e}\lesssim m_e c^2$, the timescale for inverse Comptonization (IC)
of photons to higher energies is
\begin{equation}
t_{\rm IC} \sim t_c \frac{m_e c^2}{k_b T_{\rm e}}
\end{equation}
where $t_c = (n_e c \sigma_T)^{-1}$ is the electron scattering
timescale. (This is not the energy redistribution time between gas the
electrons, that would given by $t_{\rm re} \sim E/\lambda_{\rm c}$
from Eq.~(\ref{CompG0}).) The timescale for absorption is given by the
absorption term for $dn_x/dt$ in Boltzmann's equation and computing
\begin{equation}
t_{\rm abs} \sim \frac{n_x}{\left[dn_x/dt\right]_{\rm abs}} = \frac{1}{c\alpha_x} ,
\end{equation}
where $\alpha_x=\alpha_\nu$.  From the condition $t_{\rm IC}\ge t_{\rm
  abs}$, one can thus determine those frequencies for which the
distribution is Planck because those low-energy photons are absorbed
before being upscattered.

So, a significant improvement to the BE distribution assumption is to
force $\mu_{\rm abs}\to 0$ when $t_{\rm abs}<t_{\rm flow}$ and $t_{\rm
abs}<t_{\rm IC}$ for a typical flow time or simulation timestep
$t_{\rm flow}\sim 10^{-3} (r_g/c)$.  Otherwise, IC avoids
self-absorption or insufficient time has elapsed to assume absorption
has had time to force $\mu_{\rm abs}\to 0$. This is how we modify the
BE distribution in this paper.  We do not try to model the transition
and just assume it's a discontinuous change in the distribution.  This
modified BE distribution acts as a sub-timestep model for the photon
distribution function.  We apply this modification to $\mu$ for each
process, which assumes the dominant process is the only one we need to
treat accurately.

\subsection{Absorption balancing Emission}\label{emission}

The energy emission rate is
\begin{equation}\label{lambdafromkappa}
\lambda_e = 4\pi \int_\nu j_\nu = 4\pi \int_\nu \alpha_\nu {BE}_\nu =
\kappa_{a} 4\pi \int_\nu {BE}_\nu \equiv \kappa_e 4\pi B ,
\end{equation}
with the last equation defining the so-called energy emission mean
opacity $\kappa_e$.  The number emission rate is
\begin{equation}\label{lambdanfromkappa}
\lambda_n = 4\pi \int_\nu (j_\nu/(h\nu)) = 4\pi \int_\nu \alpha_\nu
       {BN}_\nu = \kappa_{an} 4\pi \int_\nu {BN}_\nu \equiv
       \kappa_{en} 4\pi N ,
\end{equation}
where $\kappa_a$ and $\kappa_{an}$ are from the generalized absorption
mean opacity expression Eq.~(\ref{kappaafinal}) and
Eq.~(\ref{kappaanfinal}).  For these emission processes, in the
distribution functions ${BE}$ and ${BN}$ that appear in $\kappa_a$ and
$\kappa_{an}$, one replaces $T_{\rm abs}\to T_{\rm eq}$ ($T_{\rm
  eq}=T_{\rm e}$ for BR-like processes and $T_{\rm eq}=T_\gamma$ for DC-like
processes) and $\mu_{\rm abs}\to \mu_{\rm eq}=0$.

This construction of $\lambda_e$ and $\lambda_n$ just ensures that we
treat the integration limits consistently so that in the optically
thick thermal equilibrium limit the absorption balances emission.  It
also defines the so-called emission mean opacity that is just a way to
write the emission rates as a type of opacity with a simple factor $B$
or $N$ that need not be separately tabulated.

\section{Free-Free Emission}\label{sec:ff}

Following the discussion in \citet{1986rpa..book.....R} that follows
\citet{1973blho.conf..343N}, the electron-ion free-free emissivity
$j^{\rm ffei}_\nu$ is obtained from
\begin{equation}
j^{\rm ffei}_\nu = \frac{8 \sqrt{\frac{2 \pi }{3}} (n_-+n_+) n_i q^6 Z_p^2
   \sqrt{\frac{1}{k_b m_e}} e^{-\frac{h \nu}{k_b T_{\rm e}}}}{3 c^3 m_e
   \sqrt{T_{\rm e}}} \bar{g}_{\rm ffei} R_{\rm ei}(\theta),
\end{equation}
where we assume no pairs so $(n_-+n_+)\to n_e$,
and where the relativistic correction factor is
\begin{equation}
R_{\rm ei}(\theta_{\rm e}) =\begin{cases}
1 + 1.76 \theta_{\rm e}^{1.34} & \theta_{\rm e}\le 1 \\
1.4 \theta_{\rm e}^{0.5} (\ln (1.12 \theta_{\rm e}+0.48)+1.5) & \theta_{\rm e}> 1 \end{cases}
\end{equation}
\citep{1982ApJ...258..335S,1996ApJ...465..312E}. We assume
temperatures are $T\gtrsim 10^5$K, which is in the small-angle
uncertainty principle regime, so that $\bar{g}_{1,\rm ffei}=(3k_b
T_{\rm e} /(\pi h\nu))^{1/2}$ for $h\nu/(k_b T)\ge 1$ and
$\bar{g}_{2,\rm ffei}=(\sqrt{3}/\pi)\ln(2.2 k_b T_{\rm e} / (h\nu))$
for $h\nu/(k_b T)<1$, which we interpolate as $\bar{g}_{\rm
ffei}=\bar{g}_{1,\rm ffei}\exp(-1/x_e) + \bar{g}_{2,\rm
ffei}(1-\exp(-1/x_e))$.  More accurate analytical fitting formulae can
be used
\citep{1984MNRAS.209..175S,1991pav..book.....S,2000ApJS..128..125I,2001ASPC..251..268S},
although this level of accuracy is not required.

For abundances with mass fractions of Hydrogen, Helium, and
``metals'', respectively, $X$, $Y$, and $Z$, in a mostly ionized
gas $n_e/\rho = 1/(\mu_e m_u) = (1+X)/(2m_u)$ while $\sum_i n_i Z_i^2
= (\rho/m_u)\sum_i (X_i Z_i^2/A_i) = (\rho/m_u)(X+Y+B)$ for mass
fraction of species $i$ of $X_i$, where $B = \sum_{i>2} (X_i
Z_i^2/A_i)$ with $X+Y+B\approx 1-Z$ for $B\ll \{X,Y\}$.  We often
assume solar abundances (mass fractions of Hydrogen, Helium, and
``metals'', respectively, $X=0.7$, $Y=0.28$, $Z=0.02$) with electron
fraction $Y_e=(1+X)/2$ giving mean molecular weight $\bar{\mu}\approx
1/(2X + 0.75Y + 0.5Z)\approx 0.62$ for ionized gas, which enters the
gas entropy, pressure, and temperature $T_{\rm g}$.

For free-free, $\kappa_{\rm ffei} \approx 1\times 10^{24} \rho^2
T_{\rm e}^{-7/2} (1+X)(1-Z) R(\theta_{\rm e})$, for near solar abundances for the
Planck mean.  A Rosseland mean gives $\kappa_{\rm ffei,R }\approx
3.8\times 10^{22} \rho^2 T_{\rm e}^{-7/2} (1+X)(1-Z) R(\theta_{\rm e})$, which is
about $30$ times smaller than the Planck mean.  These calculations
correct the formulae used in
\citet{2015MNRAS.454L...6M}, where the free-free and free-bound
coefficients were for Rosseland while the functional form was for
Planck.

One can then compute the energy and number absorption mean from
$\alpha^{\rm ffei}_\nu = j^{\rm ffei}_\nu/B_\nu(T_{\rm e})$ from Kirchhoff's
law with a distribution based upon $T=T_{\rm e}$.

\subsection{Razin Effect for Bremsstrahlung}

The well-known infrared divergence for free-free leads to a formally
infinite number of photons generated.  Without any adjustment, the
number absorption mean opacity diverges
\citep{ascoli1956,1961AnPhy..13..379Y}.  For a given experiment size,
this divergence can be removed by adding in other non-bremsstrahlung
processes that cannot be distinguished and that can occur
\citep{akhiezer1953quantum,kaku1993quantum}.
The divergence is logarithmic, so any physical effect that would limit
the presence of low frequency radiation would likely lead to a
sufficient removal of the divergence to order unity or so.

Several processes can destroy the coherence of the emission process
over the formation length while the electron and photon continue to
distinguish themselves quantum mechanically.  This includes
Comptonization, Razin-type collective plasma effects (see below),
finite temperature effects \citep{1994PhRvD..49.1579W}, Debye-length
limit of the Coulomb potential \citep{1990ApJ...362..284G}, and
Landau-Pomeranchuk-Migdal (LPM) effect of multiple scatterings
\citep{1976JETP...44...87A,Chen:1992ne,1998JETP...86...32S,1999RvMP...71.1501K,1997STIN...9932370K,2005physics...2051F,4345536}.

Collective plasma effects at the plasma frequency scale lead to
dielectric suppression of the thermal bremsstrahlung rate that is the
lowest frequency and strongest cut-off of radiation compared to the
LPM or other effects
\citep{1960MNRAS.120..231S,dawsonoberman1962,mercier64,bekefi1966radiation,ichimaru1973,1972Ap&SS..18..267M,1994PhRvD..49.1579W,1996PhRvL..76.3550A}.
This can be accurately represented by multiplicative suppression
factor of $S = \omega^2/(\omega^2+\omega_{pe}^2) =
(1+\omega^2_{pe}/\omega^2)^{-1}$ \citep{1972Ap&SS..18..267M} for
electron plasma frequency $\omega_{pe}$.  As a contribution to the
mean opacity, a sufficiently accurate suppression factor is given by
\begin{equation}
S = (1-\omega^2_{pe}/\omega^2)^{1/2} = S = (1-x_p/x^2)^{1/2} ,
\end{equation}
\citep{bekefi1966radiation}, for which the frequency integrals then
start at $\omega=\omega_{pe}$ where $x_p=h\nu_{pe}/(k_B T_e)$ and
$\nu_{pe}=\omega_{pe}/(2\pi)$ such that
\begin{equation}
x_p = 4.3\times 10^{-7} \sqrt{n_e}/T_e = 3.3\times 10^5
\sqrt{Y_e\rho/\bar{\mu}}/T_e .
\end{equation}
So $S$ can be treated as a suppression factor that simply depends upon
the integral's lower cut-off, here $x_p$.  Note that this suppression
is not an absorption process, and instead the light wave is evanescent
below plasma frequencies.  This cut-off operates even when the system
is not in equilibrium.  In principle, this means we need to
tabulate/fit our opacities vs.  the dimensionless lower frequency
cut-off $x_{\rm cut}=x_p$.  Instead, for a given system, because the
effect of the cut-off is logarithmic, we use a fixed cut-off for
expected densities and temperatures.

\subsection{Tabulating/Fitting the Free-Free mean opacity}

After substituting $\nu\to x k_b T_{\rm e}/h$ and using $x$, ${\rm
  e}^{-\xi}$, and $\zeta=T_\gamma/T_{\rm e}$ as independent variables, and given
a dimensionfull term set as
\begin{equation}
f(n_e,n_i,T_{\rm e}) \equiv 1.2\times 10^{24} T_{\rm e}^{-7/2} \rho^2 (1+X)(1-Z) R(\theta_{\rm e}),
\end{equation}
then one can tabulate/fit the residual dimensionless factor
$\kappa_{\rm a,ffei}/f(n_e,n_i,T_{\rm e})$.  The mean opacity is quite
linear in log-log space, so such a direct look-up table can be
generated of small dimensions that covers all dimensionless space of
$\zeta=10^{-10}$--$10^{10}$ and $0\le \expf\le 1$.

For HARMRAD, we obtain a fitting function
\begin{equation}
\frac{\kappa_{\rm a, ffei}}{f(n_e,n_i,T_{\rm e})} = a \zeta^{-b}\ln(1+c\zeta) ,
\end{equation}
where in general $a(\expf),b(\expf),c(\expf)$ are functions fitted for
in that separate dimension.  Similarly, the residual for the number
opacity $\kappa_{\rm an,ffei}/f(n_e,n_i,T_{\rm e})$ can be fitted with the
same form of the expression with different constants.

The final fitted constants $a,b,c$ for the energy opacity are
\begin{eqnarray}
a &=& 0.188 (\expf)^{13.9}-0.2 (1-(\expf))^{0.565}+0.356\\
b &=& 0.0722 (\expf)^{1.36}+0.255 (1-(\expf))^{0.313}+3.06\nonumber\\
c &=& -1.41 (\expf)^{3.08}-1.44  (1-(\expf))^{0.128}+5.99\nonumber
\end{eqnarray}
and the constants $a_n,b_n,c_n$ for the number opacity are
\begin{eqnarray}
a_n &=& 21. (\expf)^{5.}-2.06 (1-(\expf))^{1.}+4.\\
b_n &=& -0.412 (\expf)^{59.1}+0.000894 (1-(\expf))^{10.2}+3.15\nonumber\\
c_n &=& 5.27 (\expf)^{69.2}+2.39 (1-(\expf))^{0.552}\nonumber
\end{eqnarray}
These fits use Razin and modified BE distribution based upon
densities and temperatures for an Eddington-accreting BH X-ray
binary.  This gives a fit accurate for the energy opacity to less
than $20\%$ relative error for $10^{-5}<\zeta<10^5$ and $0<\expf<1$
with careful attention near $\expf=1$ where significant changes
occur.  The number opacity has a similar error except at $\expf=1$
where the results are accurate to factors of two as related to sharp
changes right at $\expf=1$ for $\zeta\gtrsim 10^4$.  The fitting
coefficients at only $\expf=1$ are $\{a,b,c\} = \{0.532,3.14,4.52\}$
for the energy opacity and $\{a_n,b_n,c_n\} =\{20.0,2.67,5.00\}$ for
the number opacity. These Planck coefficients can be used for the
emission mean opacity when one also enforces $\zeta=1$.

\subsection{Electron-Electron Opacity}

The electron-electron opacity is the same as the electron-ion opacity
except $(n_-+n_+)n_i$ is replaced by $(n_-^2+n_+^2)\approx n_e^2$ and
$R_{\rm ei}$ is replaced by
\begin{equation}
R_{\rm ee}(\theta_{\rm e}) =\begin{cases}
1.7\theta_{\rm e} (1 + 1.1 \theta_{\rm e} + \theta_{\rm e}^2 - 1.06 \theta_{\rm e}^{2.5}) & \theta_{\rm e}\le 1 \\
1.7\theta_{\rm e}^{0.5} (1.46 (1.28 + \ln (1.12 \theta_{\rm e})) & \theta_{\rm e}> 1 \end{cases}
\end{equation}
\citep{1996ApJ...465..312E}. When obtaining mean opacities and
emission rates, the density and $R$ factors pull out and so any
integrations apply to both these rates. This electron-electron
opacity is added to the electron-ion opacity to give the final
free-free opacity of
\begin{equation}
\kappa_{\rm ff} = \kappa_{\rm ffei} + \kappa_{\rm ffee} ,
\end{equation}
for both energy and number mean opacities.

\subsection{Effective Opacity including Lower temperature opacities}

The bound-free (number and energy) opacity $\kappa_{\rm bf}$ is the
same as free-free with $(1+X)(1-Z)$ replaced by $\sim 750 Z
(1+X+0.75Y)$
\citep{1986rpa..book.....R}, where the $1+X+0.75Y$ term is roughly
accurate near solar abundances.  This assumes mostly ionized hydrogen
and helium.  One should also drop the high-temperature $R(\theta_{\rm e})$
factor.

Rosseland means are often provided for lower temperature opacities due
to the expectation of applications to stellar interiors with
temperature transport as the primary study.  These opacities also tend
to assume $T_{\rm e}=T_\gamma$.  Direct integration to obtain
absorption means is not easy because the required frequency-dependent
opacity data is not readily available.

To approximately convert published Rosseland means that assume
$T_{\rm e}=T_\gamma$ to absorption means that allow $T_{\rm e}\ne T_\gamma$, we
remove the electron scattering opacity (if it was included) and assume
the opacity has a generic free-free-like form of $\alpha_x\propto
(1-\exp(-x))/x^3$.  This allows us to obtain the generic ratio of
absorption mean to Rosseland mean of $\sim 30$ at $T_{\rm e}=T_\gamma$ as
consistent with the free-free calculations above, and this factor of
30 already appears in the below opacities.

At intermediate temperatures of $10^5$--$10^7$K and low densities, the
free-free and bound-free opacity change their behavior for solar
abundances to become the ``Chianti opacity'' of
\begin{equation}
\kappa_{\rm Chianti}/\rho \sim 30\times 10^{33} \rho (0.1+Z/Z_{\rm
solar})X(1+X) T_{\rm e}^{-4.7} .
\end{equation}
This behaves similarly to the free-free or free-bound opacity, but
with a steeper temperature dependence.  This accounts for the
assumed $Z=Z_{\rm solar} = 0.02$ for figure 34.1 in
\citet{2011piim.book.....D}, most applicable for baryon densities of
$n_b\sim 1{\rm cm}^{-3}$.  This gives a kink in the opacity at
$T\sim 10^7$K.  To capture the drop in opacity at even lower
temperatures and obtain a peak opacity at $T\sim 10^5$K for solar
abundances as in that figure 34.1, we use the $H^-$ opacity of
\begin{equation}
\kappa_{H^-}/\rho \approx 33\times 10^{-25}Z^{0.5} \rho^{0.5} T_{\rm g}^{7.7} ,
\end{equation}
and use the molecular opacity of $\kappa_{\rm m}/\rho \approx 3Z$.
The $H^-$ and molecular opacity should have a absorption mean that
will change with $T_\gamma$ (e.g. \citealt{2014A&A...568A..91M}), but
in our case these low temperature opacities most often occur when
$T_\gamma\sim T_{\rm e}$.

The Chianti opacity works for low densities at intermediate
temperatures, but at higher densities and lower temperatures
additional effects modify the opacity. Realistic opacities for
$T\lesssim 10^7$K at higher densities can be obtained from the OPAL
opacity tables that includes bound-bound lines and other physics
\citep{1996ApJ...464..943I}. Other opacity calculations focused on
stellar atmospheres give similar results \citep{1994MNRAS.266..805S},
while modern opacity tables use the OPAL opacity tables in some
regimes \citep{2011ApJS..192....3P}.  Particularly, the opacity is
enhanced between $T\sim 10^4$K and $T\sim 10^5$K at high densities due
to H and He as well as enhanced around $10^5$K at low densities due to
metals such like Fe.  Such opacities have been approximately fit by
\citet{1975ApJ...196..525I} with corrections for the OPAL opacities by
\citet{1995ApJ...448..905G}, but those fits do not capture the low
temperature or Fe line features.  For $T\sim 10^5$K, this opacity is
significantly enhanced beyond the electron scattering opacity and can
introduce driving of winds \citep{1998MNRAS.295..595P}. We obtain a
rough fit to the OPAL opacity Table 73 relevant for solar abundances.
We subtract out the electron scattering opacity from their Rosseland
mean to obtain a Rosseland absorption mean. The ``OPAL opacity'' fit
is given as extra terms
\begin{equation}
\kappa_{\rm COPAL} \sim 3\times 10^{-13} \kappa_{\rm Chianti}
T_{\rm e}^{1.6}\rho^{-0.4}
\end{equation}
and
\begin{equation}
\kappa_{\rm HOPAL} \sim 10^4 T_{\rm g}^{-1.2} \kappa_{\rm H^-} .
\end{equation}
The Fe line at $T\approx 1.5\times 10^5$K can be approximated by
adding (to the overall opacity) the Gaussian
\begin{equation}
\kappa_{\rm Fe}/\rho \sim 0.3\left(\frac{Z}{Z_{\rm solar}}\right)
{\rm e}^{-6(-12+\ln(T_{\rm g}))^2} .
\end{equation}
This gives an iron line bump, which can affect the structure of disks
and stars with substantial
metals \citep{2015ApJ...813...74J,2016arXiv160106836J}.  For the Fe
line we do not modify this Rosseland mean into a Planck mean, because
the line sits at a narrow band of frequencies and does not follow the
free-free frequency dependence.

Our conversion from Rosseland to absorption means assuming the
frequency-dependence is roughly free-free-like also allows us to apply
the general free-free $\zeta$ dependence to these opacities as well
estimate the energy and number mean opacities.  We take the ratio of a
given low-temperature $\kappa$ (that applies to $T_{\rm e}=T_\gamma$) to
$\kappa^{\rm ff}$ at $\zeta=1$, and then we set the final
lower-temperature opacity as $\kappa_{\rm ff}$ times that ratio.  This
is done for both energy and number mean opacities separately.  Only
for the temperature-independent molecular opacity and narrow-band Fe
line opacity do we assume a $\zeta$-independent opacity that is the
same for number and energy mean opacities.

The final effective opacity $\kappa_{\rm eff}$ that bridges between
the various opacities is given by
\begin{equation}\label{kappaeff}
(\kappa_{\rm eff}-\kappa_{\rm Fe})^{-1} \sim (k_{\rm m}+\kappa_{\rm
HOPAL})^{-1} + \kappa_{\rm COPAL}^{-1} + (\kappa_{\rm
Chianti}+\kappa_{\rm ff}+\kappa_{\rm bf})^{-1} .
\end{equation}
This gives a fit to the ''OPAL'' opacities that is accurate to less
than $30\%$ in most cases and factors of ten for $T\ll 10^5$K.  This
defines both the energy opacity and number opacity by assuming
$\kappa_{\rm m}$, $\kappa_{\rm HOPAL}$, $\kappa_{\rm COPAL}$, and
$\kappa_{\rm Chianti}$ scale with free-free (and bound-free) for
energy and number opacities.

\section{Cyclo-Synchrotron Emission}\label{sec:synch}

Cyclo-synchrotron is important when magnetic fields are strong in
diffuse plasmas.  We follow the discussion in
\citet{1996ApJ...465..327M} (see also \citealt{1965ARA&A...3..297G}),
where fits are given to cyclo-synchrotron at several discrete
temperatures from $5\times 10^8$K to $3.2\times 10^{10}$K as well as
the ultrarelativistic limit.  Other fits are considered for higher
temperatures in \citet{2011ApJ...737...21L}.  Planck mean opacities
have been considered before in the context of black hole accretion
flows \citep{1992ApJ...400..170M}.

We consider thermal electrons with temperature $T_{\rm e}$. Following
\citet{1996ApJ...465..327M} and \citet{1996ApJ...465..312E}, the emissivity is
\begin{equation}
j_{\rm syn}(\nu) = 4.43\times10^{-30} \nu_M n_e
\,\frac{x_MI'(x_M)}{K_2(1/\theta_{\rm e})} d\nu ~{\rm erg\,cm^{-3}s^{-1}},
\end{equation}
where $\theta_{\rm e}=k_b T_{\rm e}/(m_e c^2)$ and
\begin{equation}
x_M = \frac{\nu}{\nu_M}, \qquad \nu_M = \frac{3}{2} \, \frac{eB}{2\pi
m_ec}\,\theta_{\rm e}^2 = 1.19 \times 10^7 B T_{10}^2 ~{\rm Hz},
\end{equation}
where $T_{10}=T_{\rm e}/10^{10}$K. The function $I'(x_M)$ is defined in
\citet{1996ApJ...465..327M}, who provide the following fitting
function for it,
\begin{equation}
I'(x_M) = \frac{4.0505\alpha}{x_M^{1/6}} \left(1 +
\frac{0.40\beta}{x_M^{1/4}} + \frac{0.5316\gamma}{x_M^{1/2}} \right)
\exp(-1.8899x_M^{1/3}) .
\end{equation}

In the ultrarelativistic limit $\alpha=\beta=\gamma=1$ and
$K_2(1/\theta_{\rm e}) \to 2\theta_{\rm e}^2$, while table 1 in
\citet{1996ApJ...465..327M} gives these coefficients for several other
temperatures.

For synchrotron, a primary limitation of our opacities is the lack of
non-thermal electrons, which would greatly modify the opacity in
optically thin regions.

We use the low-frequency ($h\nu \ll kT_{\rm e}$) (e.g. Rayleigh-Jeans for
Planck) expansion for the distribution that appears in Kirchhoff's law
within the opacity.  Equivalently, we assume synchrotron emission at
$h\nu\gtrsim k_b T_{\rm e}$ is sub-dominant.  This approximation is
reasonable when $(kT_{\rm e})/(h\nu_M)\gg 1$, which will be valid for
simulations we consider.  One can then compute the energy and number
absorption mean from $\alpha^{\rm syn}_\nu = j^{\rm syn}_\nu/B_\nu(T_{\rm e})$.

\subsection{Razin Effect for Synchrotron}

Collective plasma effects, like the Razin effect, occur for
synchrotron
\citep{1966MNRAS.131..237H,1967ApJ...147..544M,1969ApJ...156..341S}.
This has been applied in the astrophysical case recently by
\citet{2003A&A...409..217D}. The characteristic cut-off occurs at
\begin{equation}
\nu_R \approx \nu_{pe}^2/\nu_B \approx 19n_e/B .
\end{equation}
or in terms of our dimensionless synchrotron integration variable in
$x_M^{1/3}\sim 0.01(n_e/B)^{1/3}$. Below $\nu_R$ the synchrotron
spectrum varies as $\nu^{3/2}\exp(-\nu_r/\nu)$ independent of the
electron spectrum as long as there is no high-energy cut-off in the
electron spectrum (in which case the $\nu_R$ is even higher at
$\nu_R\approx 500 \gamma_1 n_e^{3/4}/B^{1/2}$ for electron energy
cut-off at $E=\gamma_1 m_e c^2$).  This cut-off can be approximated
quite accurately by multiplying $\alpha^{\rm s}_\nu$ by ${\rm
e}^{-\nu_R/\nu}={\rm e}^{-x_R/x}$
\citep{2003A&A...409..217D}, which is the approximation we apply
to our integrals.  Like with free-free, the cut-off is small enough
that changes in the dependence of opacity vs. $\zeta=T_\gamma/T_e$
only occur for $\zeta\ll 1$, where the BE assumption is unlikely
accurate and the simulation very rarely accesses.  Hence, like with
free-free, we only tabulate/fit the region above this change in
character.

\subsection{Tabulating/Fitting Cyclo-Synchrotron}

Defining
\begin{equation}\label{phisy}
\phi \equiv \frac{kT_\gamma}{h\nu_M} ,
\end{equation}
and using a dimensionfull factor of
\begin{equation}
g(n_e,B,T_{\rm e}) = 5.85374 \times 10^{-14} n_e \phi \theta_{\rm e}^{-3}
T_\gamma^{-1} ,
\end{equation}
the residual factor $\kappa_{\rm a, syn}/g(n_e,B,T_{\rm e})$ and
$\kappa_{\rm an,syn}/g(n_e,B,T_{\rm e})$ can be tabulated.  This residual is quite
linear in log-log space, so not much resolution is required. In order
to handle regions with weak or zero magnetic field strengths or
radiation temperatures, we limit $\phi$ to no smaller than $10^{-20}$
and no larger than $10^{20}$.

For HARMRAD, we fit the energy {\it and} number opacity residual by
\begin{equation}
\left(\frac{\kappa_{\rm \{a,an\},syn}}{g(n_e,B,T_{\rm e})}\right)^{-1}
\approx  (a\phi^{-b}\ln(1+c\phi))^{-1} + (d\phi^{-e})^{-1}
\end{equation}
with different coefficients $a,b,c,d,e$ for the energy and number
opacities.  As with the free-free opacity, $a,b,c,d,e$ are
independently fitted for as functions of $\phi$ for each $\expf$.  We
obtain fits are accurate to less than $20\%$ relative error for any
temperature $T_{\rm e}$ case given in \citet{1996ApJ...465..327M} over the
range $10^{-8}<\phi<10^{15}$ and $0\le \expf\le 1$.

For densities and temperatures relevant for SgrA*, which enters the
Razin cut-off and modification of the BE distribution, the
ultrarelativistic temperature limit has coefficients for the energy
opacity of
\begin{eqnarray}
a&=&-0.0295 (\expf)^{2.29}-0.143
(1-(\expf))^{0.251}+0.236\\
b&=&0.00977 (\expf)^{730.}+0.0291
(1-(\expf))^{0.48}+2.58\nonumber\\
c&=&1.29
   (\expf)^{1.59}+3.46 (1-(\expf))^{0.234}+2.15\nonumber\\
   d&=&-78.1 (\expf)^{66.}-40.3 (1-(\expf))^{0.899}+87.4\nonumber\\
   e&=&0.415
   (\expf)^{0.399}+1.04 (1-(\expf))^{0.252}+2.68\nonumber
\end{eqnarray}
and the constants for the number opacity are
\begin{eqnarray}
a_n&=&10.8 (\expf)^{172.}-20.4
(1-(\expf))^{0.699}+29.2\\
b_n&=&-0.18(\expf)^{31.9}+0.425
(1-(\expf))^{0.179}+2.76\nonumber\\
c_n&=&0.0207
   (\expf)^{9.69}+0.0506 (1-(\expf))^{0.804}+0.0314\nonumber\\
   d_n&=&\left(1.51\times 10^6\right) (\expf)^{2830.}-1.4\times 10^5 (1-(\expf))^{3.06\times
   10^{-12}}+1.4\times 10^5\nonumber\\
   e_n&=&0.1 (\expf)^{1.95}+1.57 (1-(\expf))^{0.124}\nonumber
\end{eqnarray}
These fits are accurate to $20\%$ error over most of the domain
$10^{-4}\le \phi\le 10^{15}$ and $0\le \expf\le 1$.  For $\expf=1$,
the coefficients for the energy opacity are
$\{a,b,c,d,e\}=\{0.206,2.59,3.44,9.33,3.09\}$ and for the number
opacity are $\{a_n,b_n,c_n,d_n,e_n\}=\{40.0,2.58,0.0522,1.65\times
10^6,0.100\}$.  These Planck coefficients can be used for the
emission mean opacity when one also enforces $\phi \to
(kT_{\rm e})/(h\nu_M)$. For $\phi\le 10^{-5}$ these opacities becomes
constant for each $\expf$ vs. $\phi$ due to the Razin cut-off, so
for such values of $\phi$ one should replace the opacities with
their values at $\phi=10^{-5}$ thus overriding the above fits.

For densities and temperatures relevant for an Eddington-accreting BH
X-ray binary, we only obtain fitting accurate to order unity since
other processes usually dominate and finding fitting functions can be
difficult.  The ultrarelativistic temperature limit has coefficients
for the energy opacity of
\begin{eqnarray}
a&=&\left(-2.31\times 10^{-8}\right) (\expf)^{34.}-\left(8.24\times
10^{-9}\right)
(1-(\expf))^{2.42}+1.27\nonumber\\
b&=&-0.0261(\expf)^{738.}-0.00475 (1-(\expf))^{1.55}+1.06\\
c&=&   0.000179 (\expf)^{432.}+0.0000411 (1-(\expf))^{0.372}+0.000584\nonumber\\
d&=&   -17.7   (\expf)^{49.4}-3.33 (1-(\expf))^{2.76}+18.3\nonumber\\
e&=&   0.427 (\expf)^{0.654}+1.23 (1-(\expf))^{0.214}+2.49\nonumber
\end{eqnarray} and the constants for the number opacity are
\begin{eqnarray}
a_n&=&-0.000359 (\expf)^{1.31}-0.000552
(1-(\expf))^{0.135}+0.00209\nonumber\\
b_n&=&0.035 (\expf)^{5.43}+0.0433
(1-(\expf))^{0.159}+0.948\\
c_n&=&-0.122
   (\expf)^{37.1}-0.0685 (1-(\expf))^{2.8}+1.04\nonumber\\
d_n&=&   -8.59 (\expf)^{155.}-6.47 (1-(\expf))^{0.436}+8.71\nonumber\\
e_n&=&   -0.447
   (\expf)^{394.}+0.506 (1-(\expf))^{0.155}+2.45\nonumber
\end{eqnarray}
For $\expf=1$, the coefficients for the energy opacity are
$\{a,b,c,d,e\}=\{1.27,1.03,0.000763,0.616,2.91\}$ and for the number
opacity are
$\{a_n,b_n,c_n,d_n,e_n\}=\{0.00173,0.983,0.921,0.123,2.00\}$, to
which the fits reduce to in this limit. These Planck coefficients
can be used for the emission mean opacity when one also enforces
$\phi \to (kT_{\rm e})/(h\nu_M)$.

For the temperatures down to $T=10^8$K described
in \citet{1996ApJ...465..327M}, we have fitted the ratio of the
discrete temperature mean opacities to the ultrarelativistic mean
opacity. However, even at this lower temperature regime the fits
by \citet{1996ApJ...465..327M} are not broad-band enough to obtain an
accurate mean opacity. Further, the simulations reach to much lower
temperatures where there are no accurate fits. The fits are not too
dissimilar from the ultrarelativistic mean opacities, except when
$\phi$ is far from unity.  So, we assume the ultrarelativistic opacity
in all temperature regimes, as has been done by others \citep{fm09}.
In the future, we can use more accurate cyclo-synchrotron calculations
\citep{2011ApJ...737...21L,2016ApJ...822...34P} to compute the mean opacities.

\section{Compton Scattering}\label{sec:compton}

Comptonization modifies the spectra and cools accretion flows
\citep{2009PASJ...61..769K,2013ApJ...769..156S,2010MNRAS.403..170X,kaw12},
and Comptonization may generate useful observational signatures of
super-Eddington accretion \citep{2013MNRAS.435.1758S}.  So it is an
important process to consider.

\subsection{Thomson Scattering with Klein-Nishina}

The electron scattering opacity is
\begin{equation}
\kappa_{\rm s} \approx \kappa_{\rm es} \kappa_{\rm kn} ,
\end{equation}
where the Klein-Nishina (KN) correction for thermal electrons is
$\kappa_{\rm kn} \approx (1 + (T_{\rm e}/(4.5\times 10^8))^{0.86})^{-1}$
\citep{1976ApJ...210..440B} and $\kappa_{\rm es}=0.2 (1+X)$. This is
applicable as a Rosseland mean for a Planck distribution of photons in
non-degenerate matter (with KN correction applicable when $T_{\rm e}\sim
T_\gamma$).  However, the Rosseland mean and streaming limit of the
flux mean are similar to within $10\%$
\citep{2016arXiv160609466P}, so a fixed scattering opacity as a
function of optical depth is reasonable. A Wien distribution leads
to a faster drop in scattering opacity as $\theta_{\rm e}>1$
\citep{1984MNRAS.209..175S}, but this is a small correction as we do
not end up with solutions having large regions with $\theta_{\rm e}>1$ or
$\theta_\gamma>1$.

\subsection{Thermal Comptonization via Kompaneets Scattering}
\label{s.comptonization}

We account for energy exchange via Comptonization in the soft-photon
limit of Kompaneets equation, which we implement in a similar way to
that described in \cite{2009PASJ...61..769K}.  For a general
temperature, using the result given in equation (2.43)
in \cite{1983ASPRv...2..189P}, \citet{2015MNRAS.447...49S} obtained a
thermal Comptonization term of
\begin{eqnarray}
 \label{CompG0}
\lambda_{\rm c} &=& - c\kappa_{\rm es} E
\left[\frac{4k(T_{\rm e}-T_\gamma)}{m_{\rm e}c^2}\right] \times\\\nonumber
&&\times\left[1+3.683 \left(\frac{kT_{\rm e}}{m_{\rm e}c^2}\right) +4
\left(\frac{kT_{\rm e}}{m_{\rm e}c^2}\right)^2\right]
\left[1+\left(\frac{kT_{\rm e}}{m_{\rm e}c^2}\right)\right]^{-1} ,
\end{eqnarray}
which is consistent with the frequency-integrated energy-weighted
Kompaneets equation and is valid for a BE distribution with any
chemical potential.  We assume that the Compton-scattered radiation is
emitted isotropically in the fluid frame, which is a good
approximation in the soft photon limit or for when there are numerous
scatterings.

This scheme is tested for accuracy as in \citet{2015ApJ...807...31R}
section 4.2, except they started with a delta function for the photon
distribution and evolved the photons in a scattering region to see the
redistribution toward Wien.  Since we assume Bose-Einstein, we start
with Planck at $T_\gamma=500061$K that gives their energy density for
our Planck initial distribution.  We assume their $T_{\rm e}=5\times
10^7$K, baryon number density $n_b = 2.5\times 10^{17}$, and ideal gas
constant $\gamma=5/3$.  Then, we evolve the system and check the
timescale for equilibration as well as the final distribution and
temperature, which can be computed from the conditions of thermal
equilibrium and photon conservation.  Analytically we obtain a final
temperature of $T\approx 2.778\times 10^6$K and $\expf=0.0070$.  We
use harmrad and evolve for $t=20$s.  If we evolve for this time in a
single timestep, as possible with our implicit method, we find that
after this time, $T_{\rm e}\approx 2.786\times 10^6$K,
$T_\gamma\approx 2.779\times 10^6$K, and $\expf\approx 0.0070$
(i.e. quite Wien). If we evolve the system for about $48,000$
timesteps, then we find $T_{\rm e}=T_\gamma\approx 2.779\times 10^6$K
(with temperatures similar to machine precision) and $\expf\approx
0.0070$.  In this case, the temperatures approach each other on a
timescale of $0.02$s, as is the equilibration timescale.  Errors in
these results are dominated by the temperature and chemical potential
fits for the BE distribution that are only accurate to at worst $2\%$
as well as by the first order time error in the implicit method when
taking only a single large timestep.

Other extensions to Kompaneets can be found elsewhere that account for
relativistic
effects \citep{1998ApJ...508....1S,1998ApJ...499....1C,nagirner97,1998ApJ...499....1C,2012APh....35..742B,2015APh....62...30N},
high
temperatures \citep{1959ApJ...129..734S,2010MNRAS.403..170X,2013arXiv1307.7355G,2014arXiv1412.4593N},
and evolve the electron temperature
\citep{2009A&A...507.1243P,2012MNRAS.419.1294C}.

\subsection{Double Compton}\label{subsec:dc}

An important source of soft photons is the double (or radiative)
Compton effect, which dominates bremsstrahlung in radiation-dominated
diffuse plasmas
\citep{1981ApJ...244..392L,1981MNRAS.194..439T,1984MNRAS.209..175S,2007A&A...468..785C}.
These photons are therefore an important source for thermal
Comptonization.

The statistical factor for the DC emission process,
$e+\gamma_0\leftrightarrow e'+\gamma_1+\gamma_2$, is given by

\begin{eqnarray}
F&=&f(E) \,n_0 (1+n_1) (1+n_2)-f(E') \,n_1 n_2 (1+n_0) \\
&=& f(E) \,n_0 (1+n_1)\left[(1+n_2) - \expfun{-(E'-E)/k\Te} \frac{(1+n_0)}{n_0}\,\frac{n_1}{(1+n_1)}\,n_2\right]\nonumber
\end{eqnarray}

where a relativistic Maxwell-Boltzmann, $f(E)\propto \expfun{-E/k\Te}$, was assumed for the electrons. Inserting BE distributions for the photons around energies $h\nu_0$ and $h\nu_1$ and using $E'-E=h(\nu_0-\nu_1-\nu_2)$, one has

\begin{eqnarray}
F&=&f(E) \,n_0 (1+n_1)\left[(1+n_2) - \expfun{\frac{h(\nu_1+\nu_2-\nu_0)}{k\Te}} \,\expfun{x_0+\mu_0} \,\expfun{-x_1-\mu_1}\,n_2\right]\nonumber\\
&=&f(E) \,n_0 (1+n_1)\left[(1+n_2) - \expfun{\Delta x\left(1-\frac{\Tg}{\Te}\right)} \,\expfun{x_2}\,n_2\right],
\end{eqnarray}

where in the last step we assumed constant chemical potential, $\mu_1=\mu_0$. Let's look at the average change in the energy of the photon

\beal
\Delta x = x_0-x_1-x_2.
\end{align}

In the approximations that are used for the derivations, in fact, there is no direct energy exchange, so that all the energy comes from the scattering photons and $x_0-x_1-x_2\approx 0$. Thus, we have

\begin{eqnarray}
F&\approx& f(E) \,n_0 (1+n_1)\left[(1+n_2) - \expfun{x_2}\,n_2\right]\\
&=&f(E) \,n_0 (1+n_1)\left[1 - n_2 (\expfun{x_2}-1)\right].\nonumber
\end{eqnarray}

This shows that without direct energy exchange in the scattering event
(no recoil and Doppler boosting), the spectrum is driven towards an
equilibrium at the temperature of the photon field with $T=T_\gamma$.
Energy exchange leads to a small correction to the Compton process for
the high frequency photons ($\gamma_0$ and $\gamma_1$), which neglect
here \citep[for more discussion see][]{Chluba2005}.

For photon occupation number $n_x$, the number density of photons is
\begin{equation}
n = (8\pi (k_b T_\gamma)^3/(h^3 c^3))\int_0^\infty n_x x^2 dx ,
\end{equation}
where here we set the dimensionless energy $x=h \nu/(k_b T_\gamma)$.

The change in the occupation number per unit time due to {\it
  emission} of double Compton photons with energy $x$ is
\begin{equation}
\frac{dn^{\rm DC}_x}{dt} = t_c^{-1} \frac{4\alpha}{3\pi} \theta_\gamma^2
\frac{1}{x^3} I_{\rm DC}
\end{equation}
\citep{1981ApJ...244..392L,2007A&A...468..785C},
where $\theta_\gamma \equiv k_b T_\gamma/(m_e c^2)$, $t_c^{-1} = n_e
c \sigma_T = \kappa_{\rm es} c$, and $\alpha$ is the fine structure
constant.  Notice that the double Compton rates are dependent upon the
radiation temperature $T_\gamma$, unlike the other emission processes.
Note that \citealt{1981ApJ...244..392L} did not distinguish $T_e$ from
$T_\gamma$ when obtaining their equation 10a, but their derivation
only depends explicitly upon the radiation temperature and just
assumed $\theta_e\ll 1$ in order to drop frequency shift terms.  The
dimensionless DC emission-Gaunt factor is
\beq\label{eq:Gmono_soft_w0_T}
I_{\rm DC}(x,\expf,\theta_{\rm e},\theta_\gamma) = \int_{2x}^\infty
y^4(1+n_{y-x})n_y
\left[\frac{x}{y}H_G\left(\frac{x}{y}\right)\right]
G_m(y\theta_\gamma,\theta_{\rm e}) dy \Abst{,} \eeq where $y = h \nu/(k_b
T_\gamma)$ is a dummy variable for $x$ and $n_{y-x}$ means to
substitute $y-x$ as $x$ in $n_x$, and dimensionless
\begin{equation}
G_m(x\theta_\gamma,\theta_{\rm e})=\int_0^\infty G_{\rm m}(\omega_0, \beta_0)\,f(E_0,\theta_{\rm e})\,p_0^2\id p_0
\end{equation}
for momentum $p_0=m_e\gamma_0\beta_0$ and
$\gamma_0=1/\sqrt{1-\beta_0^2}$, photon frequency per electron
rest-mass energy $\omega_0=h\nu_0/(m_e c^2) = x \theta_\gamma$,
3-velocity $\beta_0$, and energy $E_0=m_e\gamma_0$ or $E_0^2=p_0^2 +
(m_e c^2)^2$. The relativistic Maxwell-Boltzmann distribution per $n_e$ is

\beq
\label{eq:relMBD}
f(E,\theta_{\rm e})=\frac{1}{4\pi\,\me^3\,K_2(1/\theta_{\rm e})\,\theta_{\rm e}}\,e^{-E/\me\theta_{\rm e}}
\Abst{,}
\eeq

where $K_2(1/\theta_{\rm e})$ is the modified Bessel function of the
second kind, with $\theta_{\rm e}=\kb\Te/(\me c^2)$, and where $\Ne$
is the electron number density, such that $1=\int f(E)\id^3p$, and
\bsub
\label{eq:G_mono_inv_beta_gen}
\beq
\label{eq:G_mono_inv_beta}
G_{\rm m}(\omega_0, \beta_0)
\approx \frac{\gamma_0^2 \,(1+\beta_0^2)}{1+\sum_{k=1}^4 f_k(\beta_0)\,\gamma_0^k\omega_0^k}
\Abst{,}
\eeq
with the functions $f_k(\beta_0)$
\beal
\label{eq:Gmono_soft_w0_beta_inv}
f_1(\beta_0)
&=\;\;\,\frac{1}{1+\beta_0^2}\,\left[\frac{21}{5}+\frac{42}{5}\beta_0^2+\frac{21}{25}\beta^4_0\right]
\\
f_2(\beta_0)
&=\;\;\,\frac{1}{(1+\beta_0^2)^2}\,\left[\frac{84}{25}+\frac{217}{25}\beta_0^2+\frac{1967}{125}\beta^4_0\right]
\\
f_3(\beta_0)
&=-\frac{1}{(1+\beta_0^2)^3}\,\left[\frac{2041}{875}+\frac{1306}{125}\beta_0^2\right]
\\
f_4(\beta_0)
&=\;\;\,\frac{1}{(1+\beta_0^2)^4}\,\frac{9663}{4375}
\end{align}
\esub

\citep{2007A&A...468..785C,2012MNRAS.419.1294C} (with typo fixed in
$f_3$).  As compared to previous expressions that require $\theta_{\rm e}\ll
1$ and $\theta_\gamma\ll 1$ and $x\ll 1$ (i.e. cold electrons, cold
photons, and soft photons) \citep{1981ApJ...244..392L} where $I_{\rm
  DC}\to I_{0,\rm DC}$ with
\begin{equation}
I_{0,\rm DC} = \int_0^\infty x^4 (1+n_x)n_x dx ,
\end{equation}

our version from \citet{2007A&A...468..785C} is accurate for
moderately relativistic electrons and photons (i.e. $\theta_{\rm
e}\lesssim 1$ and $\theta_\gamma\lesssim 1$ and $x\lesssim 1$).  For
$x\gtrsim 1$ double Compton is suppressed as ${\rm e}^{-2x}$, and high
energy photons are more readily generated by single Comptonization off
the double Compton photons \citep{1981MNRAS.194..439T}.

The differential number emission rate (number per unit time per unit
volume) is
\begin{equation}
d\dot{n}_{\rm DC} = (8\pi (k_b T_\gamma)^3/(h^3 c^3)) x^2 dx \frac{dn^{\rm
DC}_x}{dt} ,
\end{equation}
and the differential emission rate (energy per unit time per unit volume)
is
\begin{equation}
d\dot{E}_{\rm DC} = (k_b T_\gamma)x d\dot{n}_{\rm DC} ,
\end{equation}
such that
\begin{equation}
j_{\rm DC}(x) = d\dot{E}_{\rm DC}/(4\pi) .
\end{equation}

The energy and number mean opacities can then be computed from
$\alpha_\nu^{\rm DC} = j^{\rm DC}_\nu/B_\nu(T_\gamma) = j^{\rm
DC}(x)/B(x,T_\gamma)$ where $B(x)=B_\nu (d\nu/dx)$ with $d\nu/dx=\kb
T_\gamma/h$.  As shown in \S\ref{subsec:dc}, here Kirchhoff's law is
based upon a distribution at $T=T_\gamma$, because these DC
expressions include no recoil or Doppler shifting, and they presume
energy balance between the incoming photon and the two outgoing
photons.  That is, DC here includes no energy exchange with electrons
and so absorption only drives the photon distribution to Planck at its
own radiation temperature.  Energy exchange from single Comptonization
dominates that one would obtain from recoil/Doppler for double
Compton, so the energy exchange from Eq.~(\ref{CompG0}) is sufficient
as a independent (non-DC) mechanism to drive thermal equilibrium
between electrons and photons.  Unlike free-free, electrons can be
completely cold and DC emission still occurs if there are seed
photons \citep{1981ApJ...244..392L}.

We are not aware of work done to establish the Razin effect with
Double Compton, so we apply the same approach used for bremsstrahlung.

\subsection{Tabulating/Fitting Double Compton}

The dimensionfull factor for the DC opacity is
\begin{equation}
h(n_e,T_\gamma,\expf) = 7.360\times 10^{-46} n_e T_\gamma^2 \expf
\end{equation}
where the $\expf$ factor was pulled from the dimensionless integrals.
Now the residual factor $\kappa_{\rm a, DC}/(p(\theta_{\rm e})
h(n_e,T_\gamma,\expf))$ and $\kappa_{\rm an, DC}/(p(\theta_{\rm e})
h(n_e,T_\gamma,\expf))$ can be tabulated.  The DC emission opacity is
obtained by choosing $T_{\rm eq}=T_\gamma$ and $\mu_{\rm eq}=0$, but
that still leaves the emission opacity dependent upon the radiation
$\mu$ from the DC $j_\nu$.

For HARMRAD, we fit the opacity residuals.  For DC our approach is to
first fit the regime $\theta_{\rm e}\ll 1$ where the residual is constant
vs. $\theta_{\rm e}$, and then find a rough fit vs. $\theta_{\rm e}\gtrsim 0.01$
when the opacity begins to change by more than $20\%$ vs. $\theta_{\rm e}$.

For any $\theta_{\rm e}$, the residual factors $\kappa_{\rm a,
DC}/(p(\theta_{\rm e}) h(n_e,T_\gamma,\expf))$ and $\kappa_{\rm an,
DC}/(p(\theta_{\rm e}) h(n_e,T_\gamma,\expf))$ can be fit by the form
\begin{equation}
\left(\frac{\kappa_{\rm \{a,an\}, DC}}{p(\theta_{\rm e})
h(n_e,T_\gamma,\expf)}\right)^{-1} \approx a^{-1} +
\left(b\theta_\gamma^{-c}\right)^{-1}+
\left(d\theta_\gamma^{-c/3}\right)^{-1} ,
\end{equation}
with a relative error less than $20\%$.  For DC, we choose densities
and temperatures relevant for an Eddington-accreting BH X-ray binary,
which enters the Razin cut-off and modification of the BE
distribution.  DC for diffuse systems like SgrA* is much lower than
synchrotron due to the very low radiation temperature of thermal
synchrotron, so it is not considered.

For $\theta_{\rm e}\ll 1$, we obtain fits are accurate to less than $20\%$
relative error for all $10^{-4}\le \theta_\gamma\le 10^2$ and
$0\le\expf\le 1$.  The energy absorption opacity coefficients are
\begin{eqnarray}
a &=& 4.16 (1-(\expf))^{1.69}+6.7 (\expf)^{0.942}+3.1\times 10^{\text{-8}}\\
b &=& -0.0334 (1-(\expf))^{0.469}-0.0021 (\expf)^{0.0217}+0.042\nonumber\\
c &=& -0.18 (\expf)^{33.}+0.201 (1-(\expf))^{0.258}+3.8\nonumber\\
d &=& 0.0169  (\expf)^{35.4}-0.0626
(1-(\expf))^{0.35}+0.118\nonumber
\end{eqnarray}
The energy emission opacity coefficients are
\begin{eqnarray}
a &=& -0.0589 (1-(\expf))^{10.7}+0.488 (\expf)^{1.75}+6.34\\
b &=& 0.0282 (\expf)^{1.56}+0.0142 (1-(\expf))^{0.361}+0.00875\nonumber\\
c &=& -0.16 (\expf)^{15.4}+0.184 (1-(\expf))^{0.366}+3.78\nonumber\\
d &=& 0.015 (\expf)^{26.3}-0.0256
(1-(\expf))^{0.398}+0.119\nonumber
\end{eqnarray}
The Planck limit gives coefficients $\{a,b,c,d\}=\{6.83,0.0374,3.63,0.134\}$ for the energy
absorption and emission opacities.

The number absorption opacity coefficients are
\begin{eqnarray}
a_n &=& 29.4 (\expf)^{285.}-76.4 (1-(\expf))^{0.136}+87.5\\
b_n &=& 0.196 (\expf)^{18.1}-1.12 (1-(\expf))^{0.134}+1.16\nonumber\\
c_n &=& 0.0427 (1-(\expf))^{182.}-0.8 (\expf)^{21.6}+3.93\nonumber\\
d_n &=& 1.87 (\expf)^{309.}-2.72  (1-(\expf))^{0.106}+2.86\nonumber
\end{eqnarray}
The number emission mean opacity coefficients are
\begin{eqnarray}
a_n &=& -81.7 (\expf)^{1.01}-94.8 (1-(\expf))^{0.925}+198.\\
b_n &=& 1.31 (\expf)^{1.12}+1.05 (1-(\expf))^{0.249}+4.8\times 10^{\text{-11}}\nonumber\\
c_n &=& -0.418 (\expf)^{14.8}+0.442 (1-(\expf))^{0.361}+3.44\nonumber\\
d_n &=& 1.37 (\expf)^{31.3}-1.38 (1-(\expf))^{0.316}+3.38\nonumber
\end{eqnarray}
The Planck limit gives coefficients
$\{a_n,b_n,c_n,d_n\}=\{116.,1.34,3.03,4.72\}$.

To obtain a rough fit that will be accurate for $\theta_{\rm e}\gtrsim
0.01$, we fit the ratio of the higher $\theta_{\rm e}$ opacities to the
$\theta_{\rm e}\ll 1$ opacity.  This gives a fit to the ratio $p(\theta_{\rm e})$
of the general opacity to the $\theta_{\rm e}\ll 1$ opacity
\begin{equation}
p(\theta_{\rm e}) \approx (1+\theta_{\rm e})^{-3}
\end{equation}
which does not change the accuracy at low $\theta_{\rm e}$ but
improves the accuracy for $\theta_{\rm e}\le 1$ to be within a factor
of three or better and $\theta_{\rm e}\le 0.1$ to be within a factor
of two or better.  This factor $p(\theta_{\rm e})$ gives the same
correction and accuracy for all opacities (energy absorption, number
absorption, energy emission, and number emission).  The DC calculation
is only accurate for $\theta_{\rm e}\lesssim 1$, beyond which pair
production processes (not included) become important.

\label{lastpage}

{\small \begin{thebibliography}{144}
\expandafter\ifx\csname natexlab\endcsname\relax\def\natexlab#1{#1}\fi

\bibitem[{Abramowicz} et~al.(1988){Abramowicz}, {Czerny}, {Lasota} \&
  {Szuszkiewicz}]{abr88}
{Abramowicz} M.~A., {Czerny} B., {Lasota} J.~P., {Szuszkiewicz} E., 1988, \apj,
  332, 646

\bibitem[Akhiezer \& Berestetsky(1953)]{akhiezer1953quantum}
Akhiezer A., Berestetsky V., 1953, Quantum electrodynamics, no. v. 1 in Quantum
  Electrodynamics, Techical information service extension [1953?]

\bibitem[{Akopyan} \& {Tsytovich}(1976)]{1976JETP...44...87A}
{Akopyan} A.~V., {Tsytovich} V.~N., 1976, Soviet Journal of Experimental and
  Theoretical Physics, 44, 87

\bibitem[{Anthony} et~al.(1996){Anthony}, {Becker-Szendy}, {Bosted}
  et~al.]{1996PhRvL..76.3550A}
{Anthony} P.~L., {Becker-Szendy} R., {Bosted} P.~E., et~al., 1996, Physical
  Review Letters, 76, 3550

\bibitem[Ascoli \& Bussetti(1956)]{ascoli1956}
Ascoli R., Bussetti G., 1956, Il Nuovo Cimento, 4, 2, 189

\bibitem[{Avara} et~al.(2015){Avara}, {McKinney} \&
  {Reynolds}]{2015arXiv150805323A}
{Avara} M.~J., {McKinney} J.~C., {Reynolds} C.~S., 2015, ArXiv e-prints

\bibitem[{Balbus} \& {Hawley}(1998)]{1998RvMP...70....1B}
{Balbus} S.~A., {Hawley} J.~F., 1998, Reviews of Modern Physics, 70, 1

\bibitem[{B{\'e}gu{\'e}} \& {Pe'er}(2015)]{2015ApJ...802..134B}
{B{\'e}gu{\'e}} D., {Pe'er} A., 2015, \apj, 802, 134

\bibitem[Bekefi(1966)]{bekefi1966radiation}
Bekefi G., 1966, Radiation processes in plasmas, Wiley series in plasma
  physics, Wiley

\bibitem[{Blaes} et~al.(2006){Blaes}, {Davis}, {Hirose}, {Krolik} \&
  {Stone}]{2006ApJ...645.1402B}
{Blaes} O.~M., {Davis} S.~W., {Hirose} S., {Krolik} J.~H., {Stone} J.~M., 2006,
  \apj, 645, 1402

\bibitem[{Brown} \& {Preston}(2012)]{2012APh....35..742B}
{Brown} L.~S., {Preston} D.~L., 2012, Astroparticle Physics, 35, 742

\bibitem[{Buchler} \& {Yueh}(1976)]{1976ApJ...210..440B}
{Buchler} J.~R., {Yueh} W.~R., 1976, \apj, 210, 440

\bibitem[{Burigana} et~al.(1991){Burigana}, {Danese} \& {de
  Zotti}]{1991A&A...246...49B}
{Burigana} C., {Danese} L., {de Zotti} G., 1991, \aap, 246, 49

\bibitem[{Burigana} et~al.(1995){Burigana}, {de Zotti} \&
  {Danese}]{1995A&A...303..323B}
{Burigana} C., {de Zotti} G., {Danese} L., 1995, \aap, 303, 323

\bibitem[{Castell{\'o}-Mor} et~al.(2016){Castell{\'o}-Mor}, {Netzer} \&
  {Kaspi}]{2016MNRAS.458.1839C}
{Castell{\'o}-Mor} N., {Netzer} H., {Kaspi} S., 2016, \mnras, 458, 1839

\bibitem[{Castor}(2004)]{2004rahy.book.....C}
{Castor} J.~I., 2004, {Radiation Hydrodynamics}

\bibitem[{Challinor} \& {Lasenby}(1998)]{1998ApJ...499....1C}
{Challinor} A., {Lasenby} A., 1998, \apj, 499, 1

\bibitem[Chen \& Klein(1993)]{Chen:1992ne}
Chen P., Klein S., 1993, AIP Conf.Proc., 279, 929

\bibitem[{Chluba}(2005)]{Chluba2005}
{Chluba} J., 2005, {Spectral Distortions of the Cosmic Microwave Background},
  Ph.D. thesis, LMU M{\"u}nchen

\bibitem[{Chluba}(2014)]{2014MNRAS.440.2544C}
{Chluba} J., 2014, \mnras, 440, 2544

\bibitem[{Chluba} et~al.(2007){Chluba}, {Sazonov} \&
  {Sunyaev}]{2007A&A...468..785C}
{Chluba} J., {Sazonov} S.~Y., {Sunyaev} R.~A., 2007, \aap, 468, 785

\bibitem[{Chluba} \& {Sunyaev}(2012)]{2012MNRAS.419.1294C}
{Chluba} J., {Sunyaev} R.~A., 2012, \mnras, 419, 1294

\bibitem[{Christen} \& {Kassubek}(2014)]{2014JPhD...47J3001C}
{Christen} T., {Kassubek} F., 2014, Journal of Physics D Applied Physics, 47,
  363001

\bibitem[{Dai} et~al.(2015){Dai}, {McKinney} \& {Miller}]{2015ApJ...812L..39D}
{Dai} L., {McKinney} J.~C., {Miller} M.~C., 2015, \apjl, 812, L39

\bibitem[{Davis} et~al.(2014){Davis}, {Jiang}, {Stone} \&
  {Murray}]{2014ApJ...796..107D}
{Davis} S.~W., {Jiang} Y.-F., {Stone} J.~M., {Murray} N., 2014, \apj, 796, 107

\bibitem[Dawson \& Oberman(1962)]{dawsonoberman1962}
Dawson J., Oberman C., 1962, Physics of Fluids (1958-1988), 5, 5, 517

\bibitem[{Di Matteo} et~al.(1997){Di Matteo}, {Celotti} \&
  {Fabian}]{1997MNRAS.291..805D}
{Di Matteo} T., {Celotti} A., {Fabian} A.~C., 1997, \mnras, 291, 805

\bibitem[{Dougherty} et~al.(2003){Dougherty}, {Pittard}, {Kasian}, {Coker},
  {Williams} \& {Lloyd}]{2003A&A...409..217D}
{Dougherty} S.~M., {Pittard} J.~M., {Kasian} L., {Coker} R.~F., {Williams}
  P.~M., {Lloyd} H.~M., 2003, \aap, 409, 217

\bibitem[{Draine}(2011)]{2011piim.book.....D}
{Draine} B.~T., 2011, {Physics of the Interstellar and Intergalactic Medium}

\bibitem[{Esin} et~al.(1996){Esin}, {Narayan}, {Ostriker} \&
  {Yi}]{1996ApJ...465..312E}
{Esin} A.~A., {Narayan} R., {Ostriker} E., {Yi} I., 1996, \apj, 465, 312

\bibitem[{Fabian} et~al.(2015){Fabian}, {Lohfink}, {Kara}, {Parker},
  {Vasudevan} \& {Reynolds}]{2015MNRAS.451.4375F}
{Fabian} A.~C., {Lohfink} A., {Kara} E., {Parker} M.~L., {Vasudevan} R.,
  {Reynolds} C.~S., 2015, \mnras, 451, 4375

\bibitem[{Fortmann} et~al.(2005){Fortmann}, {Reinholz}, {Roepke} \&
  {Wierling}]{2005physics...2051F}
{Fortmann} C., {Reinholz} H., {Roepke} G., {Wierling} A., 2005, ArXiv Physics
  e-prints

\bibitem[Fortmann et~al.(2007)Fortmann, Roepke \& Wierling]{4345536}
Fortmann C., Roepke G., Wierling A., 2007, in { Plasma Science, 2007. ICOPS
  2007. IEEE 34th International Conference on\/},  230--230

\bibitem[{Fragile} et~al.(2012){Fragile}, {Gillespie}, {Monahan}, {Rodriguez}
  \& {Anninos}]{2012ApJS..201....9F}
{Fragile} P.~C., {Gillespie} A., {Monahan} T., {Rodriguez} M., {Anninos} P.,
  2012, \apjs, 201, 9

\bibitem[{Fragile} \& {Meier}(2009)]{fm09}
{Fragile} P.~C., {Meier} D.~L., 2009, \apj, 693, 771

\bibitem[{Fragile} et~al.(2014){Fragile}, {Olejar} \&
  {Anninos}]{2014ApJ...796...22F}
{Fragile} P.~C., {Olejar} A., {Anninos} P., 2014, \apj, 796, 22

\bibitem[{Gammie}(1999)]{1999ApJ...522L..57G}
{Gammie} C.~F., 1999, \apjl, 522, L57

\bibitem[{Gammie} et~al.(2003){Gammie}, {McKinney} \&
  {T{\'o}th}]{2003ApJ...589..444G}
{Gammie} C.~F., {McKinney} J.~C., {T{\'o}th} G., 2003, \apj, 589, 444

\bibitem[{Garain} \& {Chakrabarti}(2013)]{2013arXiv1307.7355G}
{Garain} S.~K., {Chakrabarti} H.~G.~S.~K., 2013, ArXiv e-prints

\bibitem[{Ginzburg} \& {Syrovatskii}(1965)]{1965ARA&A...3..297G}
{Ginzburg} V.~L., {Syrovatskii} S.~I., 1965, \araa, 3, 297

\bibitem[{Gould}(1990)]{1990ApJ...362..284G}
{Gould} R.~J., 1990, \apj, 362, 284

\bibitem[{Guzik} \& {Cox}(1995)]{1995ApJ...448..905G}
{Guzik} J.~A., {Cox} A.~N., 1995, \apj, 448, 905

\bibitem[{Heng} et~al.(2012){Heng}, {Hayek}, {Pont} \&
  {Sing}]{2012MNRAS.420...20H}
{Heng} K., {Hayek} W., {Pont} F., {Sing} D.~K., 2012, \mnras, 420, 20

\bibitem[{Hirose} et~al.(2009){Hirose}, {Krolik} \&
  {Blaes}]{2009ApJ...691...16H}
{Hirose} S., {Krolik} J.~H., {Blaes} O., 2009, \apj, 691, 16

\bibitem[{Hornby} \& {Williams}(1966)]{1966MNRAS.131..237H}
{Hornby} J.~M., {Williams} P.~J.~S., 1966, \mnras, 131, 237

\bibitem[{Hubeny} et~al.(2003){Hubeny}, {Burrows} \&
  {Sudarsky}]{2003ApJ...594.1011H}
{Hubeny} I., {Burrows} A., {Sudarsky} D., 2003, \apj, 594, 1011

\bibitem[Huebner \& Barfield(2014)]{huebner2014opacity}
Huebner W., Barfield W., 2014, Opacity, Astrophysics and Space Science Library,
  Springer New York

\bibitem[{Iben}(1975)]{1975ApJ...196..525I}
{Iben} Jr. I., 1975, \apj, 196, 525

\bibitem[Ichimaru(1973)]{ichimaru1973}
Ichimaru S., 1973, Basic Principles Of Plasma Physics: A Statistical Approach
  (Frontiers in Physics), Frontiers in Physics, W. A. Benjamin

\bibitem[{Iglesias} \& {Rogers}(1996)]{1996ApJ...464..943I}
{Iglesias} C.~A., {Rogers} F.~J., 1996, \apj, 464, 943

\bibitem[{Igumenshchev} et~al.(2003){Igumenshchev}, {Narayan} \&
  {Abramowicz}]{2003ApJ...592.1042I}
{Igumenshchev} I.~V., {Narayan} R., {Abramowicz} M.~A., 2003, \apj, 592, 1042

\bibitem[{Itoh} et~al.(2000){Itoh}, {Sakamoto}, {Kusano}, {Nozawa} \&
  {Kohyama}]{2000ApJS..128..125I}
{Itoh} N., {Sakamoto} T., {Kusano} S., {Nozawa} S., {Kohyama} Y., 2000, \apjs,
  128, 125

\bibitem[{Jiang} et~al.(2015){Jiang}, {Cantiello}, {Bildsten}, {Quataert} \&
  {Blaes}]{2015ApJ...813...74J}
{Jiang} Y.-F., {Cantiello} M., {Bildsten} L., {Quataert} E., {Blaes} O., 2015,
  \apj, 813, 74

\bibitem[{Jiang} et~al.(2016){Jiang}, {Davis} \& {Stone}]{2016arXiv160106836J}
{Jiang} Y.-F., {Davis} S., {Stone} J., 2016, ArXiv e-prints

\bibitem[{Jiang} et~al.(2013){Jiang}, {Stone} \& {Davis}]{2013ApJ...778...65J}
{Jiang} Y.-F., {Stone} J.~M., {Davis} S.~W., 2013, \apj, 778, 65

\bibitem[{Jiang} et~al.(2014{\natexlab{a}}){Jiang}, {Stone} \&
  {Davis}]{2014ApJ...796..106J}
{Jiang} Y.-F., {Stone} J.~M., {Davis} S.~W., 2014{\natexlab{a}}, \apj, 796, 106

\bibitem[{Jiang} et~al.(2014{\natexlab{b}}){Jiang}, {Stone} \&
  {Davis}]{2014ApJ...784..169J}
{Jiang} Y.-F., {Stone} J.~M., {Davis} S.~W., 2014{\natexlab{b}}, \apj, 784, 169

\bibitem[{Just} et~al.(2007){Just}, {Brandt}, {Shemmer}
  et~al.]{2007ApJ...665.1004J}
{Just} D.~W., {Brandt} W.~N., {Shemmer} O., et~al., 2007, \apj, 665, 1004

\bibitem[Kaku(1993)]{kaku1993quantum}
Kaku M., 1993, Quantum Field Theory: A Modern Introduction, Oxford University
  Press

\bibitem[{Kawashima} et~al.(2012)]{kaw12}
{Kawashima} T., et~al., 2012, \apj, 752, 18

\bibitem[{Kawashima} et~al.(2009){Kawashima}, {Ohsuga}, {Mineshige},
  {Heinzeller}, {Takabe} \& {Matsumoto}]{2009PASJ...61..769K}
{Kawashima} T., {Ohsuga} K., {Mineshige} S., {Heinzeller} D., {Takabe} H.,
  {Matsumoto} R., 2009, \pasj, 61, 769

\bibitem[{Klein}(1999)]{1999RvMP...71.1501K}
{Klein} S., 1999, Reviews of Modern Physics, 71, 1501

\bibitem[{Klein}(1997)]{1997STIN...9932370K}
{Klein} S.~R., 1997, NASA STI/Recon Technical Report N, 99, 32370

\bibitem[{Kubota} \& {Done}(2004)]{2004MNRAS.353..980K}
{Kubota} A., {Done} C., 2004, \mnras, 353, 980

\bibitem[{Kulkarni} et~al.(2011){Kulkarni}, {Penna}, {Shcherbakov}
  et~al.]{2011MNRAS.414.1183K}
{Kulkarni} A.~K., {Penna} R.~F., {Shcherbakov} R.~V., et~al., 2011, \mnras,
  414, 1183

\bibitem[{Leung} et~al.(2011){Leung}, {Gammie} \& {Noble}]{2011ApJ...737...21L}
{Leung} P.~K., {Gammie} C.~F., {Noble} S.~C., 2011, \apj, 737, 21

\bibitem[{Lightman}(1981)]{1981ApJ...244..392L}
{Lightman} A.~P., 1981, \apj, 244, 392

\bibitem[{Ludwig} et~al.(1994){Ludwig}, {Jordan} \&
  {Steffen}]{1994A&A...284..105L}
{Ludwig} H.-G., {Jordan} S., {Steffen} M., 1994, \aap, 284, 105

\bibitem[{Mahadevan} et~al.(1996){Mahadevan}, {Narayan} \&
  {Yi}]{1996ApJ...465..327M}
{Mahadevan} R., {Narayan} R., {Yi} I., 1996, \apj, 465, 327

\bibitem[{Malygin} et~al.(2014){Malygin}, {Kuiper}, {Klahr}, {Dullemond} \&
  {Henning}]{2014A&A...568A..91M}
{Malygin} M.~G., {Kuiper} R., {Klahr} H., {Dullemond} C.~P., {Henning} T.,
  2014, \aap, 568, A91

\bibitem[{Mason} \& {Turolla}(1992)]{1992ApJ...400..170M}
{Mason} A., {Turolla} R., 1992, \apj, 400, 170

\bibitem[{McCray}(1967)]{1967ApJ...147..544M}
{McCray} R., 1967, \apj, 147, 544

\bibitem[{McKinney}(2006)]{2006MNRAS.368.1561M}
{McKinney} J.~C., 2006, \mnras, 368, 1561

\bibitem[{McKinney} \& {Blandford}(2009)]{2009MNRAS.394L.126M}
{McKinney} J.~C., {Blandford} R.~D., 2009, \mnras, 394, L126

\bibitem[{McKinney} et~al.(2015){McKinney}, {Dai} \&
  {Avara}]{2015MNRAS.454L...6M}
{McKinney} J.~C., {Dai} L., {Avara} M.~J., 2015, \mnras, 454, L6

\bibitem[{McKinney} \& {Gammie}(2004)]{2004ApJ...611..977M}
{McKinney} J.~C., {Gammie} C.~F., 2004, \apj, 611, 977

\bibitem[{McKinney} et~al.(2012){McKinney}, {Tchekhovskoy} \&
  {Blandford}]{2012MNRAS.423.3083M}
{McKinney} J.~C., {Tchekhovskoy} A., {Blandford} R.~D., 2012, \mnras, 423, 3083

\bibitem[{McKinney} et~al.(2014){McKinney}, {Tchekhovskoy}, {Sadowski} \&
  {Narayan}]{2014MNRAS.441.3177M}
{McKinney} J.~C., {Tchekhovskoy} A., {Sadowski} A., {Narayan} R., 2014, \mnras,
  441, 3177

\bibitem[{McKinney} \& {Uzdensky}(2012)]{2012MNRAS.419..573M}
{McKinney} J.~C., {Uzdensky} D.~A., 2012, \mnras, 419, 573

\bibitem[{Melrose}(1972)]{1972Ap&SS..18..267M}
{Melrose} D.~B., 1972, \apss, 18, 267

\bibitem[Mercier(1964)]{mercier64}
Mercier R.~P., 1964, Proceedings of the Physical Society, 83, 5, 819

\bibitem[{Mihalas} \& {Mihalas}(1984)]{1984oup..book.....M}
{Mihalas} D., {Mihalas} B.~W., 1984, {Foundations of radiation hydrodynamics}

\bibitem[{Miller} et~al.(2011){Miller}, {Boutloukos}, {Lo} \&
  {Lamb}]{2011fxts.confE..24M}
{Miller} M.~C., {Boutloukos} S., {Lo} K.~H., {Lamb} F.~K., 2011, in { Fast
  X-ray Timing and Spectroscopy at Extreme Count Rates (HTRS 2011)\/}, ~24

\bibitem[{Mishra} et~al.(2016){Mishra}, {Fragile}, {Johnson} \&
  {Klu{\'z}niak}]{2016arXiv160304082M}
{Mishra} B., {Fragile} P.~C., {Johnson} L.~C., {Klu{\'z}niak} W., 2016, ArXiv
  e-prints

\bibitem[Modest(2013)]{modest2013}
Modest M.~F., ed., 2013, Front matter, Academic Press, Boston, third edition
  edn.

\bibitem[Nagirner et~al.(1997)Nagirner, Loskutov \& Grachev]{nagirner97}
Nagirner D., Loskutov V., Grachev S., 1997, Astrophysics, 40, 3, 227

\bibitem[{Narayan} et~al.(2003){Narayan}, {Igumenshchev} \&
  {Abramowicz}]{2003PASJ...55L..69N}
{Narayan} R., {Igumenshchev} I.~V., {Abramowicz} M.~A., 2003, \pasj, 55, L69

\bibitem[{Narayan} et~al.(2016){Narayan}, {Zhu}, {Psaltis} \& {Sa{\c
  d}owski}]{2016MNRAS.457..608N}
{Narayan} R., {Zhu} Y., {Psaltis} D., {Sa{\c d}owski} A., 2016, \mnras, 457,
  608

\bibitem[{Niedzwiecki} et~al.(2014){Niedzwiecki}, {Stepnik} \&
  {Xie}]{2014arXiv1412.4593N}
{Niedzwiecki} A., {Stepnik} A., {Xie} F.-G., 2014, ArXiv e-prints

\bibitem[{Novikov} \& {Thorne}(1973)]{1973blho.conf..343N}
{Novikov} I.~D., {Thorne} K.~S., 1973, in { Black Holes (Les Astres Occlus)\/},
  edited by C.~{Dewitt}, B.~S. {Dewitt},  343--450

\bibitem[{Nozawa} \& {Kohyama}(2015)]{2015APh....62...30N}
{Nozawa} S., {Kohyama} Y., 2015, Astroparticle Physics, 62, 30

\bibitem[{O' Riordan} et~al.(2016{\natexlab{a}}){O' Riordan}, {Pe'er} \&
  {McKinney}]{2016arXiv160701060O}
{O' Riordan} M., {Pe'er} A., {McKinney} J.~C., 2016{\natexlab{a}}, ArXiv
  e-prints

\bibitem[{O' Riordan} et~al.(2016{\natexlab{b}}){O' Riordan}, {Pe'er} \&
  {McKinney}]{2016ApJ...819...95O}
{O' Riordan} M., {Pe'er} A., {McKinney} J.~C., 2016{\natexlab{b}}, \apj, 819,
  95

\bibitem[{Page} \& {Thorne}(1974)]{pt74}
{Page} D.~N., {Thorne} K.~S., 1974, \apj, 191, 499

\bibitem[{Pandya} et~al.(2016){Pandya}, {Zhang}, {Chandra} \&
  {Gammie}]{2016ApJ...822...34P}
{Pandya} A., {Zhang} Z., {Chandra} M., {Gammie} C.~F., 2016, \apj, 822, 34

\bibitem[{Patch}(1967)]{1967JQSRT...7..611P}
{Patch} R., 1967, \jqsrt, 7, 611

\bibitem[{Paxton} et~al.(2011){Paxton}, {Bildsten}, {Dotter}, {Herwig},
  {Lesaffre} \& {Timmes}]{2011ApJS..192....3P}
{Paxton} B., {Bildsten} L., {Dotter} A., {Herwig} F., {Lesaffre} P., {Timmes}
  F., 2011, \apjs, 192, 3

\bibitem[{Penna} et~al.(2010){Penna}, {McKinney}, {Narayan}, {Tchekhovskoy},
  {Shafee} \& {McClintock}]{pmntsm10}
{Penna} R.~F., {McKinney} J.~C., {Narayan} R., {Tchekhovskoy} A., {Shafee} R.,
  {McClintock} J.~E., 2010, \mnras, 408, 752

\bibitem[{Pessah} \& {Psaltis}(2005)]{pp05}
{Pessah} M.~E., {Psaltis} D., 2005, \apj, 628, 879

\bibitem[{Poutanen}(2016)]{2016arXiv160609466P}
{Poutanen} J., 2016, ArXiv e-prints

\bibitem[{Poutanen} \& {Svensson}(1996)]{1996ApJ...470..249P}
{Poutanen} J., {Svensson} R., 1996, \apj, 470, 249

\bibitem[{Pozdnyakov} et~al.(1983){Pozdnyakov}, {Sobol} \&
  {Syunyaev}]{1983ASPRv...2..189P}
{Pozdnyakov} L.~A., {Sobol} I.~M., {Syunyaev} R.~A., 1983, Astrophysics and
  Space Physics Reviews, 2, 189

\bibitem[{Procopio} \& {Burigana}(2009)]{2009A&A...507.1243P}
{Procopio} P., {Burigana} C., 2009, \aap, 507, 1243

\bibitem[{Proga} et~al.(1998){Proga}, {Stone} \& {Drew}]{1998MNRAS.295..595P}
{Proga} D., {Stone} J.~M., {Drew} J.~E., 1998, \mnras, 295, 595

\bibitem[{Roberts} et~al.(2016){Roberts}, {Ott}, {Haas}, {O'Connor}, {Diener}
  \& {Schnetter}]{2016arXiv160407848R}
{Roberts} L.~F., {Ott} C.~D., {Haas} R., {O'Connor} E.~P., {Diener} P.,
  {Schnetter} E., 2016, ArXiv e-prints

\bibitem[{Rosdahl} \& {Teyssier}(2015)]{2015MNRAS.449.4380R}
{Rosdahl} J., {Teyssier} R., 2015, \mnras, 449, 4380

\bibitem[{Ryan} et~al.(2015){Ryan}, {Dolence} \& {Gammie}]{2015ApJ...807...31R}
{Ryan} B.~R., {Dolence} J.~C., {Gammie} C.~F., 2015, \apj, 807, 31

\bibitem[{Rybicki} \& {Lightman}(1986)]{1986rpa..book.....R}
{Rybicki} G.~B., {Lightman} A.~P., 1986, {Radiative Processes in Astrophysics}

\bibitem[{Sadowski} \& {Narayan}(2015{\natexlab{a}})]{2015arXiv150804980S}
{Sadowski} A., {Narayan} R., 2015{\natexlab{a}}, ArXiv e-prints

\bibitem[{Sadowski} \& {Narayan}(2015{\natexlab{b}})]{2015arXiv150300654S}
{Sadowski} A., {Narayan} R., 2015{\natexlab{b}}, arxiv: 1503.00654

\bibitem[{Sadowski} et~al.(2016){Sadowski}, {Wielgus}, {Narayan}, {Abarca} \&
  {McKinney}]{2016arXiv160503184S}
{Sadowski} A., {Wielgus} M., {Narayan} R., {Abarca} D., {McKinney} J.~C., 2016,
  ArXiv e-prints

\bibitem[{Sakamoto} et~al.(2001){Sakamoto}, {Itoh}, {Kusano}, {Nozawa} \&
  {Kohyama}]{2001ASPC..251..268S}
{Sakamoto} T., {Itoh} N., {Kusano} S., {Nozawa} S., {Kohyama} Y., 2001, in {
  New Century of X-ray Astronomy\/}, edited by H.~{Inoue}, H.~{Kunieda}, vol.
  251 of { Astronomical Society of the Pacific Conference Series\/},  268

\bibitem[{Sampson}(1965)]{1965JQSRT...5..211S}
{Sampson} D., 1965, \jqsrt, 5, 211

\bibitem[{Sampson}(1959)]{1959ApJ...129..734S}
{Sampson} D.~H., 1959, \apj, 129, 734

\bibitem[{Sazonov} \& {Sunyaev}(1998)]{1998ApJ...508....1S}
{Sazonov} S.~Y., {Sunyaev} R.~A., 1998, \apj, 508, 1

\bibitem[{S{\c a}dowski}(2016)]{2016MNRAS.459.4397S}
{S{\c a}dowski} A., 2016, \mnras, 459, 4397

\bibitem[{S{\c a}dowski} et~al.(2014){S{\c a}dowski}, {Narayan}, {McKinney} \&
  {Tchekhovskoy}]{2014MNRAS.439..503S}
{S{\c a}dowski} A., {Narayan} R., {McKinney} J.~C., {Tchekhovskoy} A., 2014,
  \mnras, 439, 503

\bibitem[{S{\c a}dowski} et~al.(2015){S{\c a}dowski}, {Narayan},
  {Tchekhovskoy}, {Abarca}, {Zhu} \& {McKinney}]{2015MNRAS.447...49S}
{S{\c a}dowski} A., {Narayan} R., {Tchekhovskoy} A., {Abarca} D., {Zhu} Y.,
  {McKinney} J.~C., 2015, \mnras, 447, 49

\bibitem[{Scheuer}(1960)]{1960MNRAS.120..231S}
{Scheuer} P.~A.~G., 1960, \mnras, 120, 231

\bibitem[{Schnittman} et~al.(2013){Schnittman}, {Krolik} \&
  {Noble}]{2013ApJ...769..156S}
{Schnittman} J.~D., {Krolik} J.~H., {Noble} S.~C., 2013, \apj, 769, 156

\bibitem[{Seaton}(1993)]{1993MNRAS.265L..25S}
{Seaton} M.~J., 1993, \mnras, 265, L25

\bibitem[{Seaton} et~al.(1994){Seaton}, {Yan}, {Mihalas} \&
  {Pradhan}]{1994MNRAS.266..805S}
{Seaton} M.~J., {Yan} Y., {Mihalas} D., {Pradhan} A.~K., 1994, \mnras, 266, 805

\bibitem[{Shu}(1991)]{1991pav..book.....S}
{Shu} F.~H., 1991, {The physics of astrophysics. Volume 1: Radiation.}

\bibitem[{Shul'Ga} \& {Fomin}(1998)]{1998JETP...86...32S}
{Shul'Ga} N.~F., {Fomin} S.~P., 1998, Soviet Journal of Experimental and
  Theoretical Physics, 86, 32

\bibitem[{Simon}(1969)]{1969ApJ...156..341S}
{Simon} M., 1969, \apj, 156, 341

\bibitem[{Steffen} et~al.(2006){Steffen}, {Strateva}, {Brandt}
  et~al.]{2006AJ....131.2826S}
{Steffen} A.~T., {Strateva} I., {Brandt} W.~N., et~al., 2006, \aj, 131, 2826

\bibitem[{Struchtrup}(1997)]{1997AnPhy.257..111S}
{Struchtrup} H., 1997, Annals of Physics, 257, 111

\bibitem[{Sturrock} et~al.(1986){Sturrock}, {Holzer}, {Mihalas} \&
  {Ulrich}]{1986psun....1.....S}
{Sturrock} P.~A., {Holzer} T.~E., {Mihalas} D.~M., {Ulrich} R.~K., eds., 1986,
  {Physics of the sun. Volume 1 - The solar interior. Volume 2 The solar
  atmosphere. Volume 3 - Astrophysics and solar-terrestrial relations}, vol.~1

\bibitem[{Sunyaev} \& {Zeldovich}(1970)]{1970Ap&SS...7....3S}
{Sunyaev} R.~A., {Zeldovich} Y.~B., 1970, \apss, 7, 3

\bibitem[{Sutton} et~al.(2013){Sutton}, {Roberts} \&
  {Middleton}]{2013MNRAS.435.1758S}
{Sutton} A.~D., {Roberts} T.~P., {Middleton} M.~J., 2013, \mnras, 435, 1758

\bibitem[{Svensson}(1982)]{1982ApJ...258..335S}
{Svensson} R., 1982, \apj, 258, 335

\bibitem[{Svensson}(1984)]{1984MNRAS.209..175S}
{Svensson} R., 1984, \mnras, 209, 175

\bibitem[{Takahashi} et~al.(2016){Takahashi}, {Ohsuga}, {Kawashima} \&
  {Sekiguchi}]{2016ApJ...826...23T}
{Takahashi} H.~R., {Ohsuga} K., {Kawashima} T., {Sekiguchi} Y., 2016, \apj,
  826, 23

\bibitem[{Takahashi} et~al.(2013){Takahashi}, {Ohsuga}, {Sekiguchi}, {Inoue} \&
  {Tomida}]{2013ApJ...764..122T}
{Takahashi} H.~R., {Ohsuga} K., {Sekiguchi} Y., {Inoue} T., {Tomida} K., 2013,
  \apj, 764, 122

\bibitem[{Tchekhovskoy} et~al.(2012){Tchekhovskoy}, {McKinney} \&
  {Narayan}]{2012JPhCS.372a2040T}
{Tchekhovskoy} A., {McKinney} J.~C., {Narayan} R., 2012, Journal of Physics
  Conference Series, 372, 1, 012040

\bibitem[{Tchekhovskoy} et~al.(2011){Tchekhovskoy}, {Narayan} \&
  {McKinney}]{2011MNRAS.418L..79T}
{Tchekhovskoy} A., {Narayan} R., {McKinney} J.~C., 2011, \mnras, 418, L79

\bibitem[{Thorne}(1981)]{1981MNRAS.194..439T}
{Thorne} K.~S., 1981, \mnras, 194, 439

\bibitem[{Tsang} \& {Milosavljevi{\'c}}(2015)]{2015MNRAS.453.1108T}
{Tsang} B.~T.-H., {Milosavljevi{\'c}} M., 2015, \mnras, 453, 1108

\bibitem[{Uzdensky} \& {McKinney}(2011)]{2011PhPl...18d2105U}
{Uzdensky} D.~A., {McKinney} J.~C., 2011, Physics of Plasmas, 18, 4, 042105

\bibitem[{V{\"o}gler} et~al.(2004){V{\"o}gler}, {Bruls} \&
  {Sch{\"u}ssler}]{2004A&A...421..741V}
{V{\"o}gler} A., {Bruls} J.~H.~M.~J., {Sch{\"u}ssler} M., 2004, \aap, 421, 741

\bibitem[{Vurm} et~al.(2013){Vurm}, {Lyubarsky} \&
  {Piran}]{2013ApJ...764..143V}
{Vurm} I., {Lyubarsky} Y., {Piran} T., 2013, \apj, 764, 143

\bibitem[{Weldon}(1994)]{1994PhRvD..49.1579W}
{Weldon} H.~A., 1994, \prd, 49, 1579

\bibitem[{Xie} et~al.(2010){Xie}, {Nied{\'z}wiecki}, {Zdziarski} \&
  {Yuan}]{2010MNRAS.403..170X}
{Xie} F.-G., {Nied{\'z}wiecki} A., {Zdziarski} A.~A., {Yuan} F., 2010, \mnras,
  403, 170

\bibitem[{Yennie} et~al.(1961){Yennie}, {Frautschi} \&
  {Suura}]{1961AnPhy..13..379Y}
{Yennie} D.~R., {Frautschi} S.~C., {Suura} H., 1961, Annals of Physics, 13, 379

\end{thebibliography}
}
\end{document}